# Cycle-Based Computational Pipeline for Extracting Instantaneous Whisking Frequency

**Authors:** Guanghui Li[1*], Fangyuan Li[1*], Barbara Lykke Lind[1†], Rune W. Berg[1†]

**Affiliations:**

[1]Department of Neuroscience, Faculty of Health and Medical Science, University of Copenhagen, DK-2200, Copenhagen N, Denmark

*Authors contributed equally.

†Correspondence should be addressed to:

†Rune W. Berg, PhD.

Blegdamsvej, 3B, 2200 Copenhagen N, Denmark

E-mail: runeb@sund.ku.dk

†Barbara Lykke Lind, PhD.

Email: barbarall@sund.ku.dk

## Abstract

Whisking is a rhythmic and adaptive behavior that rodents use to probe and interact with their environment, and the frequency of movement reflects both sensorimotor processing and internal brain states. A robust and traditional method of whisker frequency estimation uses power spectral analysis of whisker position spanning several cycles. To improve temporal resolution of whisker movement, we here estimate the period for each cycle, hence indirectly extracting an instantaneous frequency. We do this uses markerless estimation of whisker position and identifying peak and trough for each cycle. The cycle period is extracted and artifacts are rejected with a ripple-exclusion validator based on peak prominence and sequential amplitude filtering. The method is compared with power spectral estimation, using Fourier transform of a temporal window of 0.5 seconds. We find that frequency estimation using a fixed window does not capture transient variability, while the cycle-by-cycle method recovers higher, time-resolved frequencies. The cycle-by-cycle approach also reveals the expected cycle-level variability. Artifact rejection through subsequence filtering removed spurious frequencies above 30 Hz, aligning refined frequencies with established physiological bounds (4–28 Hz). This pipeline provides an alternative solution for real-time compatible frequency estimation which better captures temporal variation at the expense of precision in frequency estimation.

## Introduction

Rodents, such as rats and mice, rhythmically sweep their mystacial whiskers back and forth at a frequency range of 5 to 25 Hertz (Hz) to effectively localize objects and discern textures in their environment [1-4]. This behavior is driven by an internally generated rhythm that persists in the absence of sensory reafference, consistent with activity of a central pattern generator (CPG) [5-8].

Whisker protraction is driven by intrinsic facial muscles, while retraction is driven by extrinsic facial muscles [9][10][11]. In free air, whisking typically oscillates at 5–15 Hz, with frequency increases during object palpation and adaptive sensing [10][12][18][19]. The sensory information gathered from the whiskers travel via trigeminal nerve's afferent fibers through the brainstem and the thalamus to layer IV of barrel cotex, whose columnar architerture mirrors individual whiskers [20].

Rodent whisker movements are high rhythmicity, but both the amplitude and frequency have cycle-to-cycle variability [21][22]. While displacement amplitude is widely used, it is sensitive to geometric factors (camera angle, whisker curvature, marker position), which can inflate inter-trial variability and complicate across sessions or subject's comparisons. In contrast, frequency offers a more robust and scale-invariant descriptor that more directly captures the underlying rhythmic dynamics of sensorimotor control. Frequency is also physiological meaningful: in the rodent barrel cortex, high frequency vibrational inputs can elicit prcise phase locked firing, indicating that temporal features such as frequency and phase are actively used by the brain to encode tactile information [23]. Thus, estimating whisking frequency with high temporal resolution is valuable both for behavioral quantification and for linking rhythmic sensory input to neural responses in the somatosensory system [24-28].

The frequency at which rodents whisk can be modulated by various factors, including the level of attention they are paying to a task, the difficulty of the task at hand, and the influence of neuromodulators within the brain [29][30]. Conventional method use the power spectrum to estimate the whisking frequency for each trial [5] [12-17][31]. Cite

The power spectrum can be estimated in multiple ways (periodogram, the FFT Welch's method with windowing, multitaper, or parametric AR models) [32] depending on data and goals. Any time frequency analysis based on finite windows faces a fundamental tradeoff between time and frequency precision (often framed by the Gabor uncertainly principle) [33]. Consequently, spectral estimates are inherently smeared over time, where stationarity is assumed, leading to a certain degree of temporal blur. These considerations motivate a complementary approach that provides scale-invariant, cycle-level instantaneous frequency estimates.

Conceptually, whisker motion can be treated as an oscillatory sequence. By tracking a single point on the whisker shaft with a markerless analysis software (DeepLabCut), we obtain an x,y trajectory whose time series exhibits an alternating Peak → Valley → Peak → Valley pattern. This pattern is analogous to a "wiggle sequence", a simplified oscillatory model in which the signal follows a repeating up–down–up–down trajectory. In this framework, each adjacent Peak–Valley pair represents half a whisking cycle; such identifying the time stamps of successive extrema is therefore sufficient to estimate the cycle period. Because frames are sampled at a constant frame rate, extrema indices map directly onto the physical time axis. The analytical task thus is reduced to converting the raw whisker trajectory into a validated wiggle sequence and extracting all extrema and their time stamps.

In this paper, we introduce a cycle by cycle (peak - valley) pipeline for estimating instantaneous whisking frequency at sub-cycle resolution. For comparison, we also estimate the power spectrum to extract the frequency. Our pipeline comprises robust extrema detection, cycle-to-cycle period extraction, and a ripple-exclusion validator based on peak prominence. This approach is well suited to experiments where cycle-level timing benefits downstream analyses and closed-loop control [34-37]. By complementing conventional power spectral methods with time resolved, cycle level estimates, we aim to provide a more nuanced description of whisking dynamics that is both biologically interpretable and practically useful.

# 2 Materials and Methods

## 2.1 Animals

A total of six adult male mice C57BL/6J (aged 10 to 16 weeks) were used. Animals were head-fixed atop an air-floating platform (Mobile HomeCage system, Neurotar, Helsinki, Finland), which permits voluntary locomotion while stabilizing the head for high-precision videography. Whisker movements were recorded at 200 frames per second (fps) with a high-speed camera (JAI Ltd., Copenhagen, Denmark). Recordings were acquired across over 30 trials from three mice. All data was preprocessed using DeepLabCut to extract x/y whisker coordinates. The subsequent analysis focuses on one dataset to evaluate each method. During the whisking trials, the mice were allowed to move freely while their whisker movements were recorded using high-speed videography. This allowed for detailed analysis of their whisking behavior, including the frequency, amplitude, and direction of whisker movements. Throughout the experiments, care was taken to minimize discomfort and stress for the mice. All experimental procedures adhered to the ARRIVE guidelines as well as the Council of the European Union regulations (86/609/EEC) and were authorized by the Danish Veterinary and Food Administration (Animal Research Permit No. 2024-15-0201-01739).

## 2.2 Video Acquisition and Whisker Tracking

High-speed infrared video (200 fps, 752 × 480 pixels, 0.15 mm pixel size) captured the right C-row. A point 5 mm distal to the whisker base was labeled in 200 frames and trained with open-source software (DeepLabCut 2.3) [38]. To assess labeling accuracy, a cross-validation test was performed, yielding a mean test error of 0.19 ± 0.06 pixels, indicating high spatial precision in whisker tracking. Then, the facial landmark was tracked with DeepLabCut, yielding framewise Cartesian coordinates $(x(t), y(t))$ and a frame index $coords$. Frames with a DeepLabCut likelihood of 0.9 or less (occurring less than 0.5% per session) were interpolated using a cubic-spline to preserve trajectory continuity. The labeled coordinates was subsequently converted to a polar angle using custom MATLAB scripts to preserve raw kinematics for the extraction of key features of whisker motion, including angular displacement and whisking frequency.

## 2.3 Angle construction

For a single point on a whisker, the polar angle is a scale-invariant description of whisker swing and correlates with protraction–retraction. It avoids amplitude issues due to camera zoom or marker placement. We convert the tracked pixel coordinates of one whisker point $(x(t), y(t))$ into a polar angle $\theta(t) \in [0, 360°]$ with the arctangent function and modulo operation:

$$\theta(t) = (\frac{180}{\pi} atan2(y(t), x(t))) mod 360°$$

The polar angle is converted directly from the raw Cartesian coordinates; this design intentionally avoids phase distortions and spectrum shaping, allowing both slow drifts and fast modulations to remain visible to the estimators.

## 2.4 Frequency Estimators

The whisking frequency was estimated in two independent ways: the Cycle-by-cycle frequency from detected cycles (valley–valley timing), and Sliding FFT pak, which is a conventional spectral tracking method in short, overlapping windows. For all analyses, the timebase was $t = \frac{coords}{f_s}$ .

### 2.4.1 Cycle-by-cycle (valley − valley)

To obtain a physically interpretable whisking rate as the inverse of the cycle period, we detected successive cycles on an unwrap-safe surrogate of the angle. First, we phase-unwrapped the angle in radians and mapped back to degrees to prevent wrap-around at $0°/360°$:

$$\varphi(t) = unwrap(\frac{\pi}{180^{\circ}}\theta(t)), \quad \tilde{\theta}(t) = \frac{180}{\pi}\varphi(t)$$

Valleys were detected on $\tilde{\theta}(t)$ with a robust prominence criterion and a minimum temporal spacing. We implemented a peak-valley detection algorithm that identifies rhythmic up–down patterns in the converted angular trajectory (Fig. 1). The algorithm traverses the signal by comparing adjacent values to identify inflection points, where the slope changes from positive to negative (peaks) or vice versa (valleys). This decision process is visualized as a finite-state machine with three states—start, up, and down—transitioning based on the relationship between consecutive points in the signal (nums[i-1], nums[i], and nums[i+1]). The signal begins in the start state and transitions to either an up or down state depending on whether the trend is rising or falling. From the up state, a transition to down occurs when the slope inverts, signaling a peak. Conversely, valleys are detected when transitioning from down to up. These state transitions form the basis of detecting complete whisking cycles, defined from one valley (or peak) to the next of the same kind. Additionally, a prominence threshold was used to capture the max function and includes the necessary symbols for prominence floor (*promFloor*), prominence fraction (*promFrac*), and interquartile range (*IQR*), identifying and correcting minor deviations and enabling real-time monitoring during processing:

$$prom = max(promFloor, promFrac \cdot IQR(\tilde{\theta}))$$

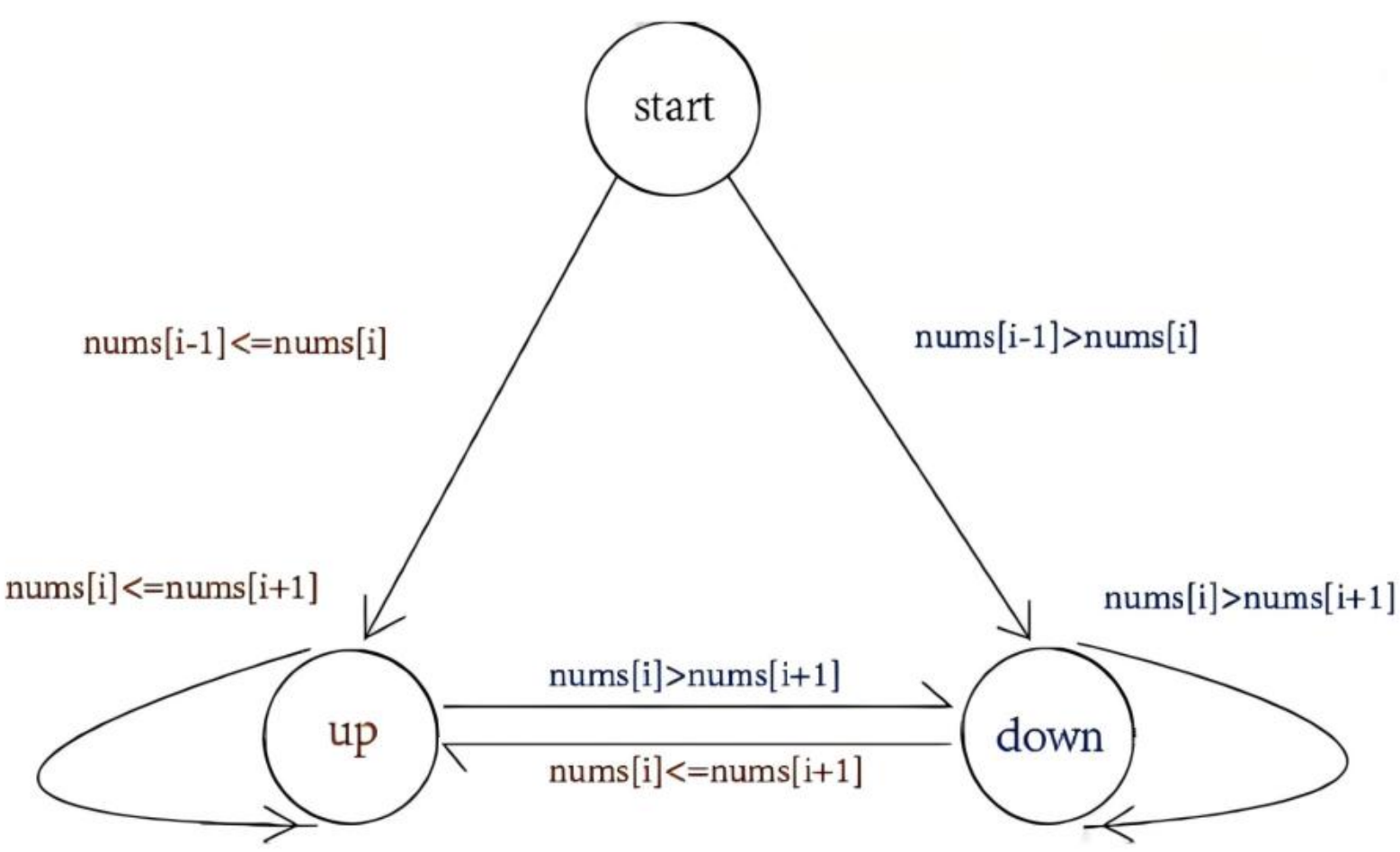

**Figure 1. Cycle detection.** Cyele detection logic using a finite-state machine. Transitions between start, up, and dow states are triggered by slope changes in the signal. Peaks are identified when the signal changes from increasing to decreasing (up - dow), and valleys when changing from decreasing to increasing (down - up).

Where promFloor (deg) prevents collapse in very quiet segments and promFrac scales the threshold by the signal’s interquartile range, improving robustness to amplitude changes across trials. The minimum peak distance was used to calculate the minimum distance ($minDist$) in samples which guards against double-counting within a single whisk:

$$minDist(samples) = max(1, \lfloor f_s \cdot \frac{minDist_{_ms}}{1000} \rfloor)$$

To estimate whisking frequency on a cycle-by-cycle basis, we first detect consecutive whisking events, typically valleys or peaks in the angle trace. When successive valleys occurred at $t_i$, cycle frequency was defined as:

$$f_i = \frac{1}{t_{i+1} - t_i}$$

The frequency is assigned to the midpoint $t_i + \frac{(t_{i+1} - t_i)}{2}$. This estimator directly reflects the kinematic rhythm (protraction-retraction), remains invariant to baseline drifts, and is readily interpretable as the fundamental whisking frequency.

### 2.4.2 Sliding FFT-derived peak

As a conventional spectrum-based comparator, we estimated a time-resolved dominant frequency within short, overlapping windows. To retain the fundamental modulation, we analyze the wrap-safe projection:

$$x_\theta(t) = cos\ (\theta(t))$$

In each window of duration winSec (samples $N_w = \lfloor f_s \cdot \text{winSec} \rfloor$ ) with fractional overlap, we linearly detrended the segment, applied a Hann window ($energy\ E_s = \sum w^2$), and computed a single-segment power spectrum:

$$P(f) = \frac{|FFT(w \cdot x_\theta)|^2}{E_s f_s}$$

Where P(f)resprents the FFT product of the Hann window w and $x_\theta$, normalized by the energy $E_s$ sampling frequency $f_s$.

We searched for the spectral global peak over $0 < f \le min(fftMaxHz, f_s/2)$ , the condition on $f$ being between 0 and the minimum of fftMaxHz and half of the sampling frequency $f_s$. The peak frequency was refined by parabolic interpolation using the peak bin and its two neighbors, yielding sub-bin resolution:

$$\delta = \frac{1}{2} \cdot \frac{P_{k-1} - P_{k+1}}{P_{k-1} - 2P_k + P_{k+1}},\quad f_{peak} = f_k + \delta \Delta f,\quad \Delta f = \frac{f_s}{N_w}$$

We intentionally avoid band-pass filtering to prevent phase distortions and to maintain an accurate spectrum when the animal transiently speeds up or slows down, or when the waveform becomes non-sinusoidal. Quality gates (based on variance and signal to noise ratio) are applied to each window as an explicit confidence criterion. Instead of generating artificial frequencies, more windows will be rejected to avoid spurious estimates when motion is weak or noisy.

# 3 Results

## 3.1 Whisker Tracking

For every mouse, we created an independent DeeplabCut project and trained a dedicated ResNet-50 network to track three target points (one on each of three whiskers) and a single stationary reference marker. One hundred frames were manually labelled in each of ten pre-cropped 500 fps videos and 95 % of the annotated frames were reserved for training and the remaining 5 % for validation. During inference, only detections with a likelihood ≥ 0.6 were accepted. Training one model per animal eliminated cross-subject variability in fur texture and whisker geometry, yielding lightweight, high-precision trackers that extracted frame-by-frame whisker trajectories for subsequent analysis.

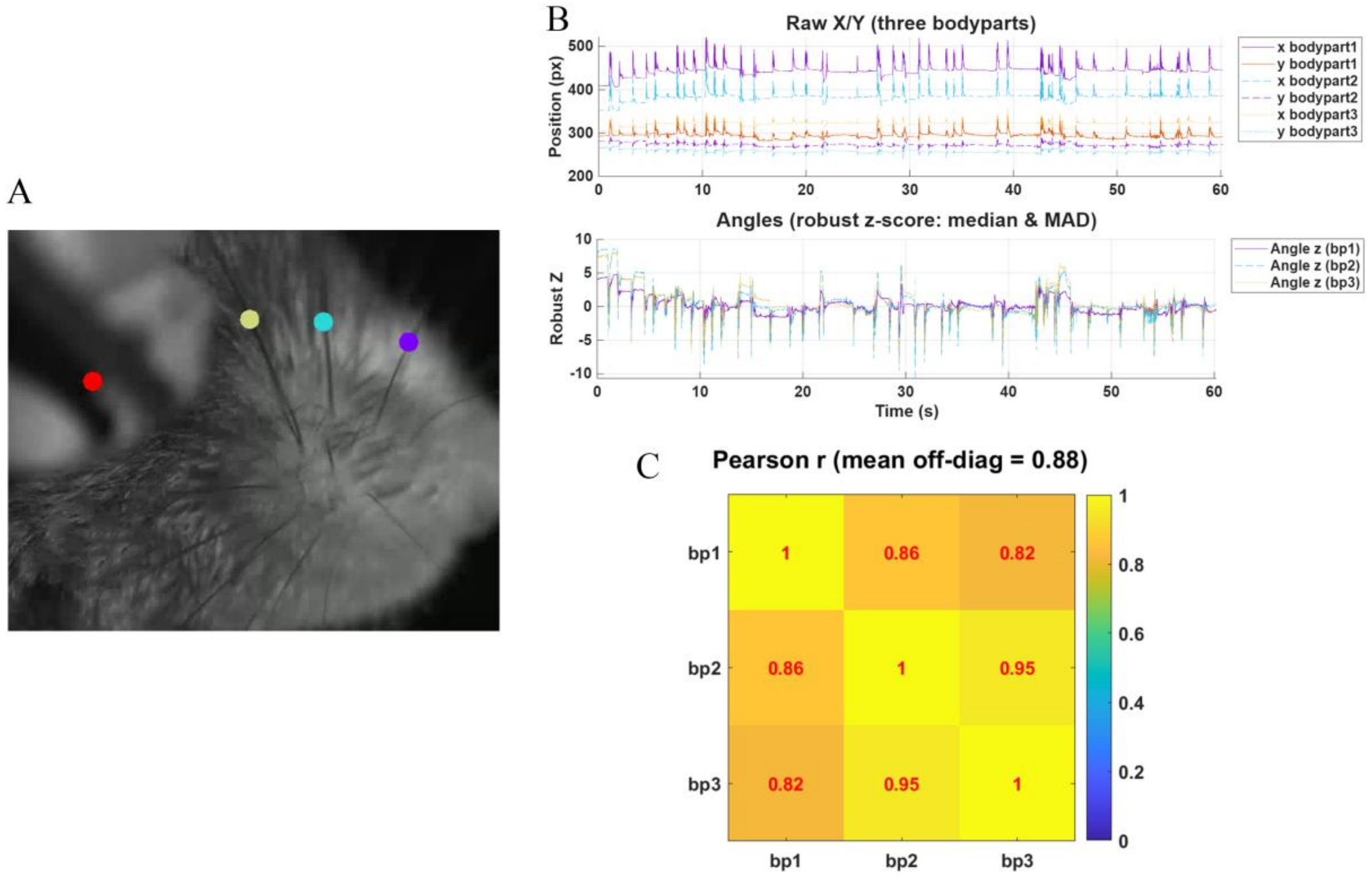

**Figure 2. Integrating the labelling strategy with the resulting kinematic read out.** A raw video frame with the reference marker (red) and the three whisker points (blue, cyan, yellow)super-imposed by the trained network (A). The oscillations that are nearly coincident in the plotsof the filtered vertical (x-y) displacements and the of the three whiskers over 10,000 frames (B) show strong temporal synchrony in their polar angle traces (C).

The customised DeeplabCut pipeline tracked three individual whiskers with sub-pixel accuracy across high-speed recordings (Fig. 2A), converting them to angles and robust-normalizing those angles using median and MAD (robust z-scores) to suppress offsets and outliers, then computes and plots time-aligned raw trajectories, normalized angles (Fig. 2B), and the results showing tightly co-fluctuating angle traces across bodyparts and a high correlation matrix (r ≈ 0.82–0.95; mean ≈ 0.88), indicating strong temporal synchrony of the three whiskers throughout the recording (Fig. 2C). Such coherence is expected because protraction and retraction are driven by the same intrinsic and extrinsic facial muscles, and it demonstrates that the trained models capture whisker motion accurately and consistently across extended high-speed recordings. Those results confirm both the biological expectation of coordinated whisking and the technical reliability of the animal-specific models, providing a robust foundation for subsequent frequency- and phase-based analyses.

## 3.2 Comparison of Algorithms

### 3.2.1 Single-whisker data used for algorithm benchmarking

Because the three tracked whiskers oscillate in near-perfect synchrony, subsequent algorithm comparisons were performed on a single representative whisker to avoid redundancy and reduce computational load.

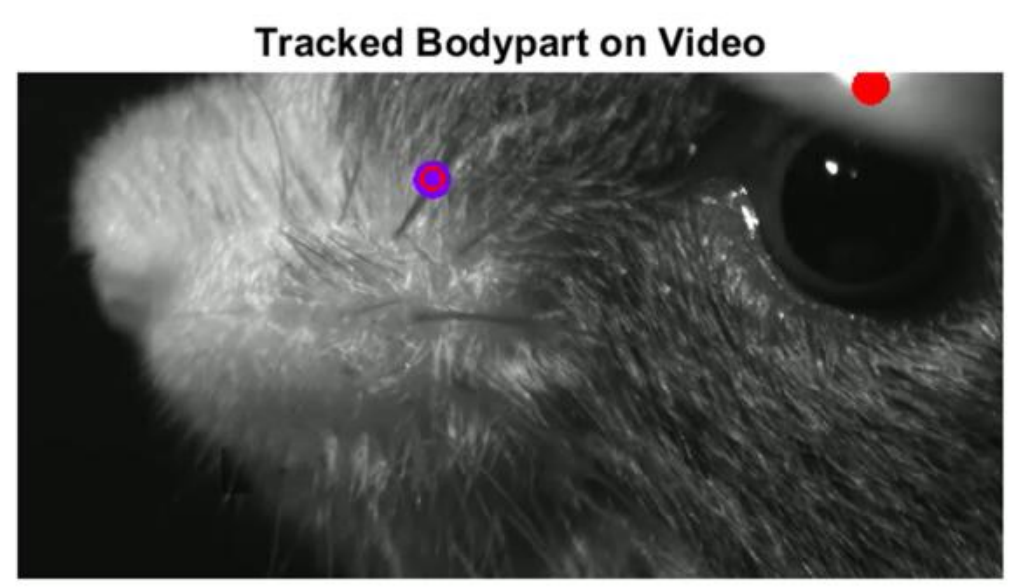

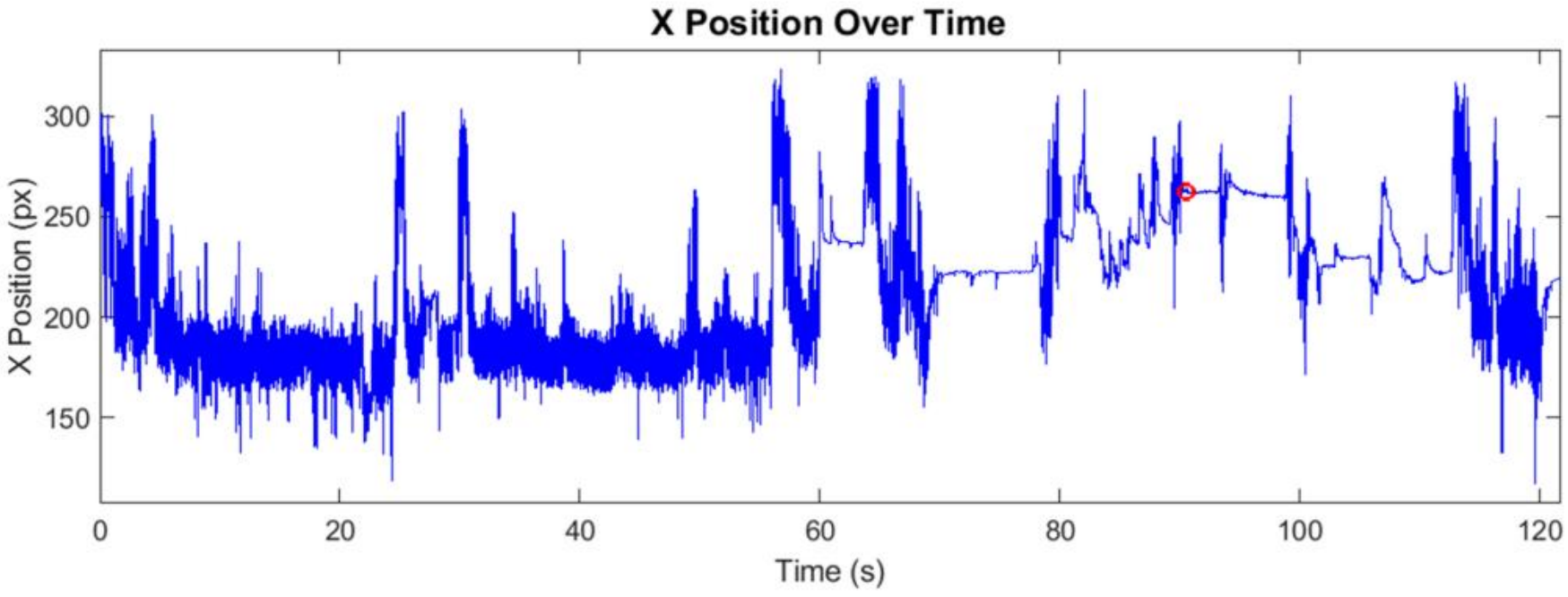

**Figure 3. Single-whisker benchmark dataset.** Top: a video frame showing the tracked whisker point (blue) and the fixed reference marker (red). Bottom: the whisker's horizontal position over time, highlighting clear peak-to-valley excursions used for algorithm benchmarking.

The data stream used in these tests is shown in Fig. 3, computed from the DeeplabCut coordinates of one whisker point together with the corresponding 200-fps video. In a tiled layout it renders, frame-by-frame, the raw video (upper panel) with the whisker point (blue) and reference marker (red) super-imposed, while simultaneously plotting the whisker's horizontal displacement x(t) over a 120-s segment (lower panel). A red cursor advances synchronously on both panes, producing an annotated MP4 file for visual verification. The pronounced peak-to-valley excursions in the x-trace make cycle boundaries unambiguous, providing a clean ground-truth signal for comparing frequency-extraction algorithms.

### 3.2.2 Whisking frequency estimation

A comprehensive analysis of whisker movement dynamics by extracting whisking frequencies from position-tracking data using the complementary signal processing techniques: peak-valley and FFT peak power spectrum detection. The pipeline begins by importing the whisker trajectory data—originally output from DeepLabCut (Fig. 4A). We unwrap $\theta(t)$ in radians to remove $0/360°$ discontinuities (Fig. 4B) and detect valleys on the unwrapped surrogate (robust to wrap jumps). Valleys are required to be at least 30 ms apart to reject implausibly short cycles at 200 Hz sampling. A robust amplitude criterion is enforced using a prominence threshold of max(0.50°, 0.50 × IQR) degrees, so only cycles with sufficient excursion are counted. Instantaneous frequency is simply $f_{cyc} = \frac{1}{\Delta t}$ between consecutive valleys and is plotted at the mid-time of each interval. The estimated frequency from cycle-by-cycle method during periods of clear whisking the frequency clusters in the 10 – 20 Hz range, while pauses or small-amplitude motion (60 – 100 s) are correctly blank or near-zero because no robust valleys pass the prominence rule (Fig. 4C).

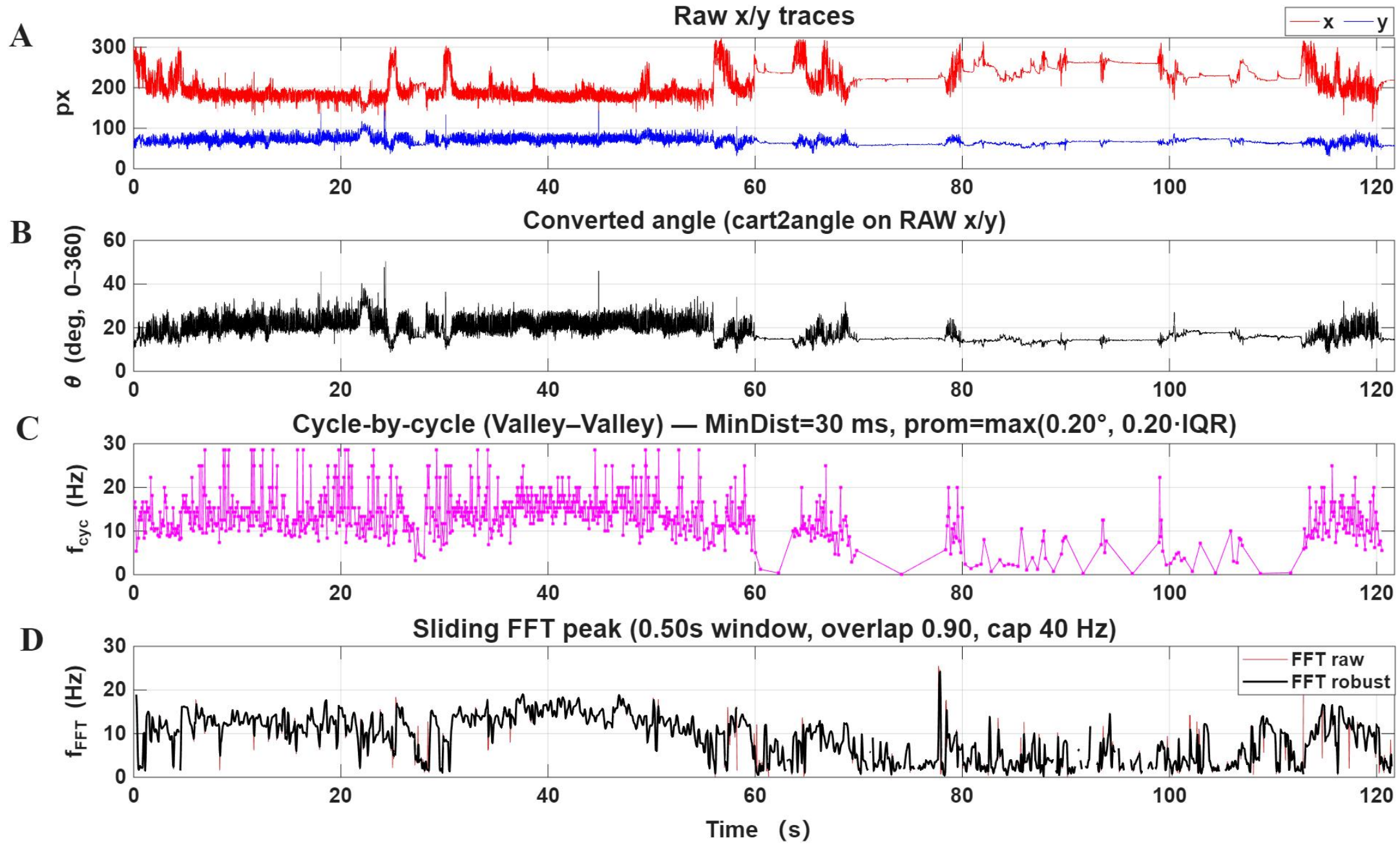

**Figure 4. Multi-method estimation of whisking frequency from tracked whisker motion.** Whisker angle was extracted from the raw 2D coordinates (5 – 30 Hz). (A) Raw x and y coordinates of the tracked whisker tip. (B) Converted polar angle of whisker orientation. (C) Cycle-by-cycle frequency computed from time intervals between detected valleys in the angular signal. (D) Frequency estimated by identifying the peak frequency in a sliding FFT window (0.5 s), representing the dominant spectral component over time.

The sliding FFT peak frequency was computed as an comparator and sampled at window centers. To make the spectrum invariant to $0/360°$ wrap and small drifts, we analyse the wrap-safe projection $x_\theta(t) = cos\,(\theta(t))$. Windows of 0.5 s with 90% overlap are detrended and Hann-tapered prior to an FFT. Within each window we select the global spectral peak below 40 Hz. To suppress spurious peaks in very low-motion epochs three guards were applied: a variance floor in the cos-domain ($1 \times 10^{-7}$) below which a window is ignored; an SNR gate requiring the peak power to exceed 3x the window's median power; a 0.10 s median filter, a max jump of 10 Hz per window, and a spike reject of 5 × MAD light temporal regularization was applied. Short gaps produced by these guards are optionally linearly filled over at most 2 windows. As shown in Fig. 4D, the sliding FFT peak provides a continuous trend that broadly agrees with the cycle metric (~10 − 15 Hz baseline, brief accelerations and slowdowns), it's intentionally suppressed in very quiet windows by the variance/SNR gates, which prevents the spurious high-frequency "overshoots" that simple peak-picking can generate when the signal is flat or dominated by noise. Small residual fluctuations are smoothed by the 0.10 s median and clipped by the 10 Hz max-jump rule (FFT raw vs FFT robust ).

### 3.2.3 Comparison of estimators

To quantify consistency between our two frequency estimators, we paired samples from the cycle-by-cycle (valley–valley) detector and the sliding FFT peak and computed agreement statistics. The time overlay of both methods follows similar low-frequency trends but differ in temporal resolution and sensitivity to signal features. As shown in Fig. 5A, the FFT trace is smoother due to window averaging and contains short gaps where variance/SNR gates rejected low quality windows (these were filled only if the gap was less than or equal to 2 windows). By contrast, the cycle-by-cycle series produces sparse or no estimates in very low-amplitude epochs (≈90 − 105 s) when valleys fail the prominence rule, yet preserves rapid excursions during fast bouts (e.g., 70 - 80 s). These results indicate that the cycle-by-cycle method is better suited for temporally precise tracking of whisking bouts and frequency dynamics, whereas the sliding FFT yields a smoother and more averaged trajectory due to its windowing nature.

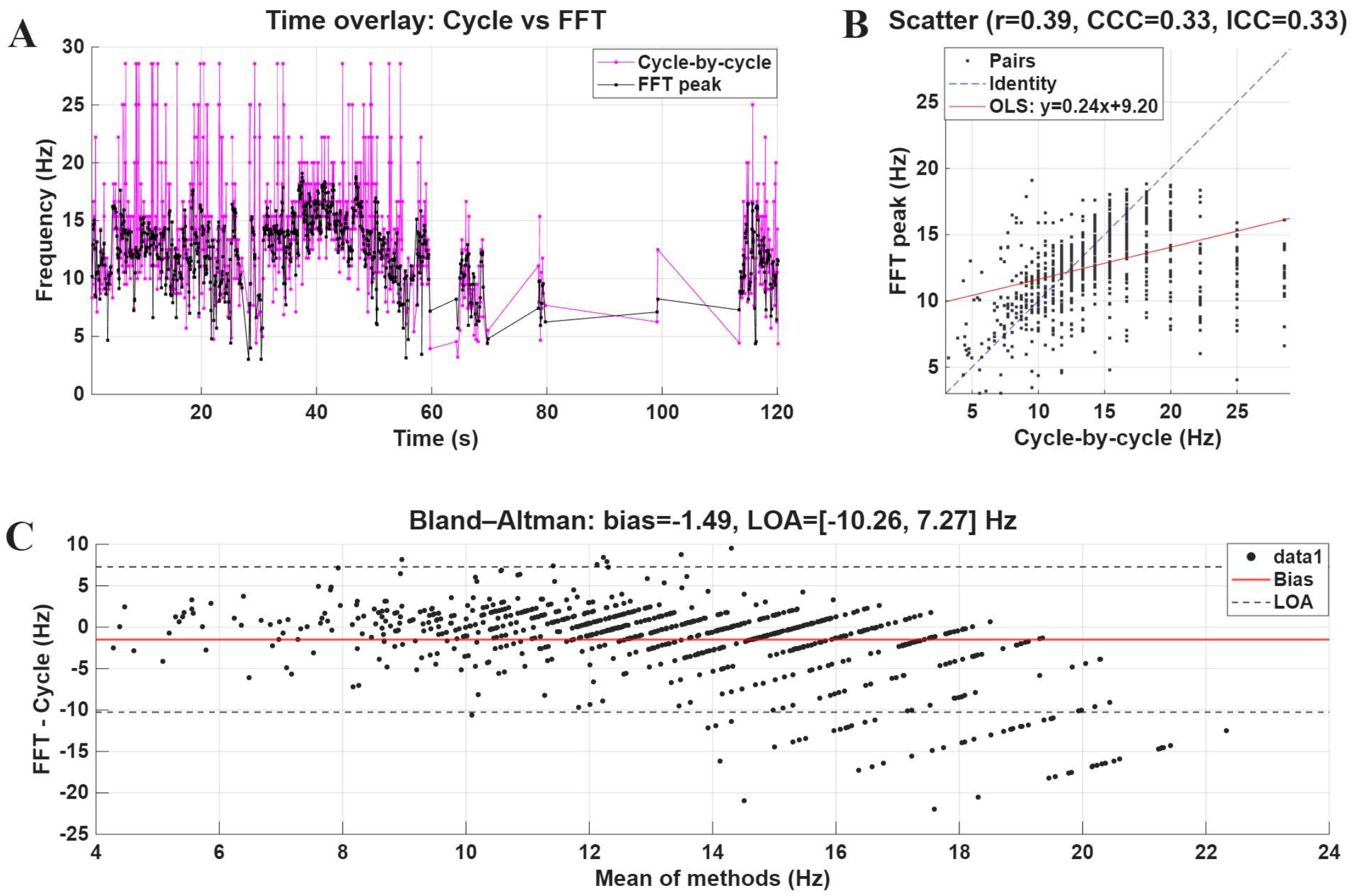

**Figure 5. Comparison between cycle-by-cycle and sliding FFT (0.5 s window) whisking frequency estimators.** Both methods exhibiting similar trends but with noticeable differences in their temporal resolution. (A) Time-series comparison shows that the sliding FFT trace is smoother but temporally coarser due to 0.5-s windowing, whereas the cycle-by-cycle method captures sharper transient changes during fast bouts but fails to estimate frequencies in low-amplitude periods. (B) Paired estimates reveal a moderate correlation (r = 0.39) but poor agreement (Lin'CCC = 0.33; ICC(2,1) = 0.33), with a regression slope < 1 indicating FFT underestimates frequency relative to cycle-based estimates. (C) Bland – Altman plot confirms a systematic bias of – 1.49 Hz (FFT lower), with wide limits of agreement ( – 10.26 to 7.27 Hz), suggesting the two methods are not interchangeable.

Further statistics with Pearson Correlation and Ordinary Least Squares (OLS) regression from Lin's concordance correlation coefficient (CCC; accuracy × precision), ICC (2,1)) corroborate these qualitative differences (Fig. 5B). Paired values show a moderate association (Pearson r = 0.39) between those two estimators indicate a poor concordance, and the Lin's CCC = 0.33 and ICC (2,1) = 0.33 indicate the two estimates are not interchangeable. The OLS line (red; FFT≈0.24×Cycle＋9.20) lies below the identity (blue), revealing a proportional bias: the FFT estimator tends to cluster around 9–15 Hz and increase only weakly as cycle-based frequency rises, suggesting that the FFT method tends to produce lower frequency values compared to the cycle-by-cycle method. Moreover, the average difference is bias = −1.49 Hz (FFT minus Cycle), meaning the FFT method reads ~1.5 Hz lower on average. The 95% limits of agreement are wide (−10.26 to 7.27 Hz), confirming that FFT underestimates the cycle-by-cycle frequency on average (Fig. 5C).

These behaviors are mechanistically expected. The FFT under-reads because as applying a 0.5-s windowing smooths rapid frequency excursions and when spectra are broad/multimodal, low variance windows or occasionally higher will be removed when a harmonic dominates a window and fall below the true cycle rate. The downward trend of differences with increasing mean frequency suggests proportional bias increases at higher whisking rates, consistent with the scatter plot. Conversely, cycle-by-cycle can over-read if the detector locks onto sub-cycles during rapid, asymmetric motions or when small inflections exceed the prominence threshold. However, this issue can be mitigated by raising prominence at the expense of sensitivity.

## 3.3 Signal Validation and Artifact Rejection for the Cycle – Cycle Whisking Frequency Estimator

### 3.3.1 Artifact identification

To further investigate the origin of abnormally high-frequency components, particularly those exceeding 30 Hz, we computed the amplitude spectrum of the angular trace over a 0 – 50 Hz range (Fig. 6). The FFT amplitude spectrum revealed a dominant frequency with a primary cluster of energy distributed below 25 Hz. The spectrum below 4 Hz may be caused by body shifts, Respiration-related head motion (typically 1–2 Hz), or baseline drift of the whiskers rather than rhythmic whisking. These components are expected in freely moving animals and are especially prominent during behavioral pauses. Meanwhile, for small amplitude components above 30 Hz, given their low amplitude and lack of concentration around specific whisking-related harmonics, these high-frequency components are likely arised from noise sources (Fig. 6C). These observations strongly suggest that the frequencies above 30 Hz are not physiologically driven, but rather artifactual contributions such as small ripples, signal noise, or instability introduced by airflow or camera jitter during recording.

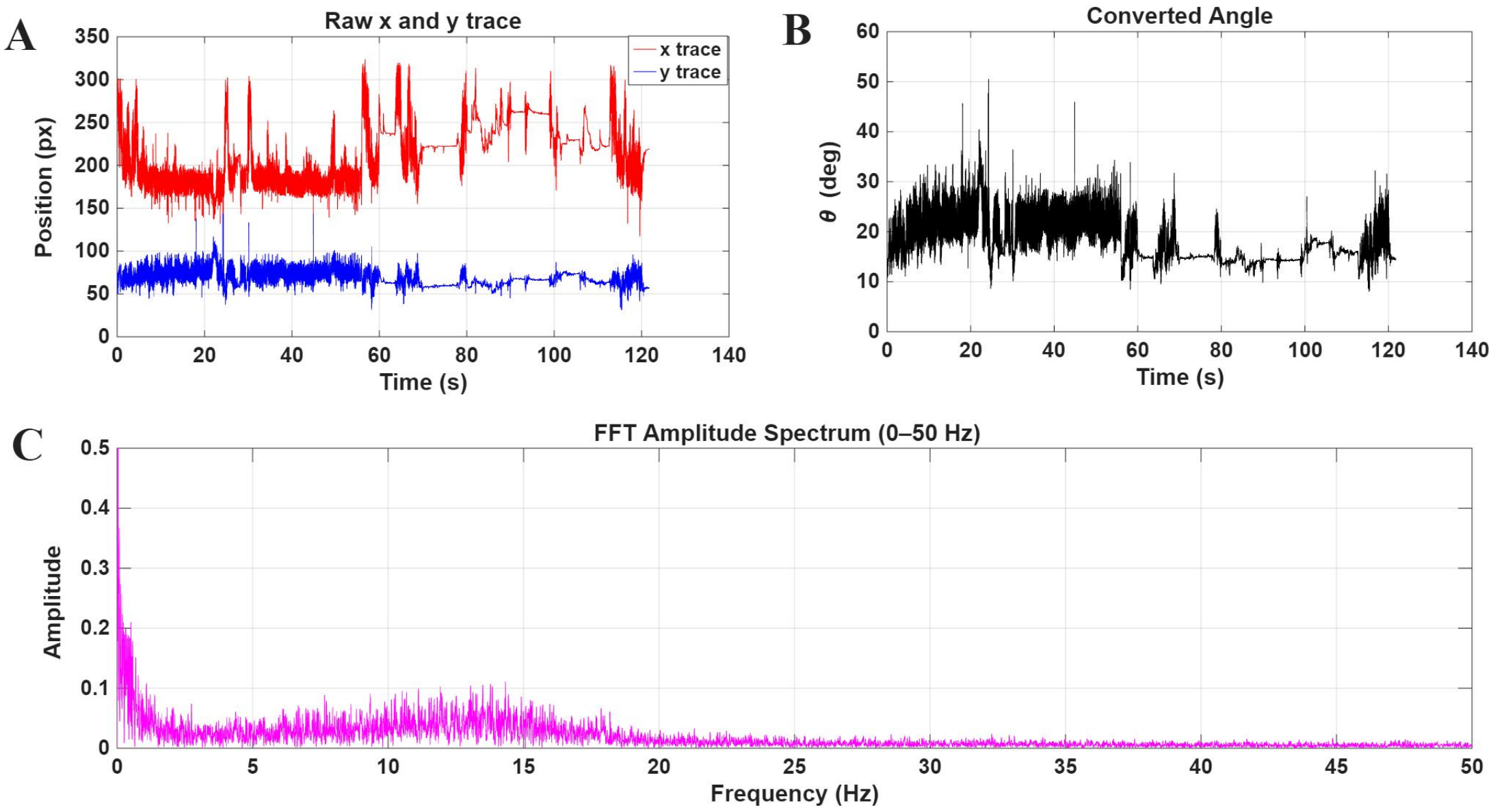

**Figure 6. Spectral evidence for excluding non-biological high-frequency whisking events.** Spectral analysis using FFT methods on whisker angle signals were examined to investigate the unexpectedly high-frequency (>30 Hz) estimates. (A and B) The raw coordinates of the tracked whisker tip and corresponding angular trajectory show bursts of motion and signal noise interspersed with clear whisking bouts. (C) The FFT spectra reveals dominant peaks near 4 Hz and 25 Hz, with sub-4 Hz components likely reflecting non-whisking activity such as head motion or baseline drift, high-frequency content above 30 Hz is minimal and likely non-physiological, arising from noise or recording artifacts.

### 3.3.2 Noise suppression via wiggle subsequence filtering

Given that frequency estimates above 30 Hz are not physiologically supported by the underlying whisker signal and are likely artifacts, we further validated the Valley-by-Valley frequency estimator by systematically excluding low amplitude cycles using prominence-based filtering.

The whisk_valley_frequency function was applied to the polar angle derived from the Cartesian coordinates via the cart2angle method. To ensure robustness against noise, a dynamic prominence threshold is applied based on the median absolute deviation (MAD), which suppresses minor fluctuations retaining genuine whisking cycles. Each of the upper plots in Fig. 7A-D shows the angular trace with detected valleys, while the bottom plots display the corresponding cycle-by-cycle frequency estimates computed from valley-to-valley intervals.

To explore the impact of parameter tuning, we applied this method to high-resolution (200 Hz) tracking data using a minimum inter-valley distance (minDist_ms) of 10 ms and systematically varied the promFloor parameter (1.0°, 0.5°, 0.3°, 0.1°), which sets the minimum amplitude required for a valley to be considered valid. Additionally, promFrac, a secondary threshold based on a fraction of the interquartile range, was used to scale the required prominence relative to signal variability. Which ensures that only significant peaks and valleys are considered in the analysis, reducing the impact of noise or minor fluctuations in the whisking signal.

As shown in Fig. 7A, applying a strict promFloor of 1.0o results in conservative valley detection, retaining only large, well-separated cycles. This yields a stable and low-variance frequency trace while effectively suppressing high-frequency artifacts. However, it may underestimate true whisking rates during fast, subtle movements due to exclusion of small-amplitude excursions. When the promFloor is relaxed to 0.5o (Fig. 7B), the algorithm becomes more sensitive to finer whisking dynamics. This increases the number of detected cycles, allowing more high-frequency bouts to be captured. A similar trend is observed with promFloor set to 0.3o (Fig. 7C), which further increases temporal resolution and variability. However, this also begins to introduce more high frequency estimates, as minor fluctuations due to small ripples or video tracking noise are now classified as valid cycles.

At the most permissive threshold of 0.1° (Fig. 7D), the cycle-by-cycle frequency trace becomes densely populated, with numerous estimates exceeding 30 Hz. While this might suggest highly dynamic whisking, these detections are more likely spurious.

The proliferation of high-frequency values, particularly those beyond biologically plausible ranges, supports our earlier conclusion that such components arise from non-physiological sources. These include mechanical vibrations, sub-pixel jitter, or small ripples in the angle trace not indicative of true whisking cycles. This comes with the trade-off of high sensitivity to noise, making the frequency estimates less reliable.

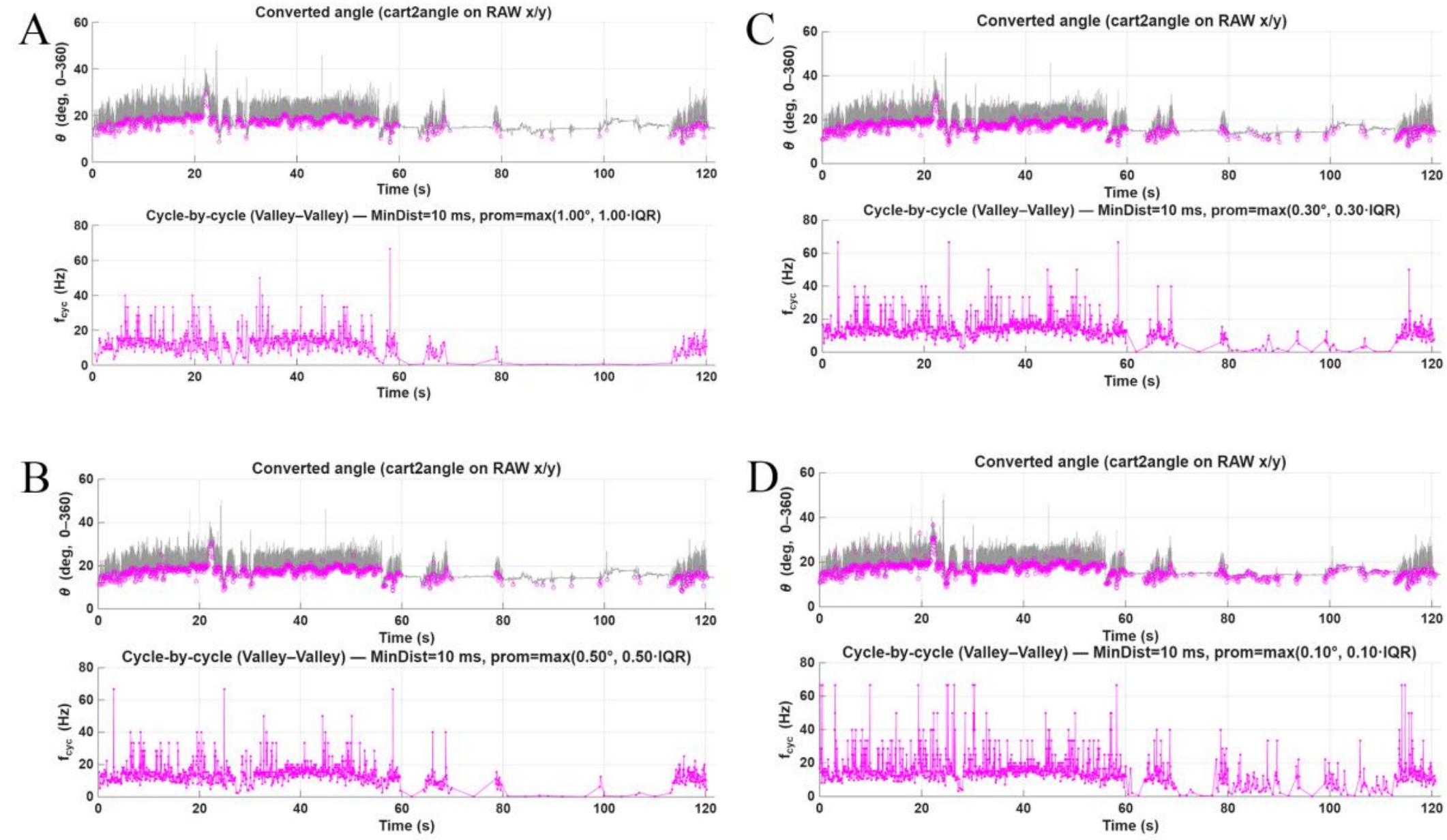

**Figure 7. Frequency extraction under different valley prominences. cycle vs FFT.** A multi-stage ripple-rejection strategy akin to peak/valley parsing with tunable thresholds: starting from stringent prominence (prom = max(1.0o, 1.0·IQR)) that removes low-amplitude oscillations (A); moderately permissive (prom = max(0.5o, 0.5·IQR)) recovers more cycles, revealing faster epochs while maintaining reasonable noise rejection (B); further relaxed criteria (prom = max(0.3o, 0.3·IQR)) increase sensitivity to small-amplitude movements, high-frequency content becomes more apparent, but spurious detections begin to appear in low-SNR segments (C); a highly permissive "ripple-passing" condition (prom = max(0.1o, 0.1·IQR)) admits many micro-deflections and the frequency trace shows dense, high-rate estimates but with reduced physiological interpretability due to noise-driven cycles (D).

The parameter sweep confirms the different parameterizations allow for a trade-off between detection sensitivity and noise in the whisker angle analysis. Loosening prominence criteria allows more high-frequency detections, but at the cost of reduced signal reliability. The detection of >30 Hz events under minimal promFloor settings convinces the assumption that these components are not behavioral but rather analytical artifacts stemming from noise sensitivity. Importantly, these results underline the importance of carefully tuning analysis parameters, particularly the prominence floor and prominence fraction, to optimize the trade-off between sensitivity and accuracy in cycle-by-cycle frequency extraction for faithful characterization of rapid whisking dynamics during sensorimotor behavior.

## 4 Discussion

In this study, we present a cycle-by-cycle frequency extraction pipeline for analyzing whisking dynamics in rodents, combining robust extrema detection and signal validation to estimate instantaneous whisking frequency. By comparing the sliding-window peak-FFT procedure against cycle-based estimates, we evaluated their ability to capture the temporal structure of whisking, resolve rapid dynamics, and reject artifacts. Each method offers unique advantages, but also trade-offs in temporal resolution, interpretability and sensitivity to noise.

The sliding FFT peak spectrum method is commonly used in behavioral neuroscience and yields a relatively narrow and distribution stable frequency at 6－12 Hz. However, its time resolution is restricted by the window size and is damping the rapid transitions in whisking frequency. This was most apparent during brief accelerations or pauses in whisking, where frequency estimates lagged or were omitted entirely due to low variance or signal-to-noise ratio (SNR) gating. Although such smoothing prevents high-frequency noise, it may miss biologically clues in transient dynamics.

In contrast, the cycle-by-cycle method produced a more temporally resolved frequency trace at the cost of a less well-defined estimation of frequency. This approach yielded higher median frequencies (~14 Hz) and wider interquartile ranges, reflecting the natural variability in rodent whisking. Importantly, its design permits sub-cycle resolution. The method identifies successive extrema (peaks and valleys) in the whisking trajectory, allowing the estimation of instantaneous frequency from both peak–peak (or valley–valley) intervals and adjacent peak–valley intervals. However, the high sensitivity of the method also makes it susceptible to spurious detections caused by micro-movements or signal ripples, particularly in low-amplitude segments.

Spectral validation by FFT amplitude spectra showed that frequencies exceeding 30 Hz, these high-frequency outliers likely stem from mechanical artifacts such as camera jitter, airflow, or tracking noise. To mitigate these effects, we implemented prominence-based ripple rejection in our cycle detection algorithm. By applying adaptive prominence thresholds scaled to the interquartile range of the signal, we selectively suppressed low-amplitude fluctuations without eliminating genuine whisking cycles. This parameterization effectively removed spurious cycles and narrowed the frequency distribution, thereby improving the biological validity of the output. Our findings emphasize the importance of careful data curation in high-resolution behavioral analyses. Without appropriate filters, temporally precise estimators risk interpreting noise as signal. Conversely, over-filtering can remove subtle but genuine variations in behavior. Our approach offers a customizable balance between sensitivity and specificity, enabling researchers to tailor the analysis to their data quality and experimental demands.

Furthermore, instantaneous whisking frequency is more than a behavioral metric, it reflects the dynamic interplay between motor output, sensory input, and brain state. Experiments using artificial whisking show that cortical whisking-related responses are strongly influenced by brain state and neuromodulatory tone, with cholinergic and noradrenergic activation altering barrel cortex activity patterns [39][40]. Computational models further indicate that whisker movement dynamics reflect shifts in spatial attention [41], while closed-loop modeling predicts how behavior and sensory feedback jointly regulate neural gain during whisking [42]. Hence, capturing instantaneous fluctuations in whisking frequency is essential to align behavior with neural dynamics in temporally precise, which is essential for causality aware analysis and closed-loop experimental designs.

Our study also highlights the importance of data cleaning in frequency-domain analysis. Without filtering out small ripples and micro-artifacts, our method will inadvertently detect false positives that bias downstream analyses. By embedding a rigorous cycle validator that combines amplitude filtering and prominence constraints, we address this challenge systematically. Moreover, the modular design of our MATLAB-based functions (whisk_valley_frequency) enables researchers to integrate and adapt components of the pipeline to suit their own data structures and experimental constraints.

Finally, this study bridges the gap between behavioral dynamics and signal processing by offering a framework that preserves both spectral and temporal information in whisking behavior. Importantly, the pipeline not only produces quantitative outputs (instantaneous frequency and amplitude) but also generates graphical validation outputs that allow for intuitive, frame-by-frame inspection. This dual layer of quantitative and visual assessment ensures scientific rigor and promotes reproducibility. By validating and comparing this pipeline across multiple estimation approaches, we offer a comprehensive framework for whisking analysis that balances temporal fidelity, biological relevance, and computational efficiency. This has broad implications for the field of systems neuroscience, particularly in studies seeking to understand how self-generated behaviors like whisking modulate cortical processing, attention and learning.

# 5 Conclusion

We present a novel, cycle-resolved pipeline for extracting instantaneous whisking frequency from rodent behavioral data by combining DeepLabCut-based tracking with cycle-by-cycle frequency analysis. We compared this method with the sliding FFT peak spectrum frequency, the sliding FFT method while computationally efficient, smooths over fluctuations but is less sensitive to high-frequency events, which may miss rapid cycles that are detected by the cycle-by-cycle method. In contrast, cycle detection yields higher-resolution frequency estimates that better capture the inherent variability and temporal dynamics of whisking behavior. The implementation of artifact rejection through ripple filtering and wiggle sequence validation, we ensured that frequency estimates remained within physiologically plausible bounds.

Overall, this method offers a robust, scalable, and open-source solution for fine-grained behavioral analysis, with direct applicability to real-time and closed-loop neuroscience experiments. In doing so, it provides a robust foundation for future investigations into the neural correlates of active sensing and the sensorimotor transformations underlying tactile exploration in rodents.

**Funding**

This work was partly supported by the Lundbeck Foundation (Grant No. R366-2021-233); and the Chinese Scholarship Council (Grant No. 202009110098).

**Ethics declaration**

All experimental procedures followed ARRIVE guidelines as well as Council of theEuropean Union regulations (86/609/EEC) and were authorized by the Danish Veterinary and Food Administration (Animal Research Permit No. 2024-15-0201-01739).

**Declaration of competing interest**

The Authors confirm that there are no conflicts of interest.

**Data availability**

MATLAB 2025a was utilized for data processing. All source data generated and analyzed, as well as the script used for data processing in this study, are available upon request to the corresponding author.

# Supplementary material

**I.** To further assess the reliability of DeepLabCut based tracking, nine datasets from the same animal were manually labeled and used to train separate DLC models for each trial. This approach allowed us to evaluate the consistency of tracking results across independently trained models derived from the same underlying behavior.

Each of the following nine plots (Figure A-I) corresponds to an independent recording trial of whisker tracking data extracted using DeepLabCut (with three whiskers).

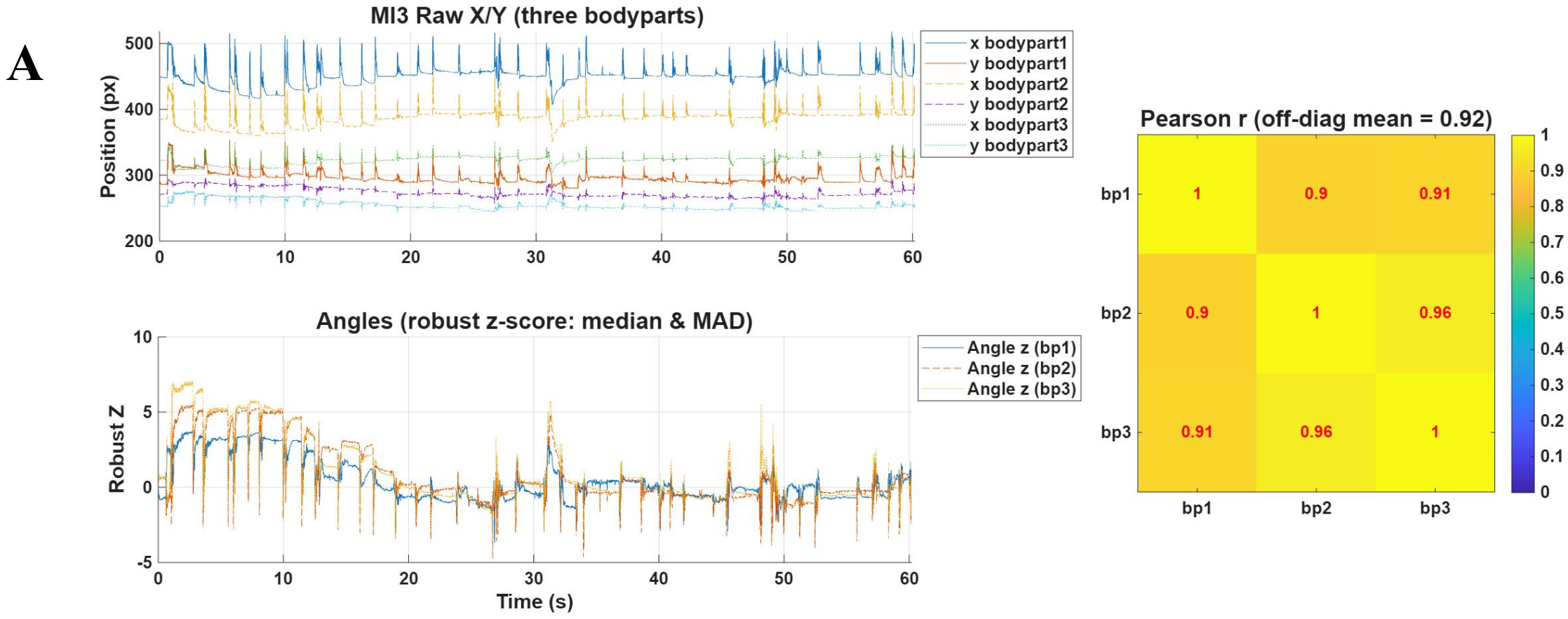

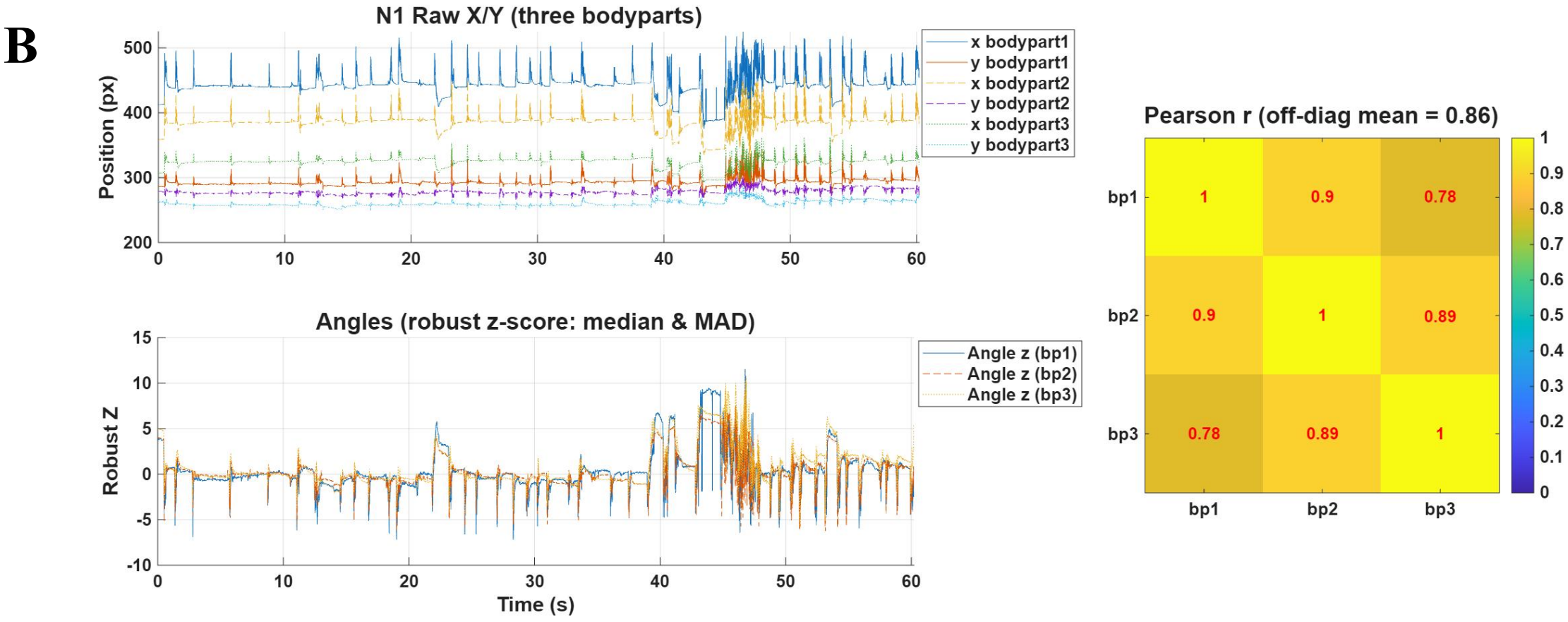

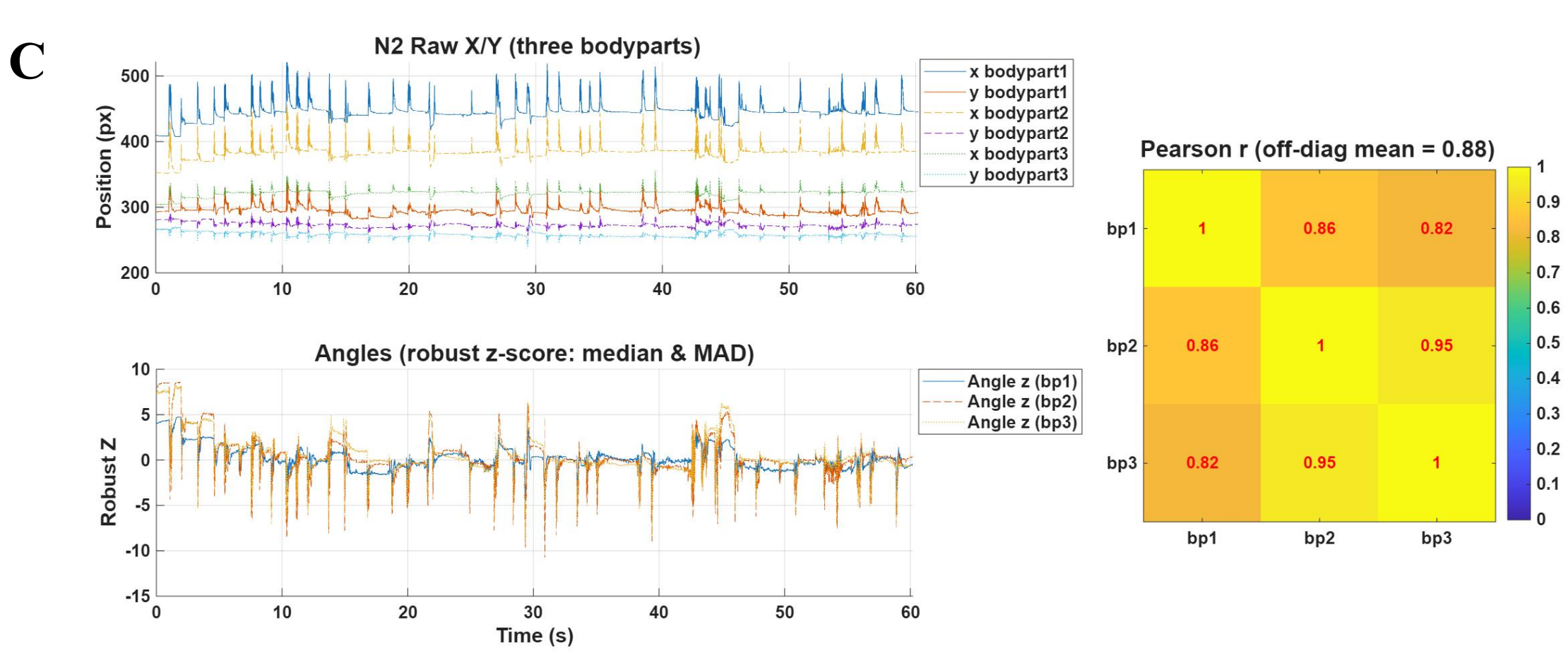

D

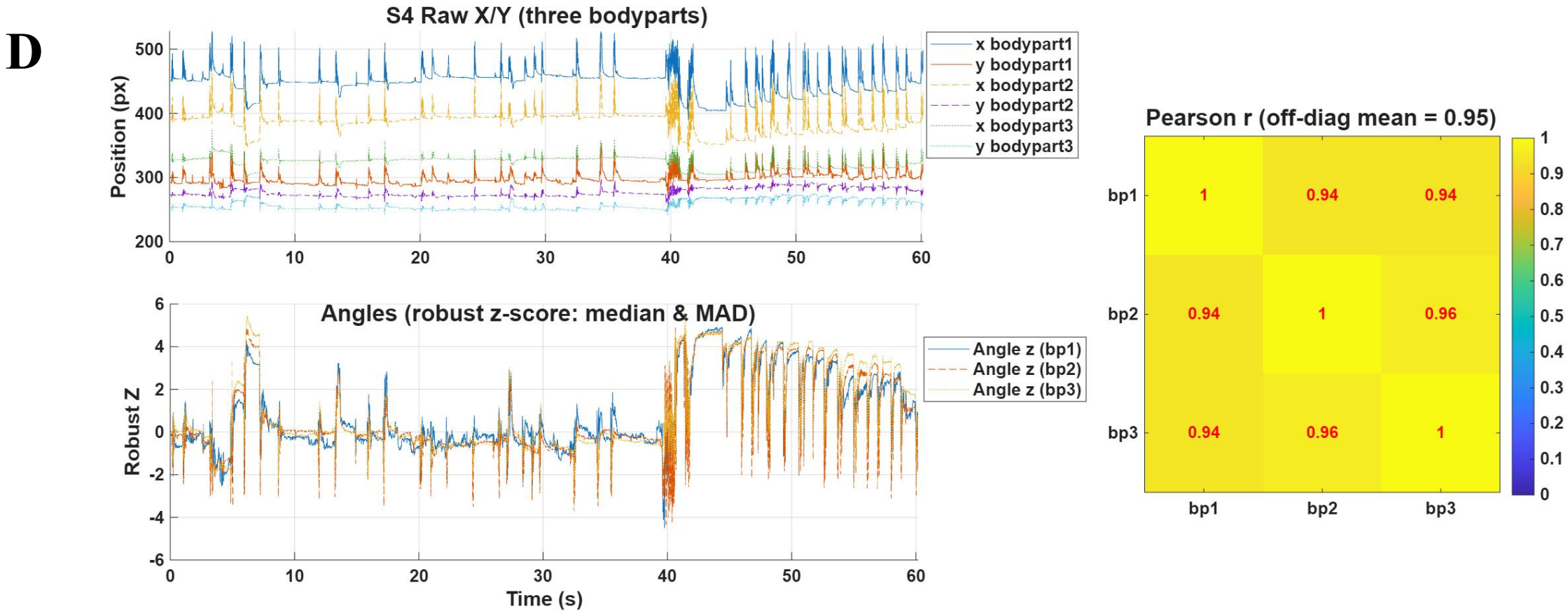

S4 Raw X/Y (three bodyparts)
Position (px)
x bodypart1
y bodypart1
x bodypart2
y bodypart2
x bodypart3
y bodypart3
Angles (robust z-score: median & MAD)
Robust Z
Angle z (bp1)
Angle z (bp2)
Angle z (bp3)
Time (s)
Pearson r (off-diag mean = 0.95)
bp1
bp2
bp3

E

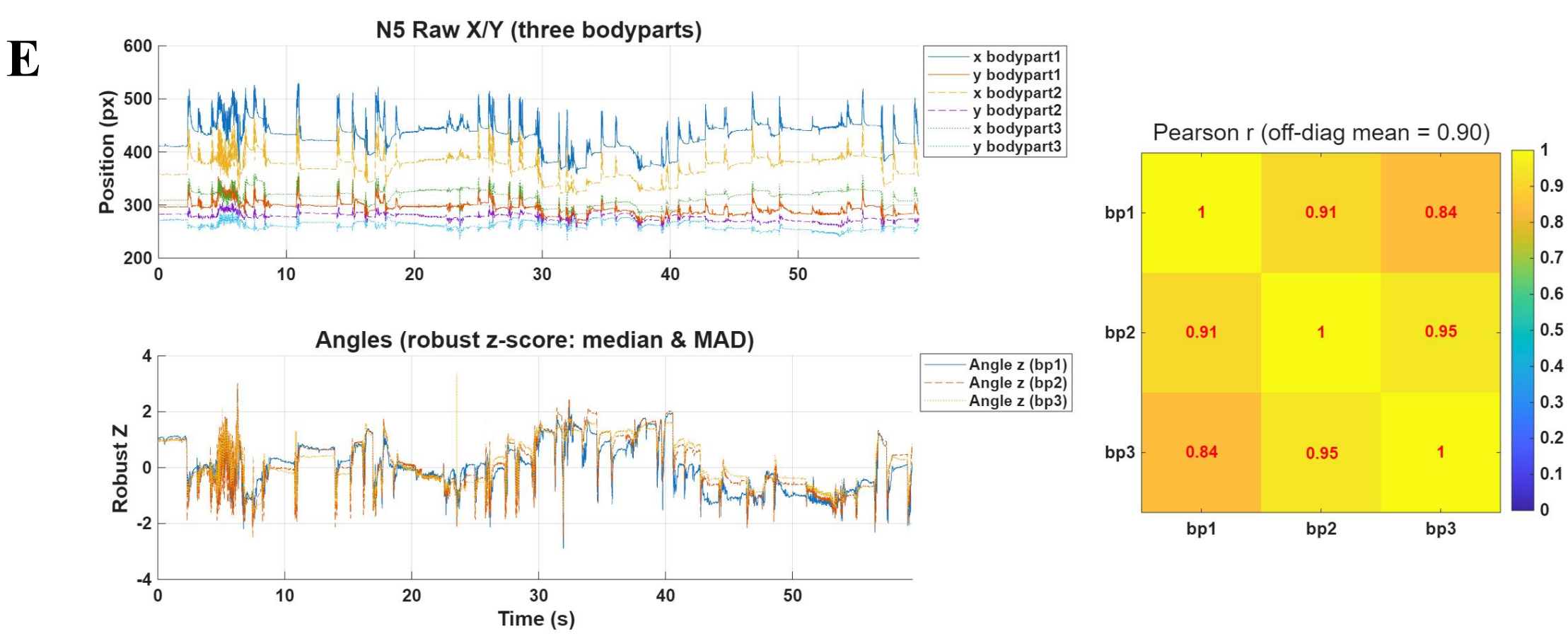

N5 Raw X/Y (three bodyparts)
Position (px)
x bodypart1
y bodypart1
x bodypart2
y bodypart2
x bodypart3
y bodypart3
Angles (robust z-score: median & MAD)
Robust Z
Angle z (bp1)
Angle z (bp2)
Angle z (bp3)
Time (s)
Pearson r (off-diag mean = 0.90)
bp1
bp2
bp3

F

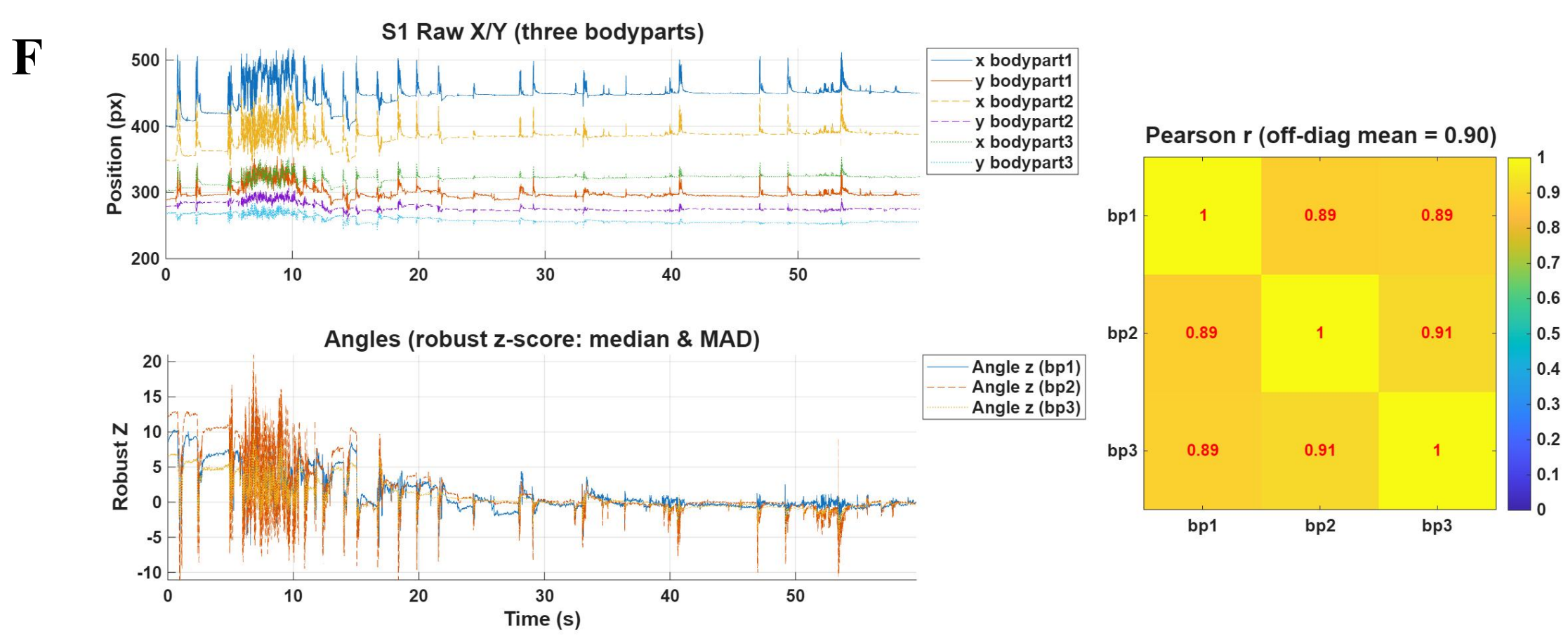

S1 Raw X/Y (three bodyparts)
Position (px)
x bodypart1
y bodypart1
x bodypart2
y bodypart2
x bodypart3
y bodypart3
Angles (robust z-score: median & MAD)
Robust Z
Angle z (bp1)
Angle z (bp2)
Angle z (bp3)
Time (s)
Pearson r (off-diag mean = 0.90)
bp1
bp2
bp3

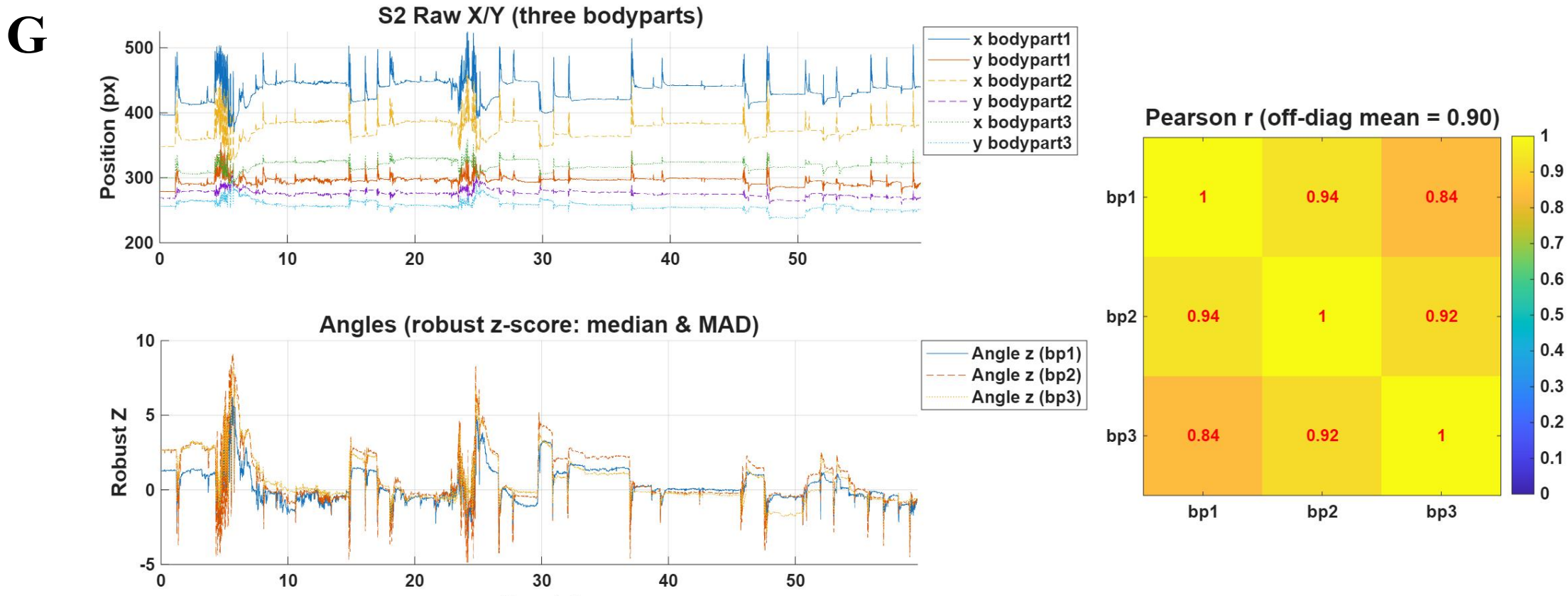
G
S2 Raw X/Y (three bodyparts)
Position (px)
x bodypart1
y bodypart1
x bodypart2
y bodypart2
x bodypart3
y bodypart3
Angles (robust z-score: median & MAD)
Robust Z
Angle z (bp1)
Angle z (bp2)
Angle z (bp3)
Time (s)
Pearson r (off-diag mean = 0.90)
bp1
bp2
bp3
1
0.94
0.84
0.94
1
0.92
0.84
0.92
1

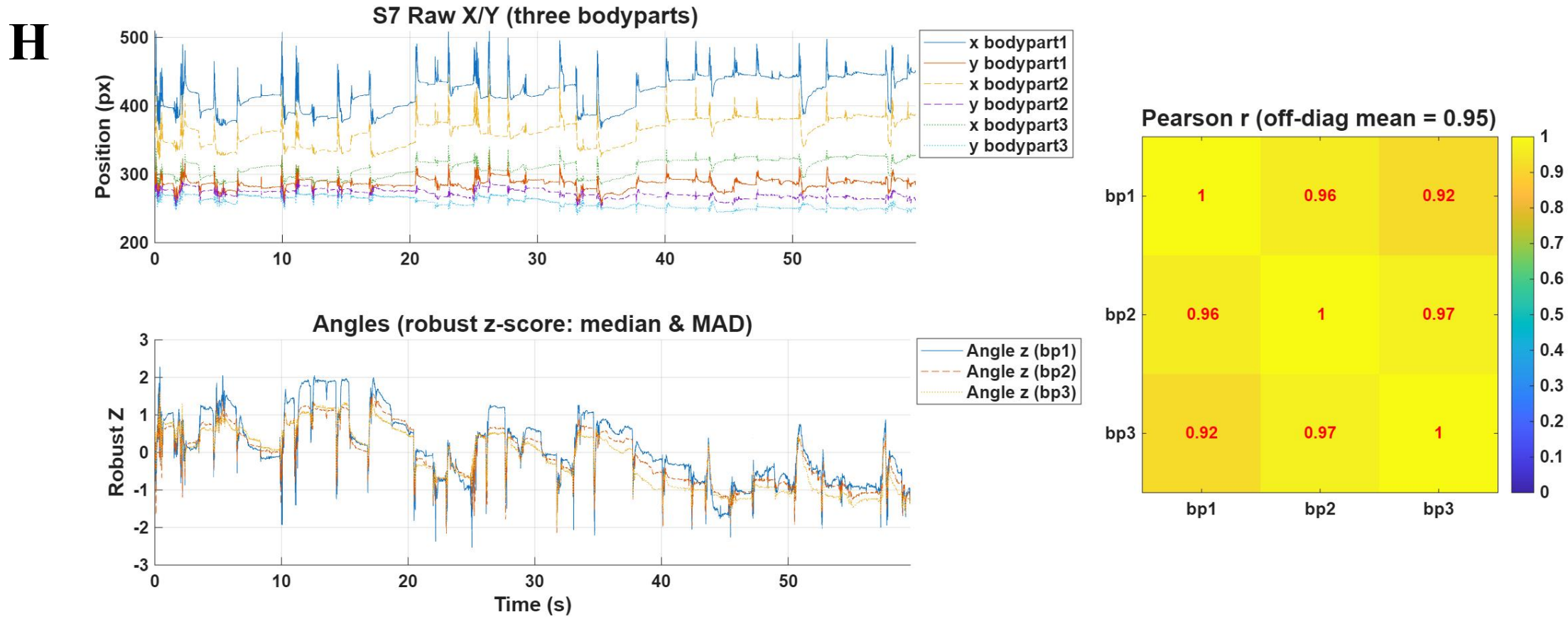
H
S7 Raw X/Y (three bodyparts)
Position (px)
x bodypart1
y bodypart1
x bodypart2
y bodypart2
x bodypart3
y bodypart3
Angles (robust z-score: median & MAD)
Robust Z
Angle z (bp1)
Angle z (bp2)
Angle z (bp3)
Time (s)
Pearson r (off-diag mean = 0.95)
bp1
bp2
bp3
1
0.96
0.92
0.96
1
0.97
0.92
0.97
1

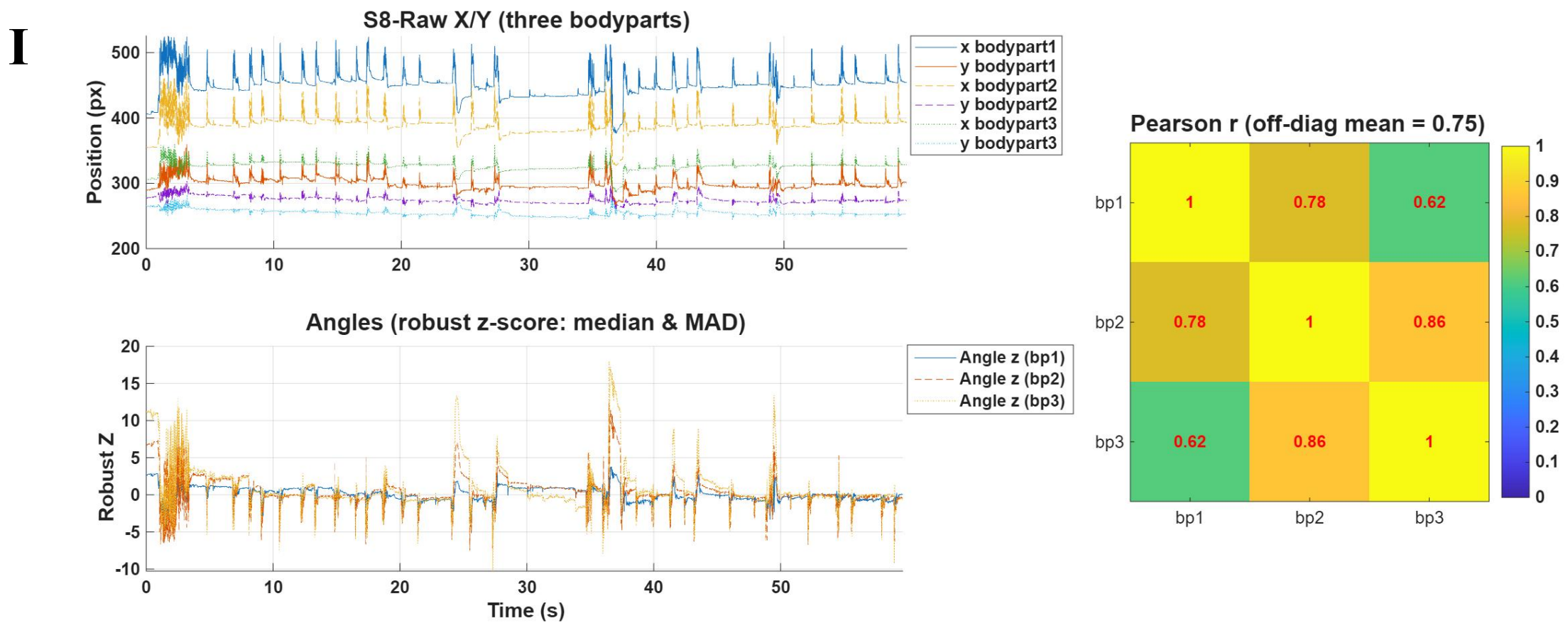
I
S8-Raw X/Y (three bodyparts)
Position (px)
x bodypart1
y bodypart1
x bodypart2
y bodypart2
x bodypart3
y bodypart3
Angles (robust z-score: median & MAD)
Robust Z
Angle z (bp1)
Angle z (bp2)
Angle z (bp3)
Time (s)
Pearson r (off-diag mean = 0.75)
bp1
bp2
bp3
1
0.78
0.62
0.78
1
0.86
0.62
0.86
1

We then quantified inter-whisker synchrony by computing Pearson correlations between angular trajectories of three whiskers. Despite being generated from independently trained models, whisker trajectories exhibited consistently high correlations (mean r = 0.889 ± 0.076, 95% CI [0.860–0.918]), which were highly significant ($t(26) = 60.66$, $p < 10^{-28}$) (Figure J). The observed high and consistent synchrony strongly indicates that DLC captures stable and biologically meaningful whisker motion.

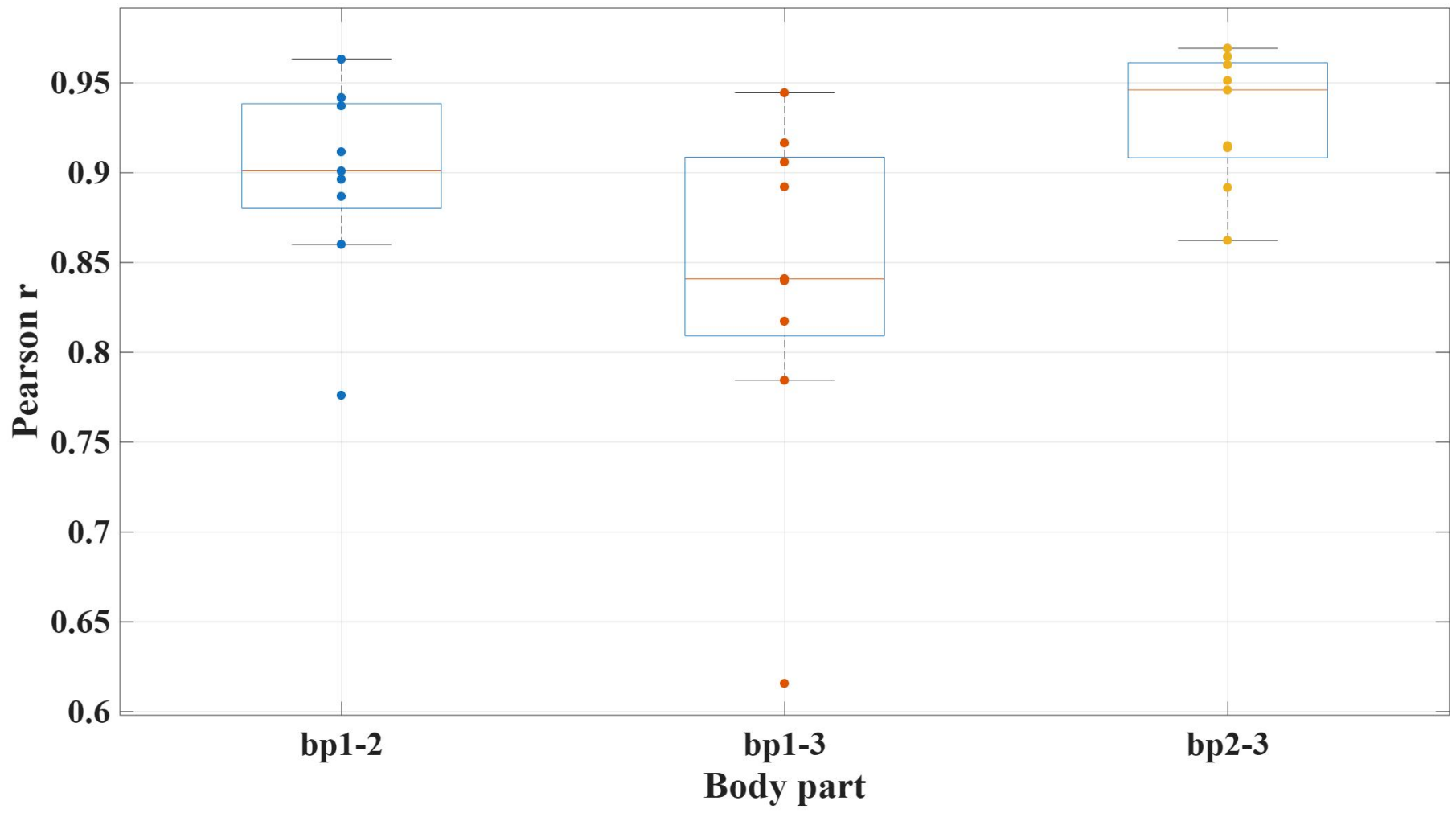

**Figure J. Inter whisker synchrony across independently trained models with DeepLabCut tracking.** The Pairwise Pearson correlations between angular trajectories of three whiskers (bp1-bp2, bp1-bp3, bp2-bp3) across nine independently processed datasets from the same animal. Across all datasets, whisker trajectories exhibited consistently high correlations (mean r = 0.889 ± 0.076), indicating strong temporal synchrony between whiskers.

Despite being trained separately, the resulting whisker trajectories showed highly consistent temporal dynamics, as reflected by strong inter-whisker correlations across all datasets. This consistency indicates that DeepLabCut reliably captures the underlying whisker motion and provides reliable and reproducible measurements of whisker kinematics across independently trained models. While not a ground truth, its consistency across independently trained models and inter-whisker synchrony reflects underlying biological coordination rather than model-specific artifacts, thereby supports its validity as a quantitative tool for whisker behavioral analysis.

**The synchrony metrics are summarized as following:**

Table I. Summary of Results Across the 9 Datasets.

| Dataset | r_1_2 | r_1_3 | r_2_3 |
|---|---|---|---|
| S4 | 0.94 | 0.94 | 0.96 |
| N2 | 0.86 | 0.82 | 0.95 |
| N1 | 0.9 | 0.78 | 0.89 |
| S2 | 0.94 | 0.84 | 0.92 |
| S7 | 0.96 | 0.92 | 0.97 |
| S8 | 0.78 | 0.62 | 0.86 |
| S1 | 0.89 | 0.89 | 0.91 |
| N5 | 0.91 | 0.84 | 0.95 |
| M13 | 0.9 | 0.91 | 0.96 |

## II. Statistical agreement between the sliding FFT and cycle-by-cycle estimators

We expanded the analysis across with nine independent datasets and summarized agreement between the sliding FFT and cycle-by-cycle estimators at the group level. Across all datasets, the FFT estimator showed a systematic negative bias relative to the cycle-based method, together with substantial error and poor concordance (low CCC and ICC). This expanded multi-dataset analysis has now been incorporated to strengthen the methodological justification and transparency of the manuscript.

Each of the following figure (A-I) corresponds to an independent recording trial of whisker tracking data extracted using DeepLabCut (with single whiskers labeling).

A

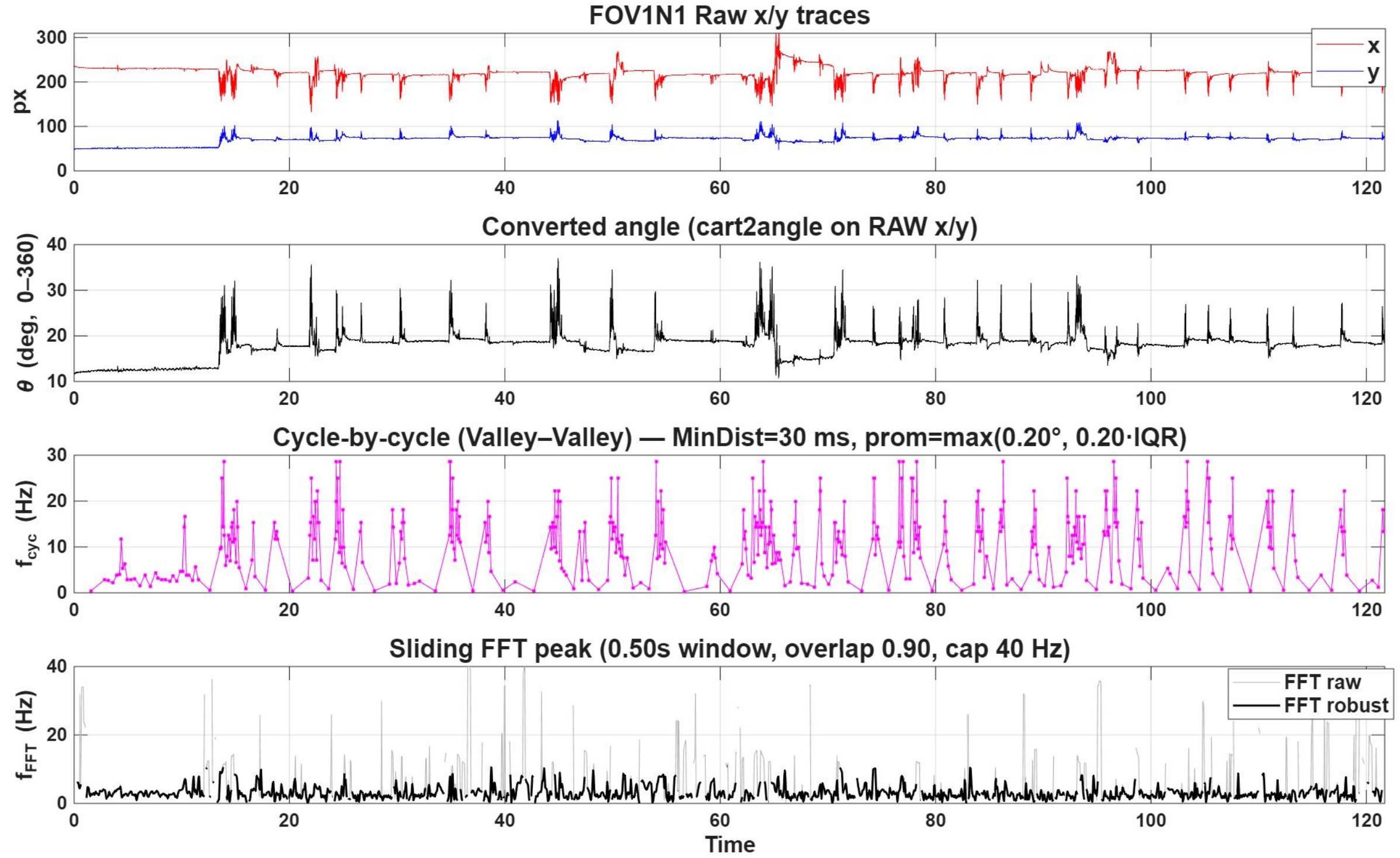

FOV1N1 Raw x/y traces
px
x
y
Converted angle (cart2angle on RAW x/y)
θ (deg, 0–360)
Cycle-by-cycle (Valley–Valley) — MinDist=30 ms, prom=max(0.20°, 0.20·IQR)
f_cyc (Hz)
Sliding FFT peak (0.50s window, overlap 0.90, cap 40 Hz)
f_FFT (Hz)
FFT raw
FFT robust
Time

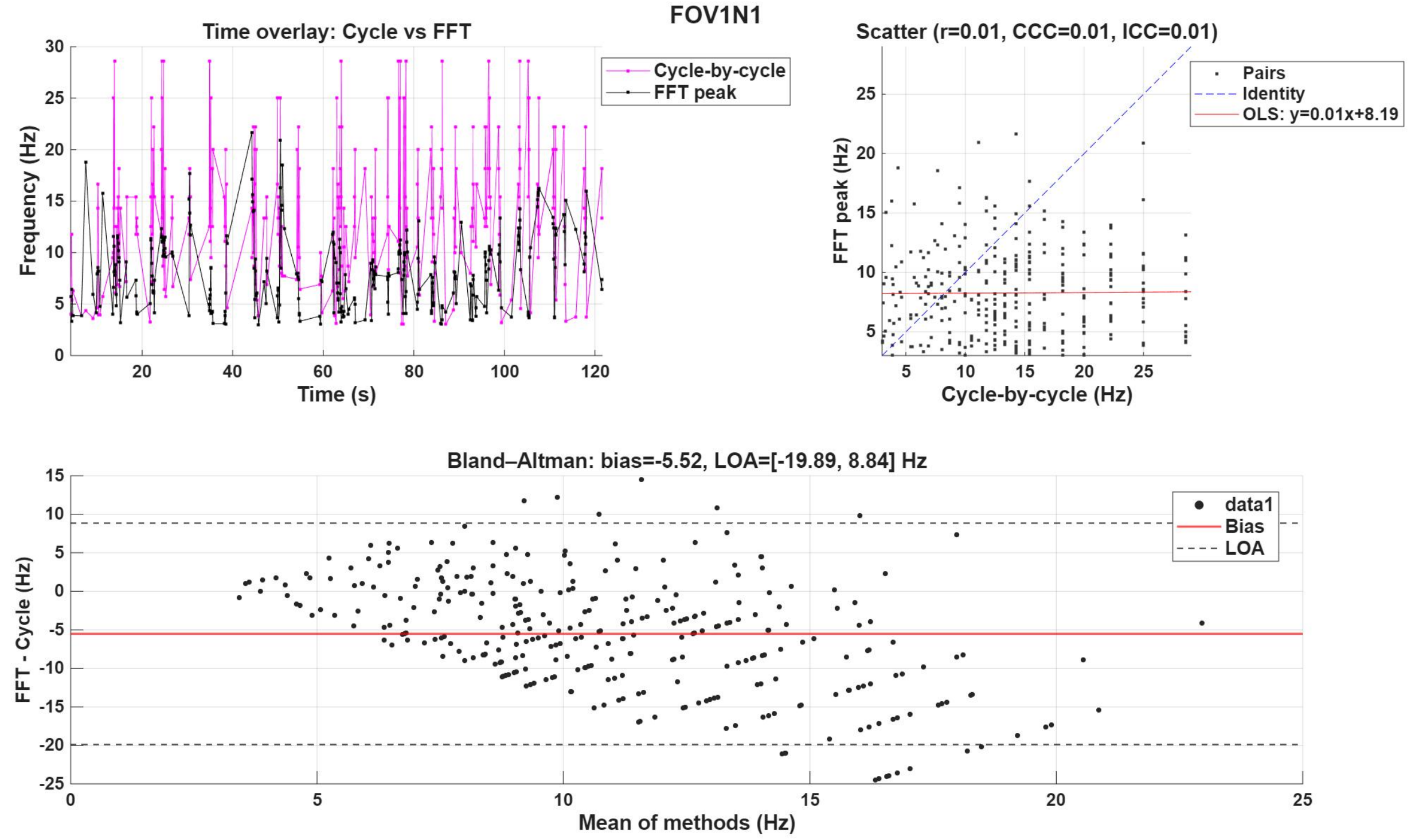

FOV1N1
Time overlay: Cycle vs FFT
Frequency (Hz)
Time (s)
Cycle-by-cycle
FFT peak
Scatter (r=0.01, CCC=0.01, ICC=0.01)
FFT peak (Hz)
Cycle-by-cycle (Hz)
Pairs
Identity
OLS: y=0.01x+8.19
Bland–Altman: bias=-5.52, LOA=[-19.89, 8.84] Hz
FFT - Cycle (Hz)
Mean of methods (Hz)
data1
Bias
LOA

**B**

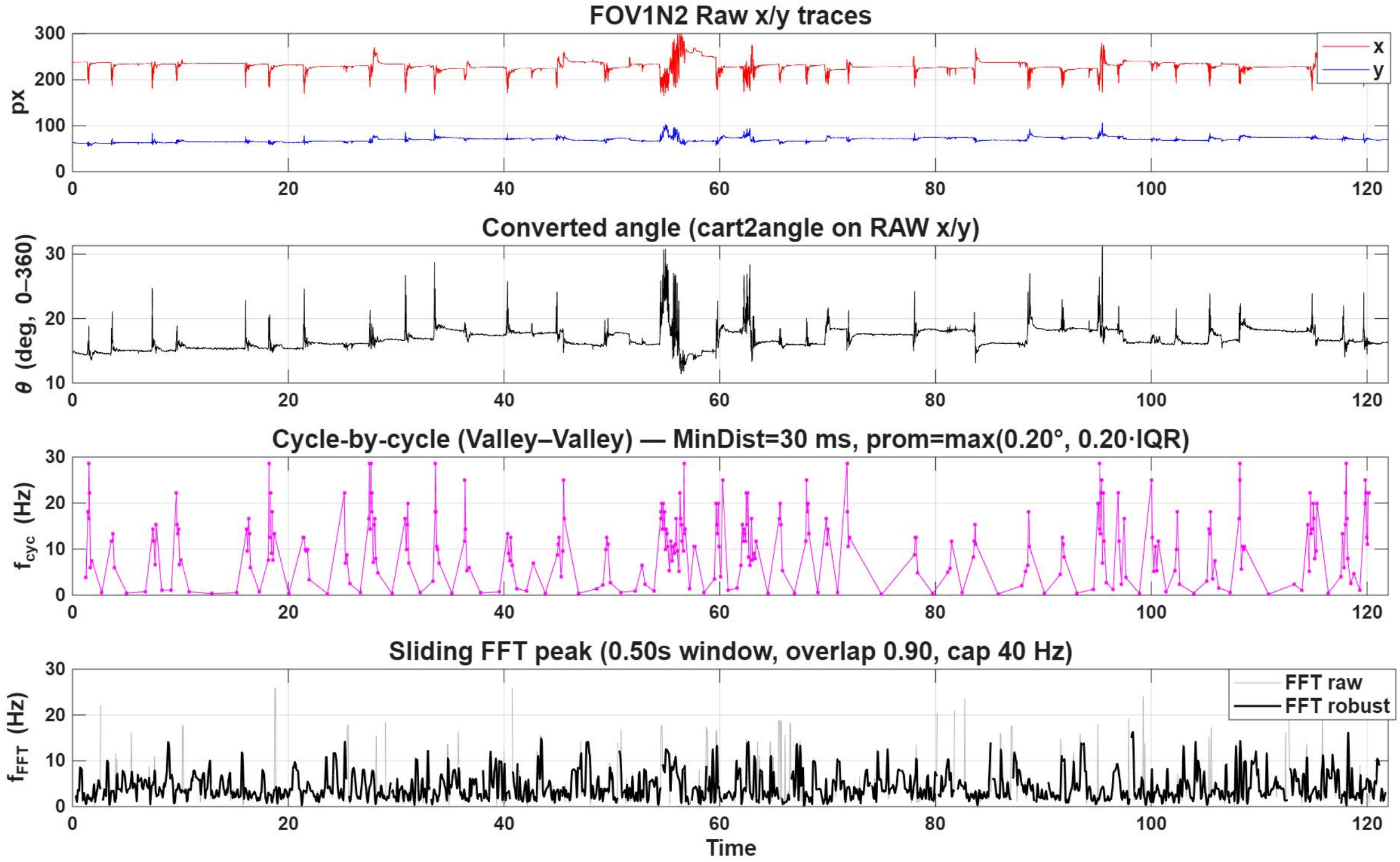

FOV1N2 Raw x/y traces
px
x
y
Converted angle (cart2angle on RAW x/y)
θ (deg, 0–360)
Cycle-by-cycle (Valley–Valley) — MinDist=30 ms, prom=max(0.20°, 0.20·IQR)
f_cyc (Hz)
Sliding FFT peak (0.50s window, overlap 0.90, cap 40 Hz)
f_FFT (Hz)
FFT raw
FFT robust
Time

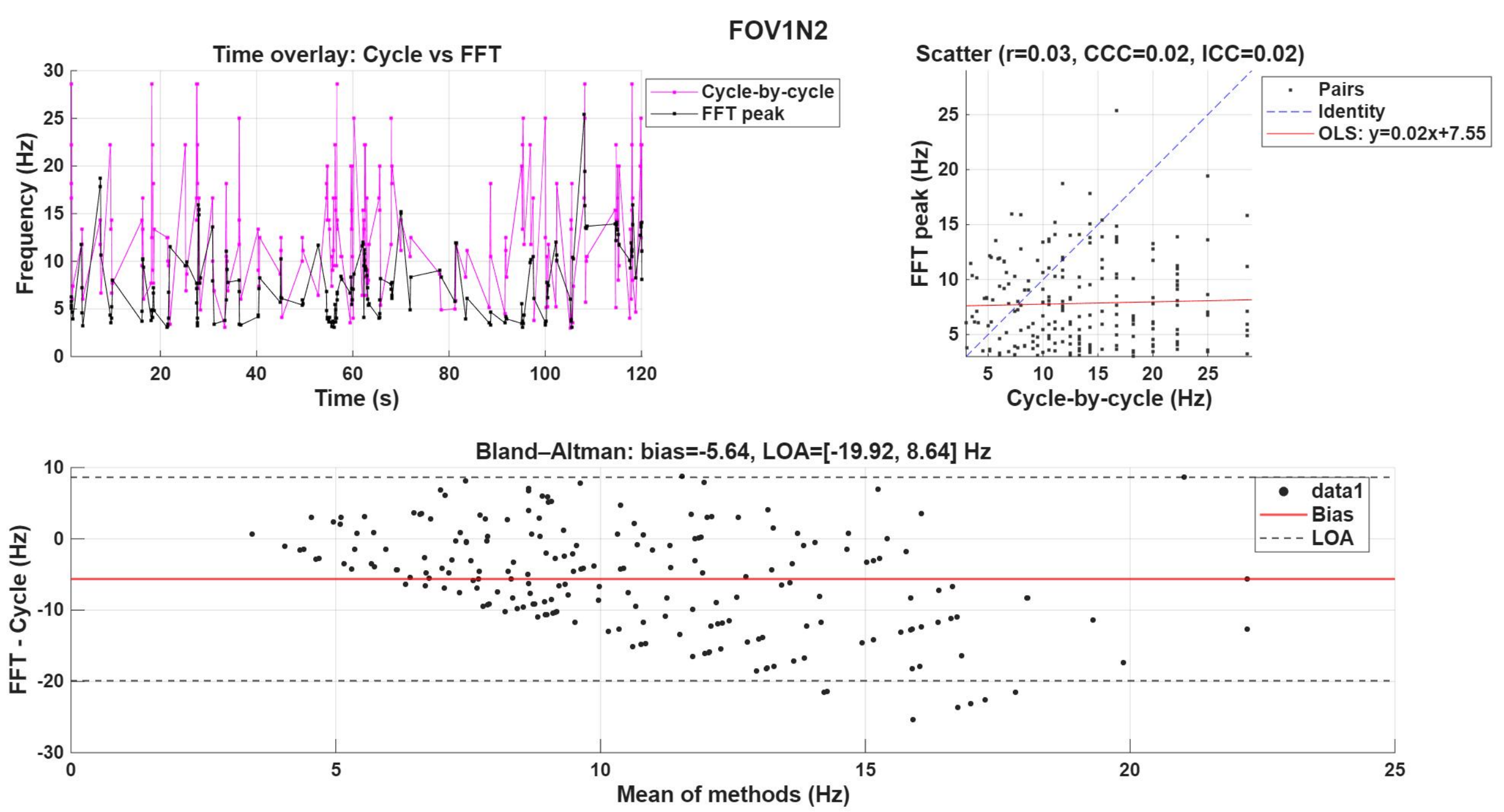

FOV1N2
Time overlay: Cycle vs FFT
Cycle-by-cycle
FFT peak
Frequency (Hz)
Time (s)
Scatter (r=0.03, CCC=0.02, ICC=0.02)
Pairs
Identity
OLS: y=0.02x+7.55
FFT peak (Hz)
Cycle-by-cycle (Hz)
Bland–Altman: bias=-5.64, LOA=[-19.92, 8.64] Hz
data1
Bias
LOA
FFT - Cycle (Hz)
Mean of methods (Hz)

C

FOV1S3 Raw x/y traces

px

x
y

Converted angle (cart2angle on RAW x/y)

θ (deg, 0–360)

Cycle-by-cycle (Valley–Valley) — MinDist=30 ms, prom=max(0.20°, 0.20·IQR)

$f_{cyc}$ (Hz)

Sliding FFT peak (0.50s window, overlap 0.90, cap 40 Hz)

$f_{FFT}$ (Hz)

FFT raw
FFT robust

Time

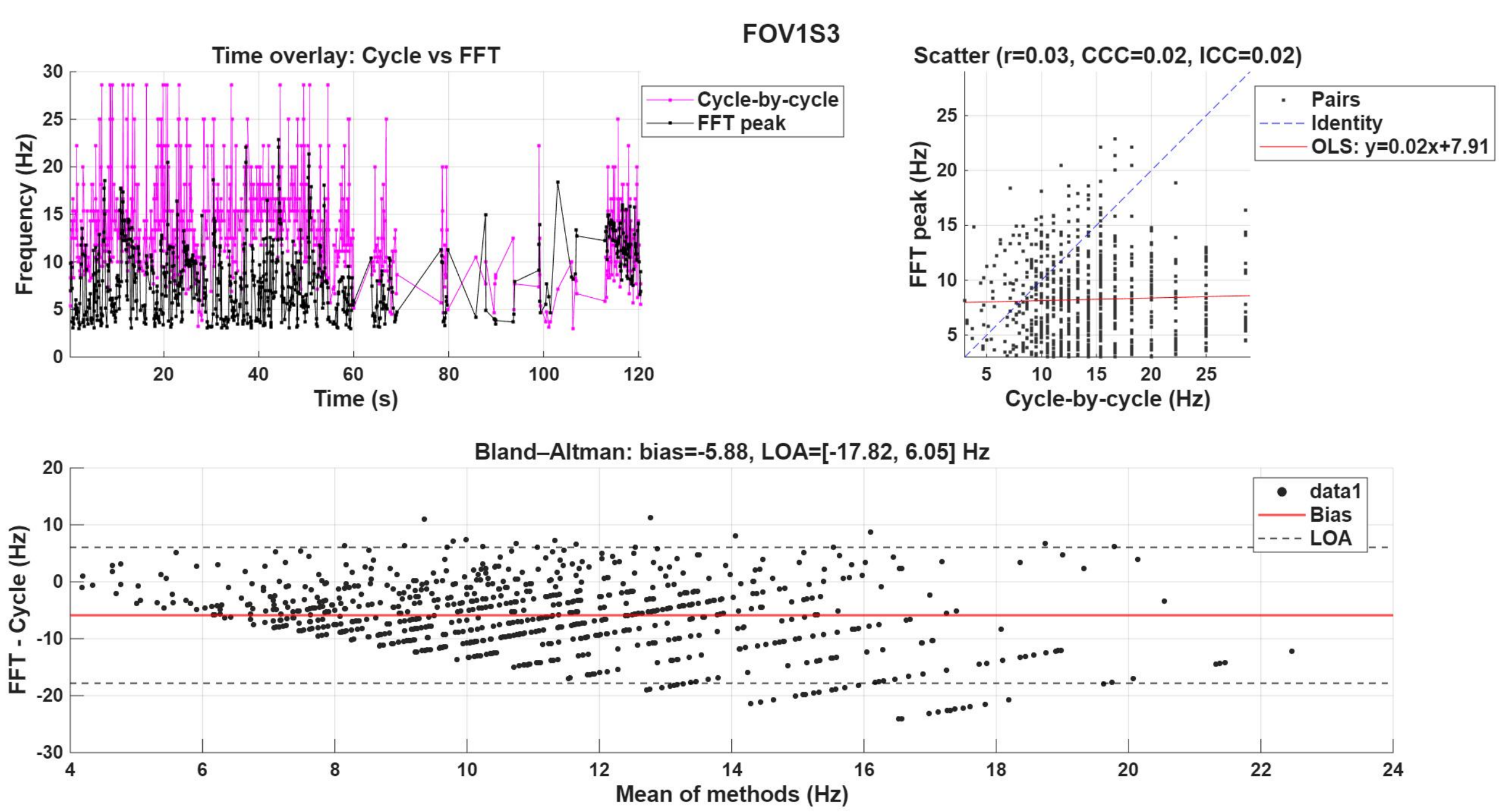

D

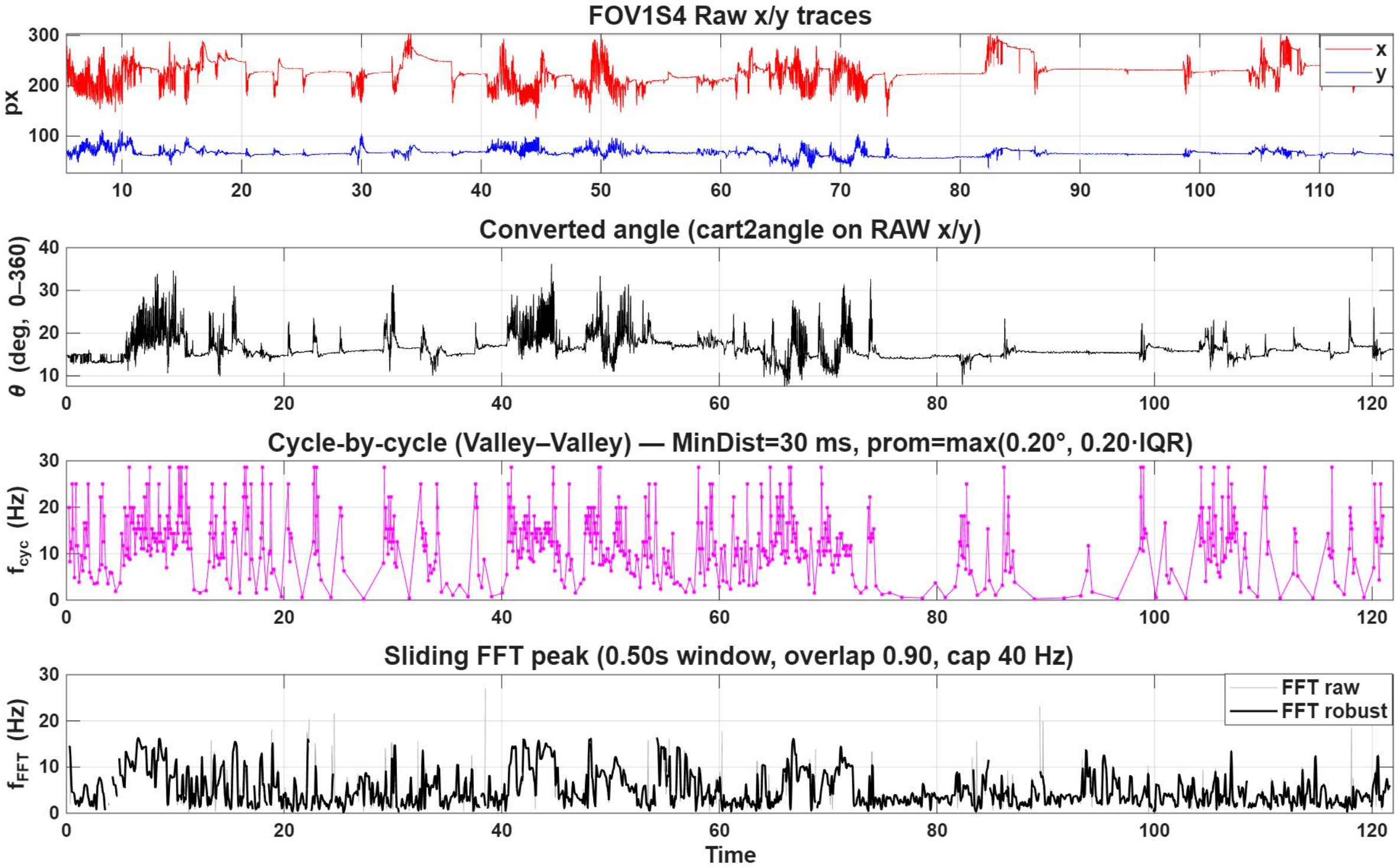
FOV1S4 Raw x/y traces
px
x
y
Converted angle (cart2angle on RAW x/y)
θ (deg, 0–360)
Cycle-by-cycle (Valley–Valley) — MinDist=30 ms, prom=max(0.20°, 0.20·IQR)
f_cyc (Hz)
Sliding FFT peak (0.50s window, overlap 0.90, cap 40 Hz)
f_FFT (Hz)
FFT raw
FFT robust
Time

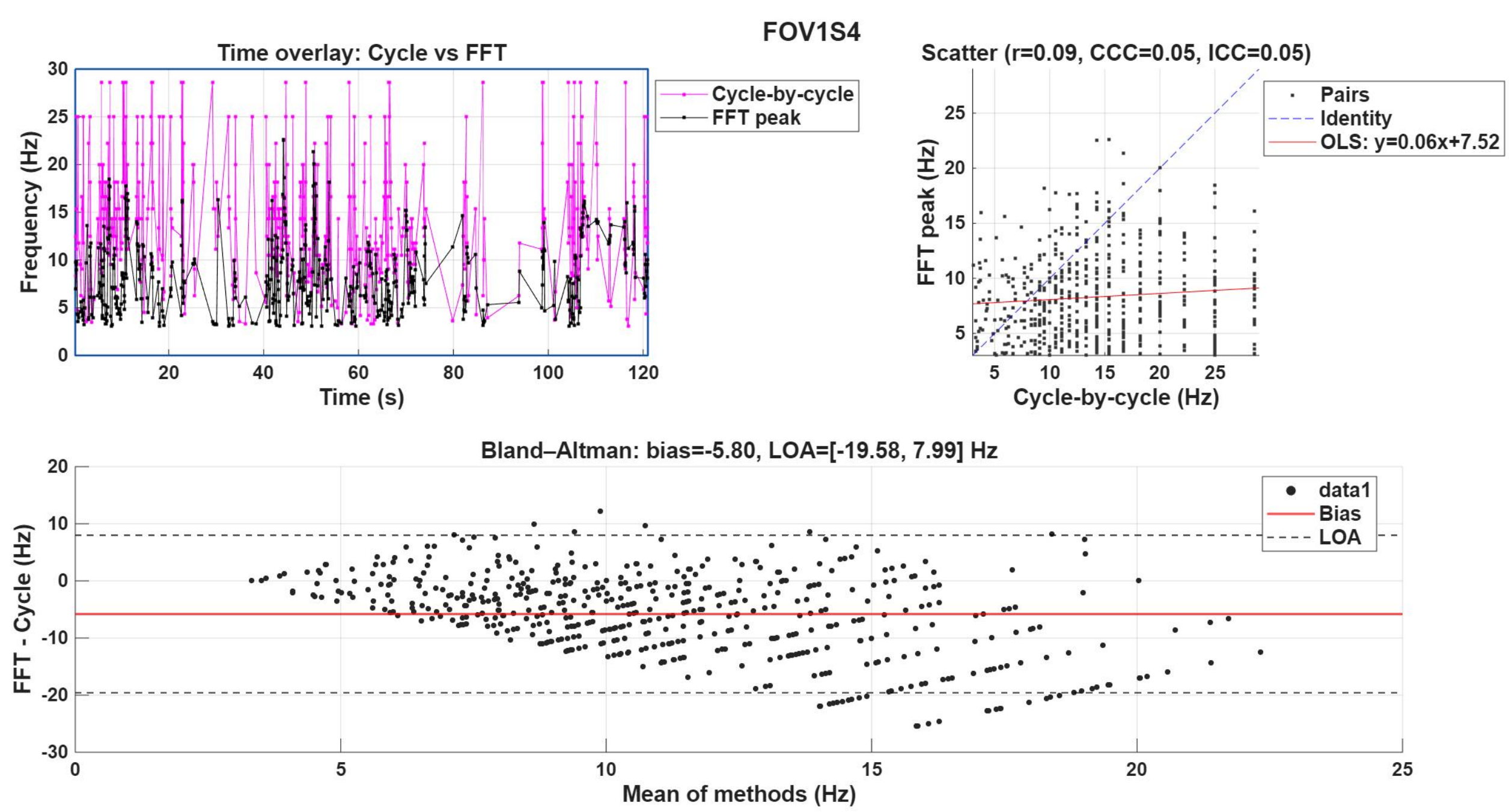
FOV1S4
Time overlay: Cycle vs FFT
Frequency (Hz)
Time (s)
Cycle-by-cycle
FFT peak
Scatter (r=0.09, CCC=0.05, ICC=0.05)
FFT peak (Hz)
Cycle-by-cycle (Hz)
Pairs
Identity
OLS: y=0.06x+7.52
Bland–Altman: bias=-5.80, LOA=[-19.58, 7.99] Hz
FFT - Cycle (Hz)
Mean of methods (Hz)
data1
Bias
LOA

**E**

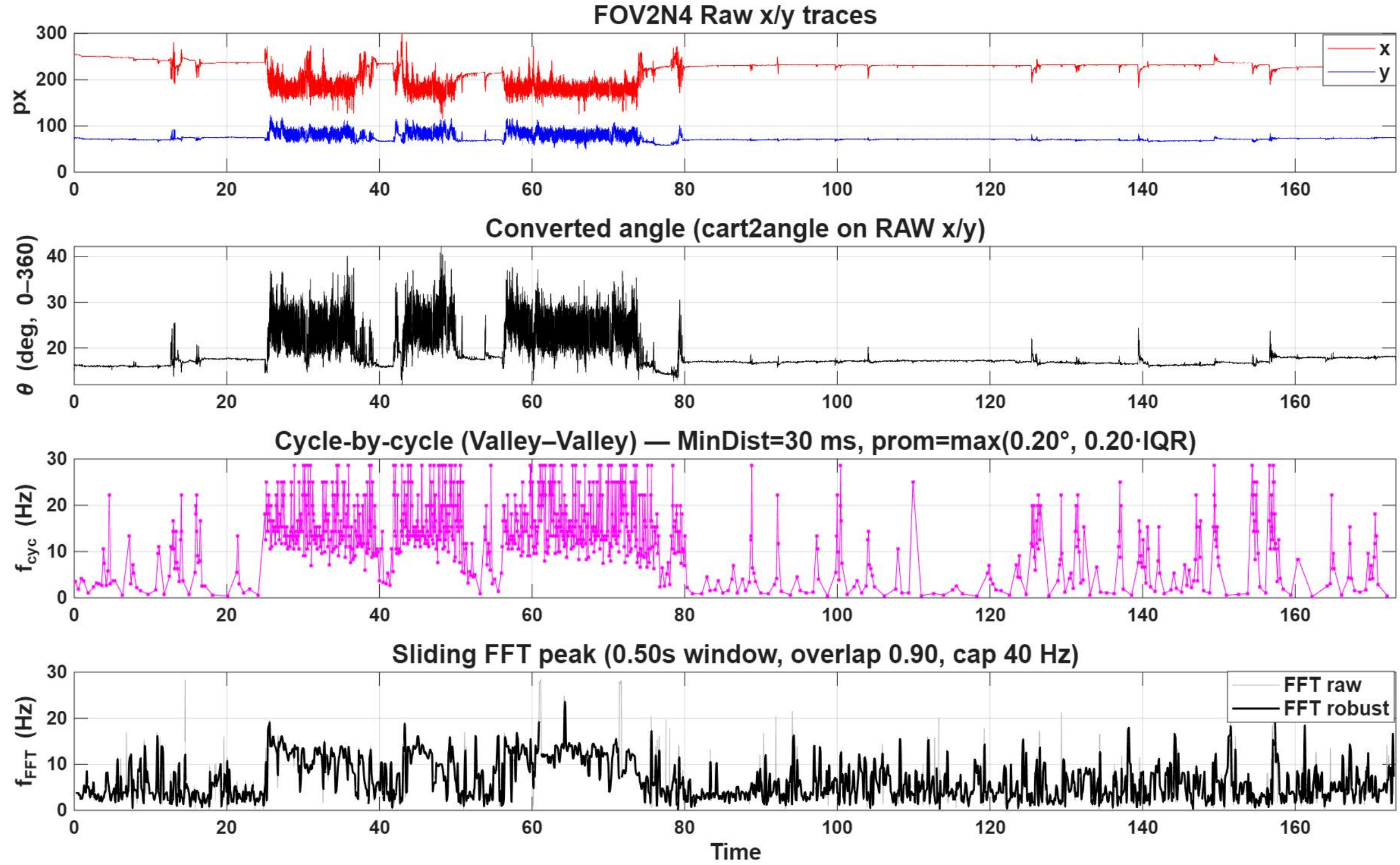

FOV2N4 Raw x/y traces
px
x
y
Converted angle (cart2angle on RAW x/y)
θ (deg, 0–360)
Cycle-by-cycle (Valley–Valley) — MinDist=30 ms, prom=max(0.20°, 0.20·IQR)
f_cyc (Hz)
Sliding FFT peak (0.50s window, overlap 0.90, cap 40 Hz)
f_FFT (Hz)
FFT raw
FFT robust
Time

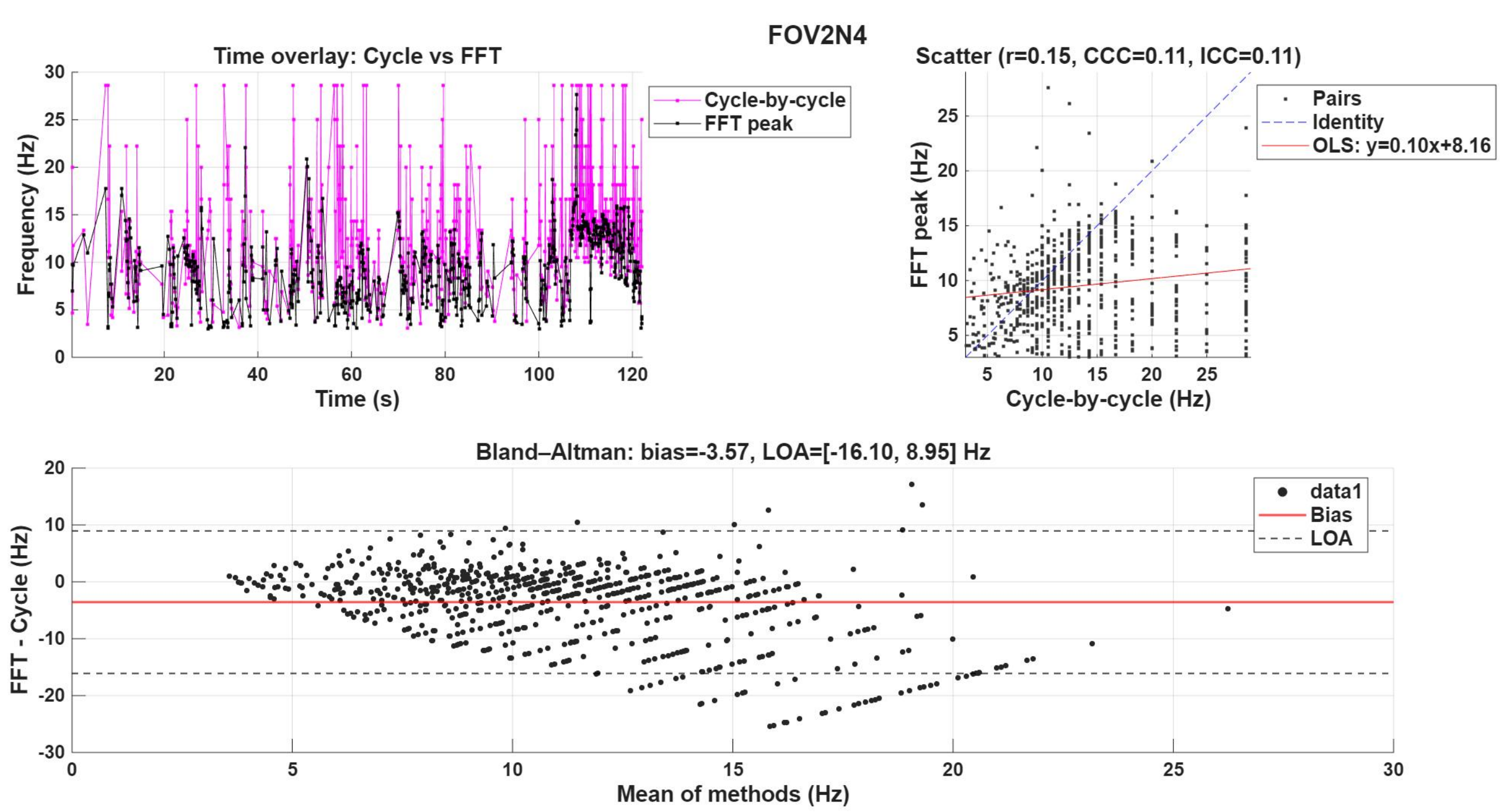

FOV2N4
Time overlay: Cycle vs FFT
Frequency (Hz)
Time (s)
Cycle-by-cycle
FFT peak
Scatter (r=0.15, CCC=0.11, ICC=0.11)
FFT peak (Hz)
Cycle-by-cycle (Hz)
Pairs
Identity
OLS: y=0.10x+8.16
Bland–Altman: bias=-3.57, LOA=[-16.10, 8.95] Hz
FFT - Cycle (Hz)
Mean of methods (Hz)
data1
Bias
LOA

**F**

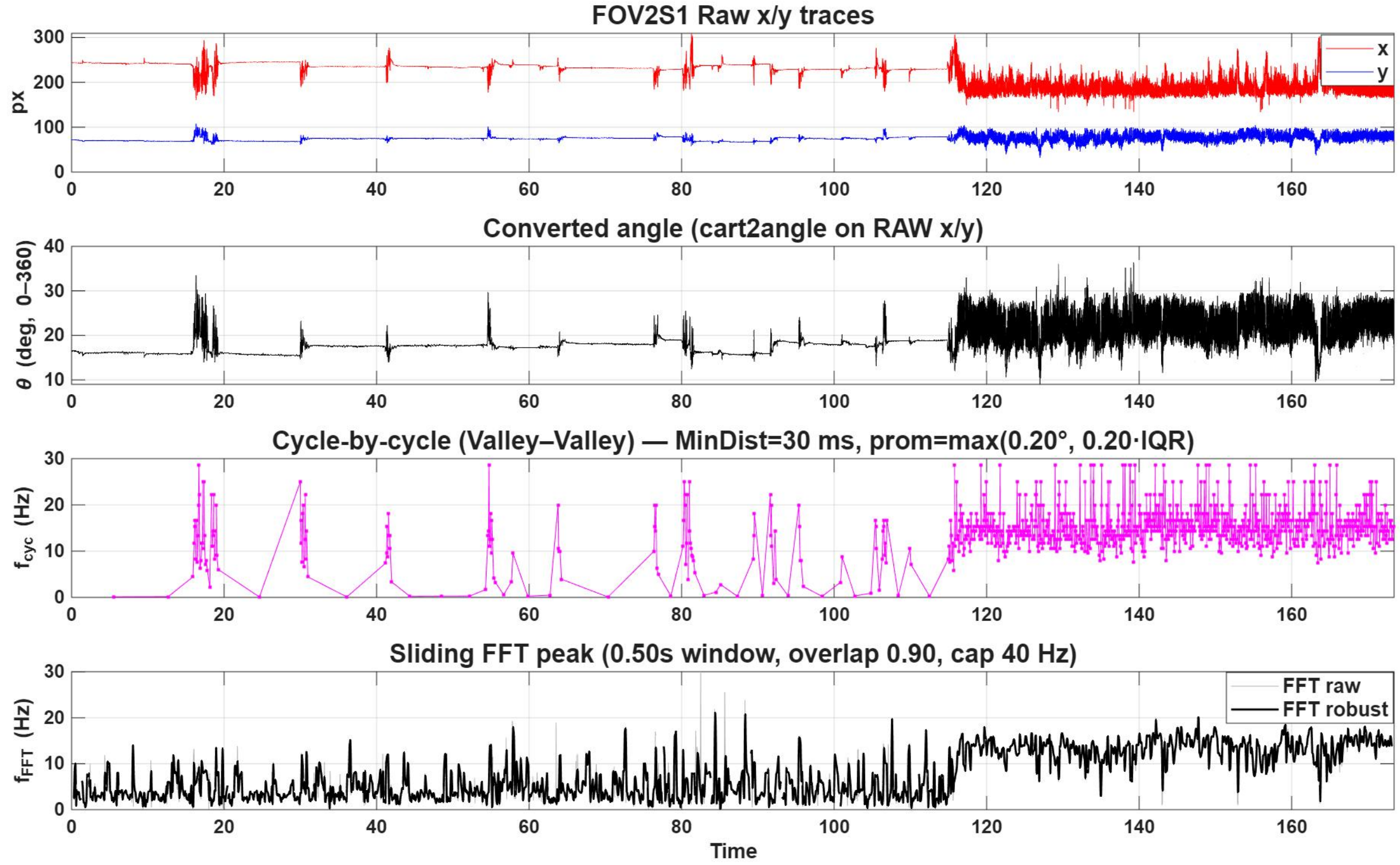

FOV2S1 Raw x/y traces
px
x
y
Converted angle (cart2angle on RAW x/y)
θ (deg, 0–360)
Cycle-by-cycle (Valley–Valley) — MinDist=30 ms, prom=max(0.20°, 0.20·IQR)
f_cyc (Hz)
Sliding FFT peak (0.50s window, overlap 0.90, cap 40 Hz)
f_FFT (Hz)
FFT raw
FFT robust
Time

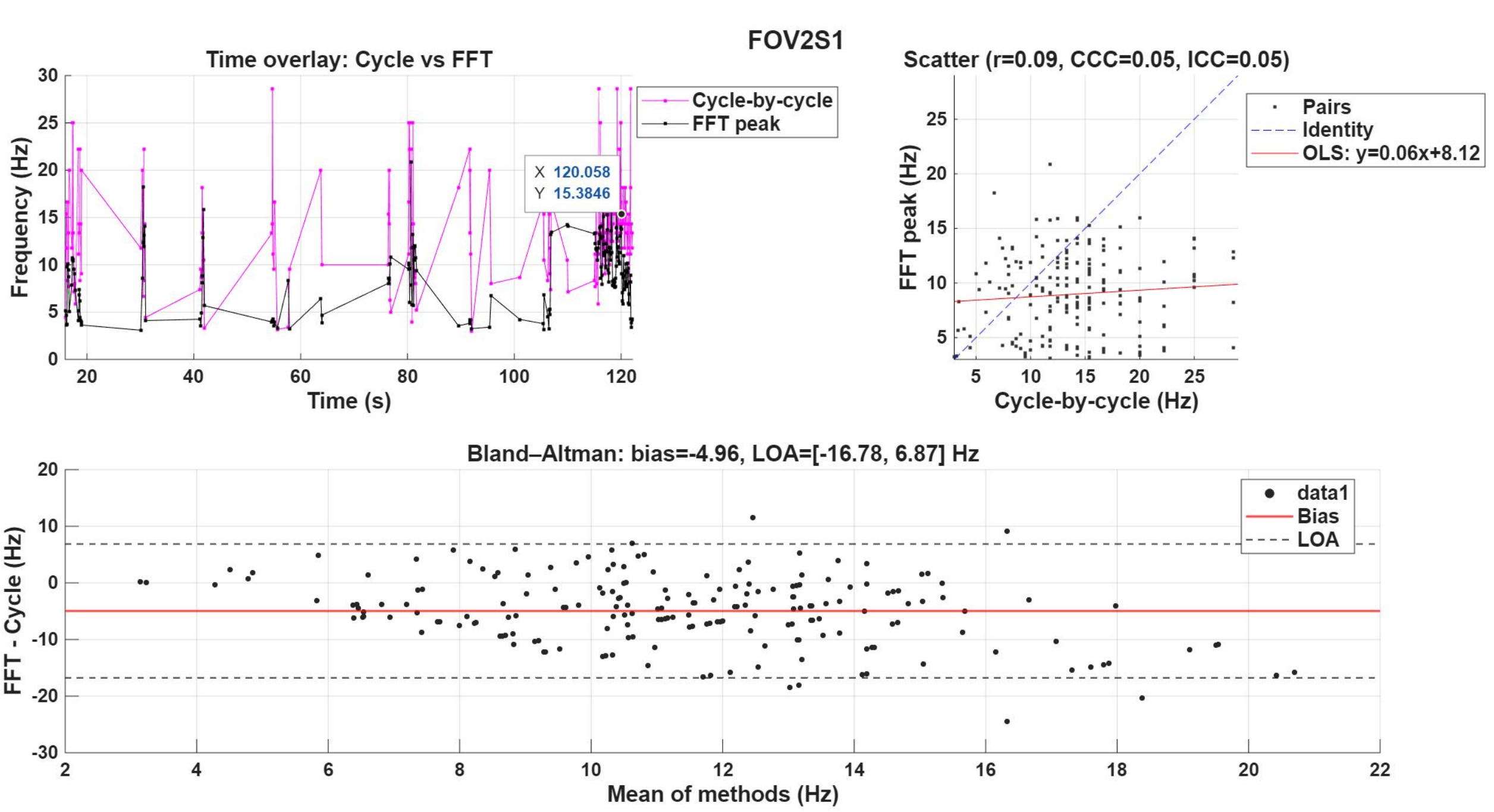

FOV2S1
Time overlay: Cycle vs FFT
Frequency (Hz)
Time (s)
Cycle-by-cycle
FFT peak
X 120.058
Y 15.3846
Scatter (r=0.09, CCC=0.05, ICC=0.05)
FFT peak (Hz)
Cycle-by-cycle (Hz)
Pairs
Identity
OLS: y=0.06x+8.12
Bland–Altman: bias=-4.96, LOA=[-16.78, 6.87] Hz
FFT - Cycle (Hz)
Mean of methods (Hz)
data1
Bias
LOA

G

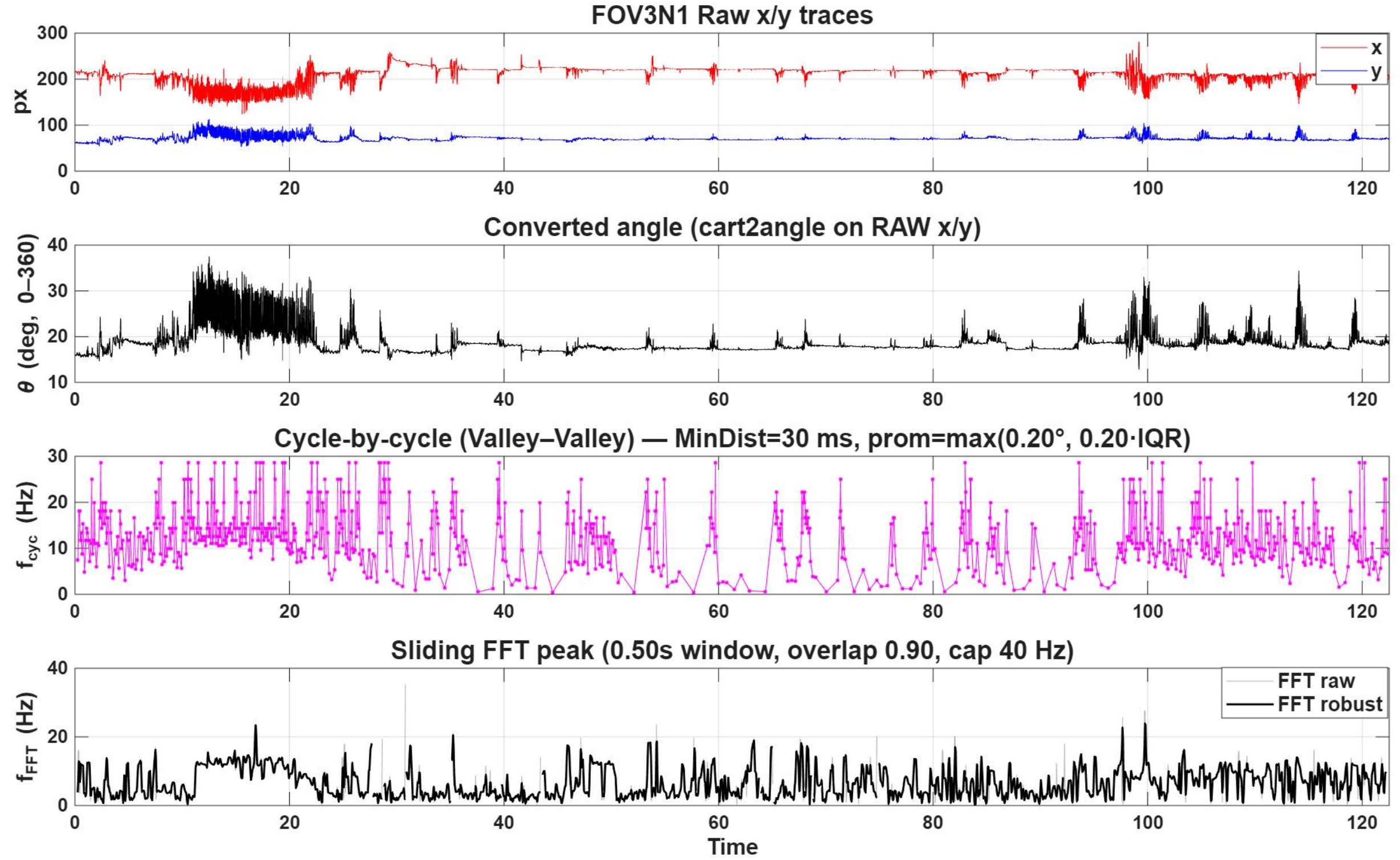
FOV3N1 Raw x/y traces
px
x
y
Converted angle (cart2angle on RAW x/y)
θ (deg, 0–360)
Cycle-by-cycle (Valley–Valley) — MinDist=30 ms, prom=max(0.20°, 0.20·IQR)
f_cyc (Hz)
Sliding FFT peak (0.50s window, overlap 0.90, cap 40 Hz)
f_FFT (Hz)
FFT raw
FFT robust
Time

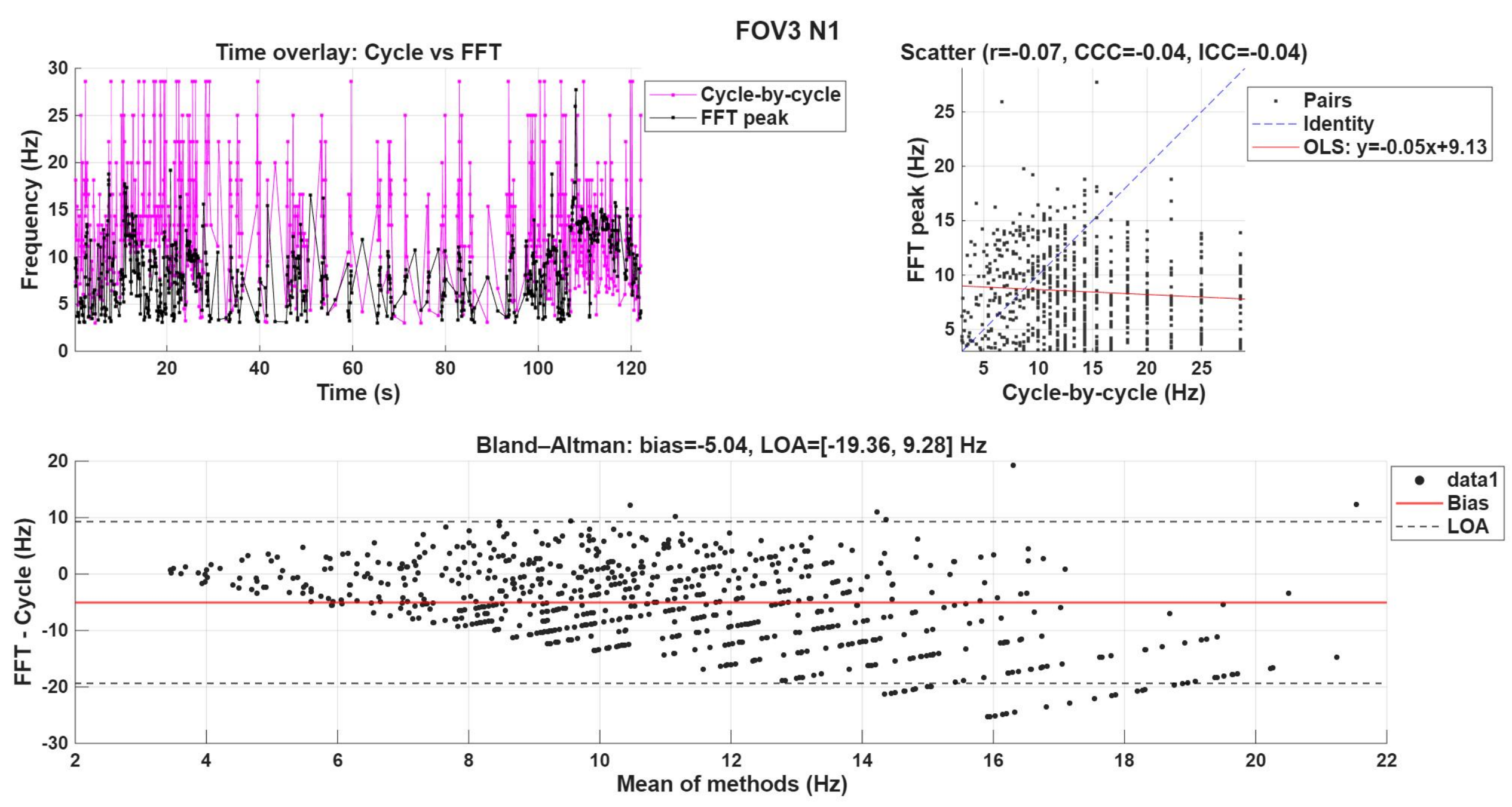
FOV3 N1
Time overlay: Cycle vs FFT
Frequency (Hz)
Time (s)
Cycle-by-cycle
FFT peak
Scatter (r=-0.07, CCC=-0.04, ICC=-0.04)
FFT peak (Hz)
Cycle-by-cycle (Hz)
Pairs
Identity
OLS: y=-0.05x+9.13
Bland–Altman: bias=-5.04, LOA=[-19.36, 9.28] Hz
FFT - Cycle (Hz)
Mean of methods (Hz)
data1
Bias
LOA

**H**

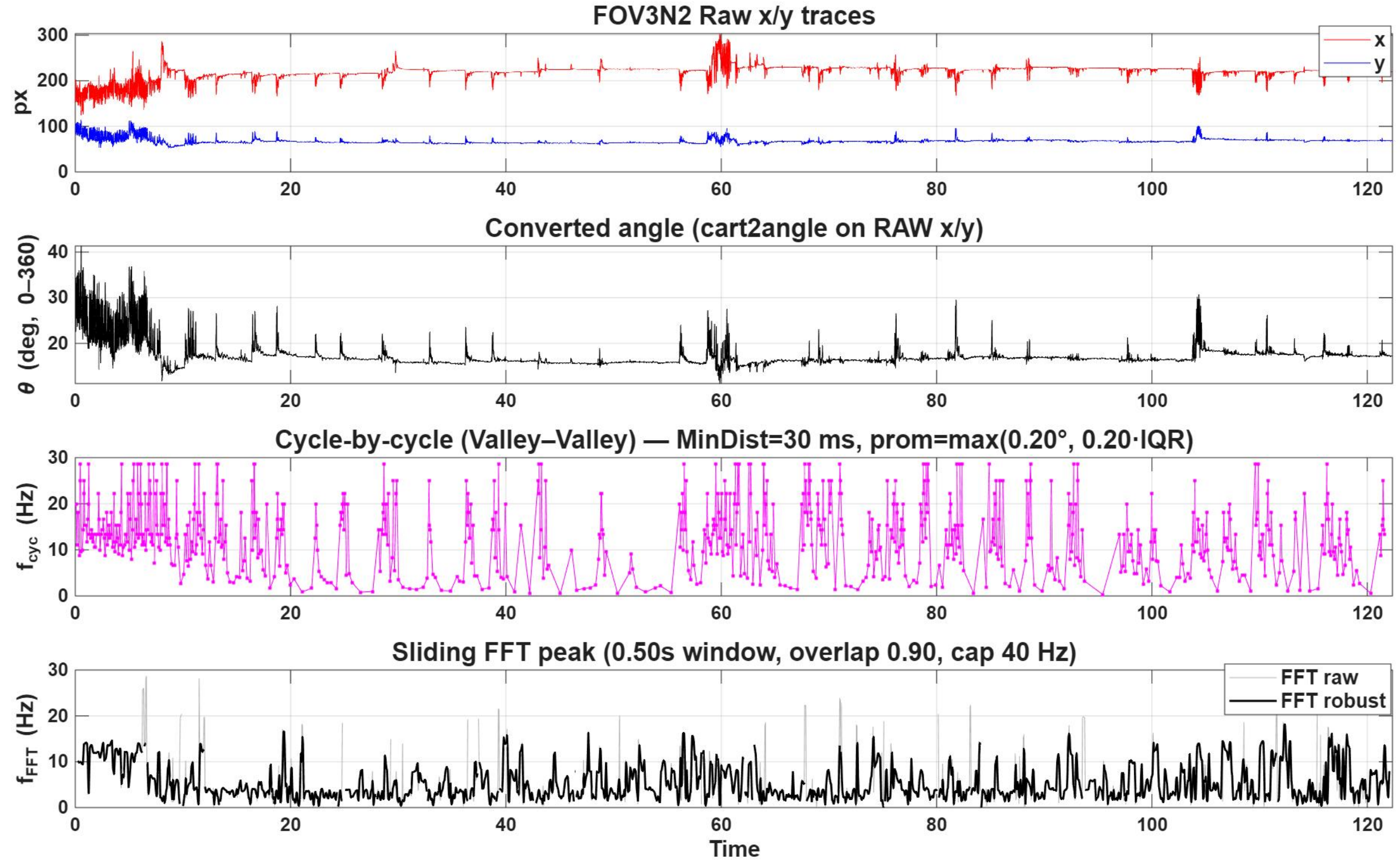
FOV3N2 Raw x/y traces
px
x
y
Converted angle (cart2angle on RAW x/y)
θ (deg, 0–360)
Cycle-by-cycle (Valley–Valley) — MinDist=30 ms, prom=max(0.20°, 0.20·IQR)
f_cyc (Hz)
Sliding FFT peak (0.50s window, overlap 0.90, cap 40 Hz)
f_FFT (Hz)
FFT raw
FFT robust
Time

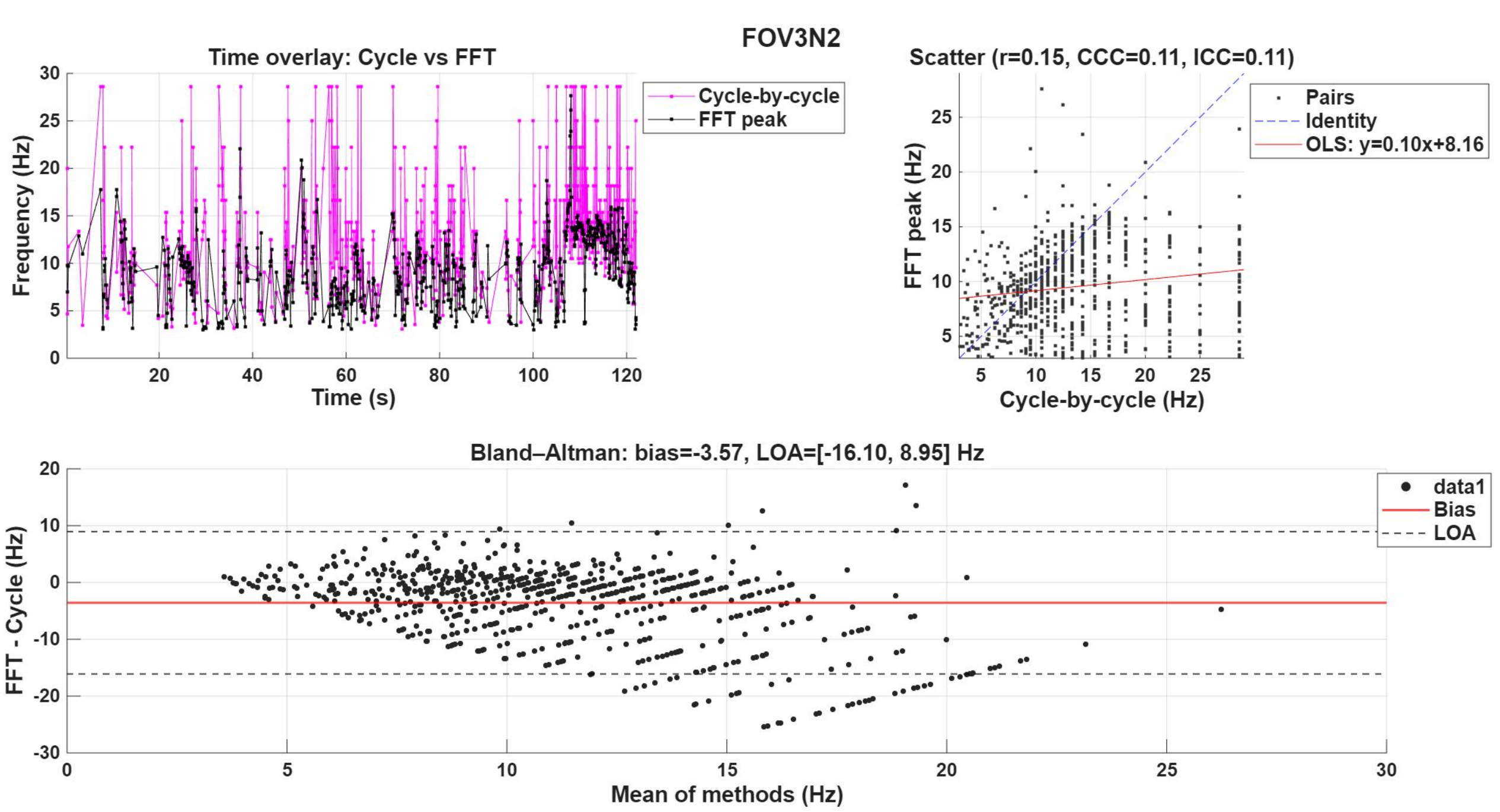
FOV3N2
Time overlay: Cycle vs FFT
Frequency (Hz)
Time (s)
Cycle-by-cycle
FFT peak
Scatter (r=0.15, CCC=0.11, ICC=0.11)
FFT peak (Hz)
Cycle-by-cycle (Hz)
Pairs
Identity
OLS: y=0.10x+8.16
Bland–Altman: bias=-3.57, LOA=[-16.10, 8.95] Hz
FFT - Cycle (Hz)
Mean of methods (Hz)
data1
Bias
LOA

**I**

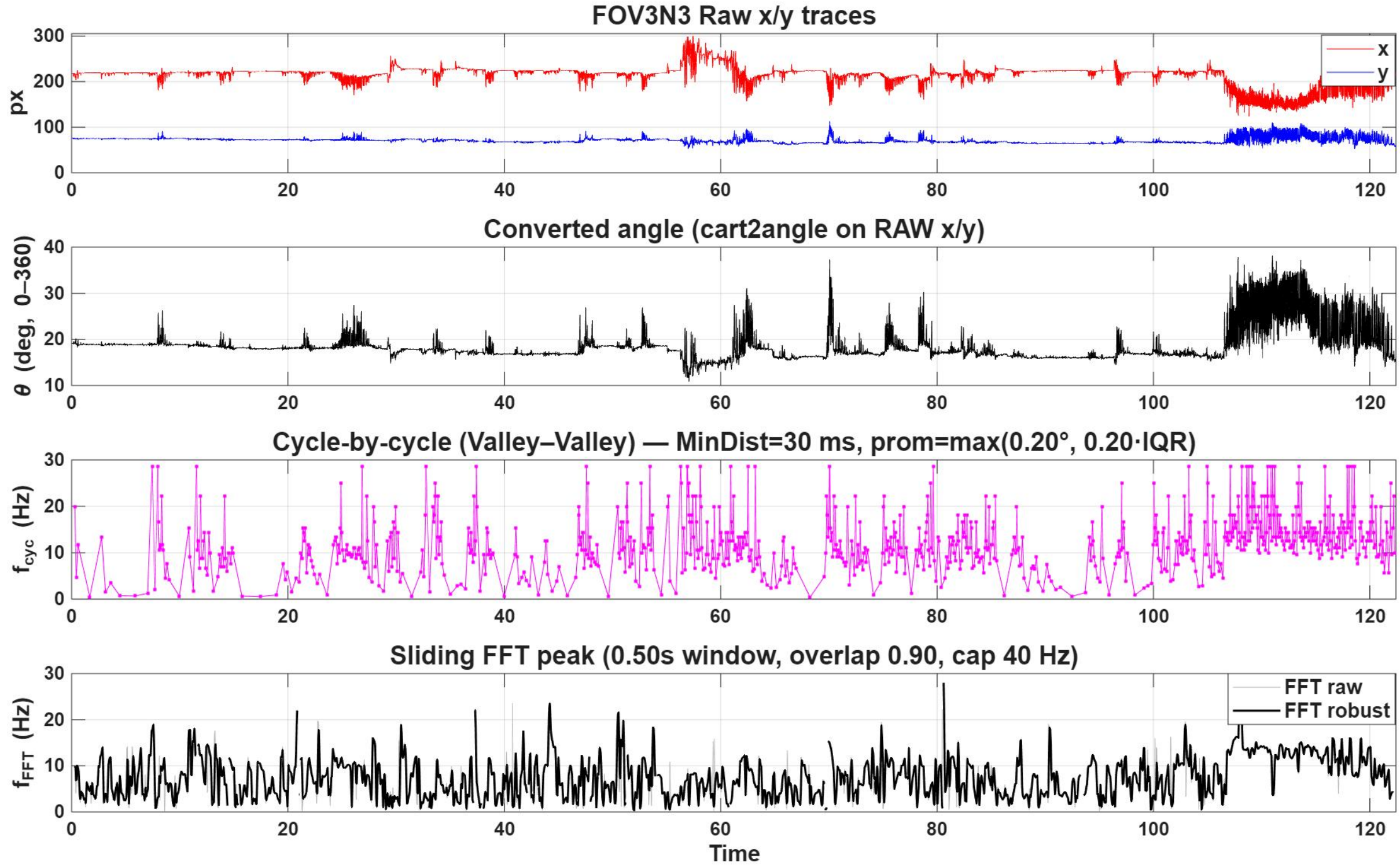

FOV3N3 Raw x/y traces
px
x
y
Converted angle (cart2angle on RAW x/y)
θ (deg, 0–360)
Cycle-by-cycle (Valley–Valley) — MinDist=30 ms, prom=max(0.20°, 0.20·IQR)
$f_{cyc}$ (Hz)
Sliding FFT peak (0.50s window, overlap 0.90, cap 40 Hz)
$f_{FFT}$ (Hz)
FFT raw
FFT robust
Time

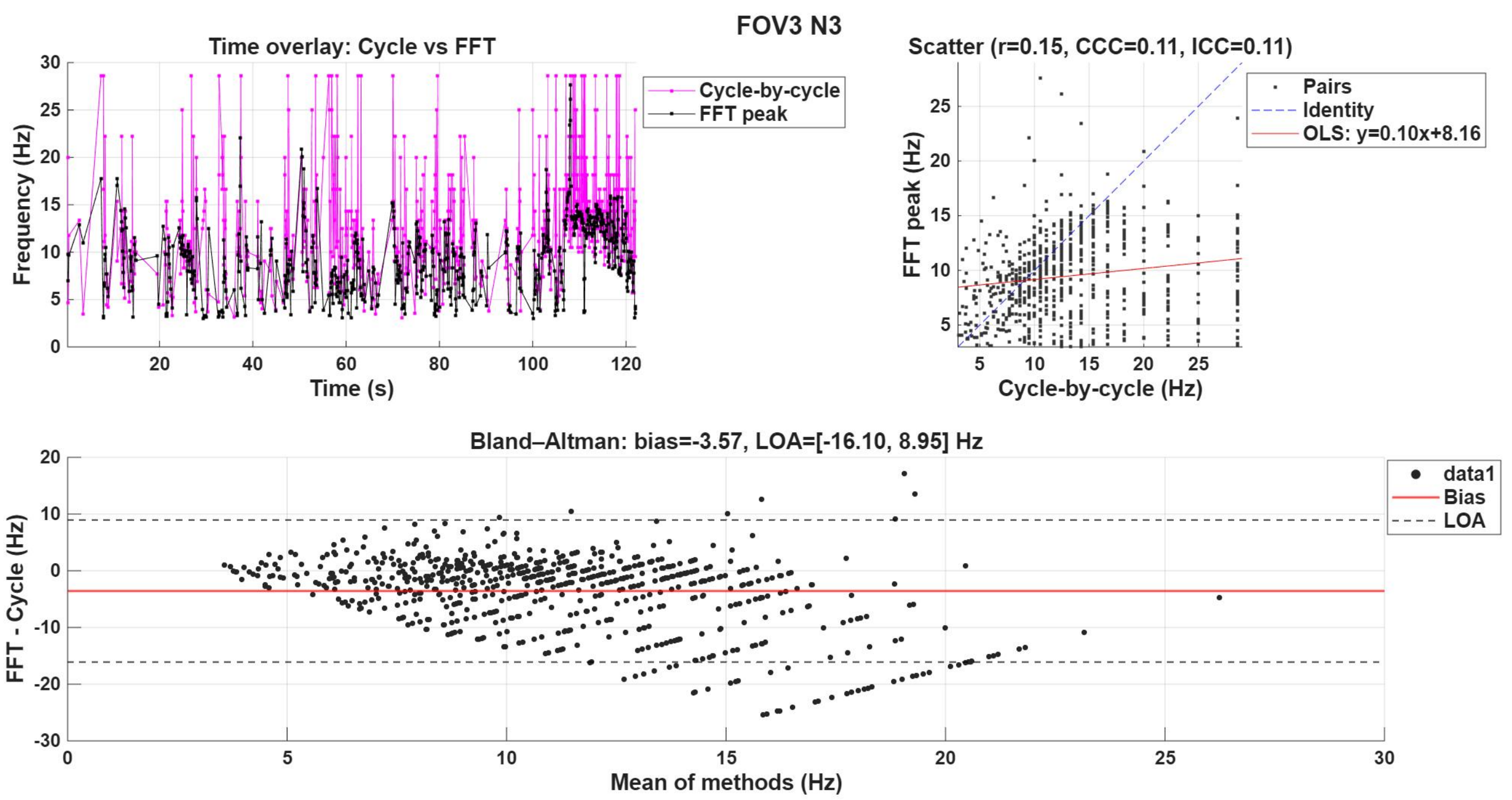

FOV3 N3
Time overlay: Cycle vs FFT
Frequency (Hz)
Time (s)
Cycle-by-cycle
FFT peak
Scatter (r=0.15, CCC=0.11, ICC=0.11)
FFT peak (Hz)
Cycle-by-cycle (Hz)
Pairs
Identity
OLS: y=0.10x+8.16
Bland–Altman: bias=-3.57, LOA=[-16.10, 8.95] Hz
FFT - Cycle (Hz)
Mean of methods (Hz)
data1
Bias
LOA

As shown in Figure J, across datasets, the sliding FFT estimator exhibited a consistent negative bias relative to the cycle-by-cycle method (~−5 Hz), accompanied by substantial error (RMSE ≈ 7–9 Hz; MAE ≈ 5–7 Hz). Correlation between the two estimators remained weak (Pearson r < 0.2), and agreement metrics were extremely low (CCC ≈ 0.05; ICC ≈ 0.05), indicating that the methods are not interchangeable. And those results showed that FFT estimator systematically underestimates whisking frequency relative to the cycle-based method, the error magnitude is large relative to the physiological range (4–28 Hz). The detailed statistics are summarized in Table I.

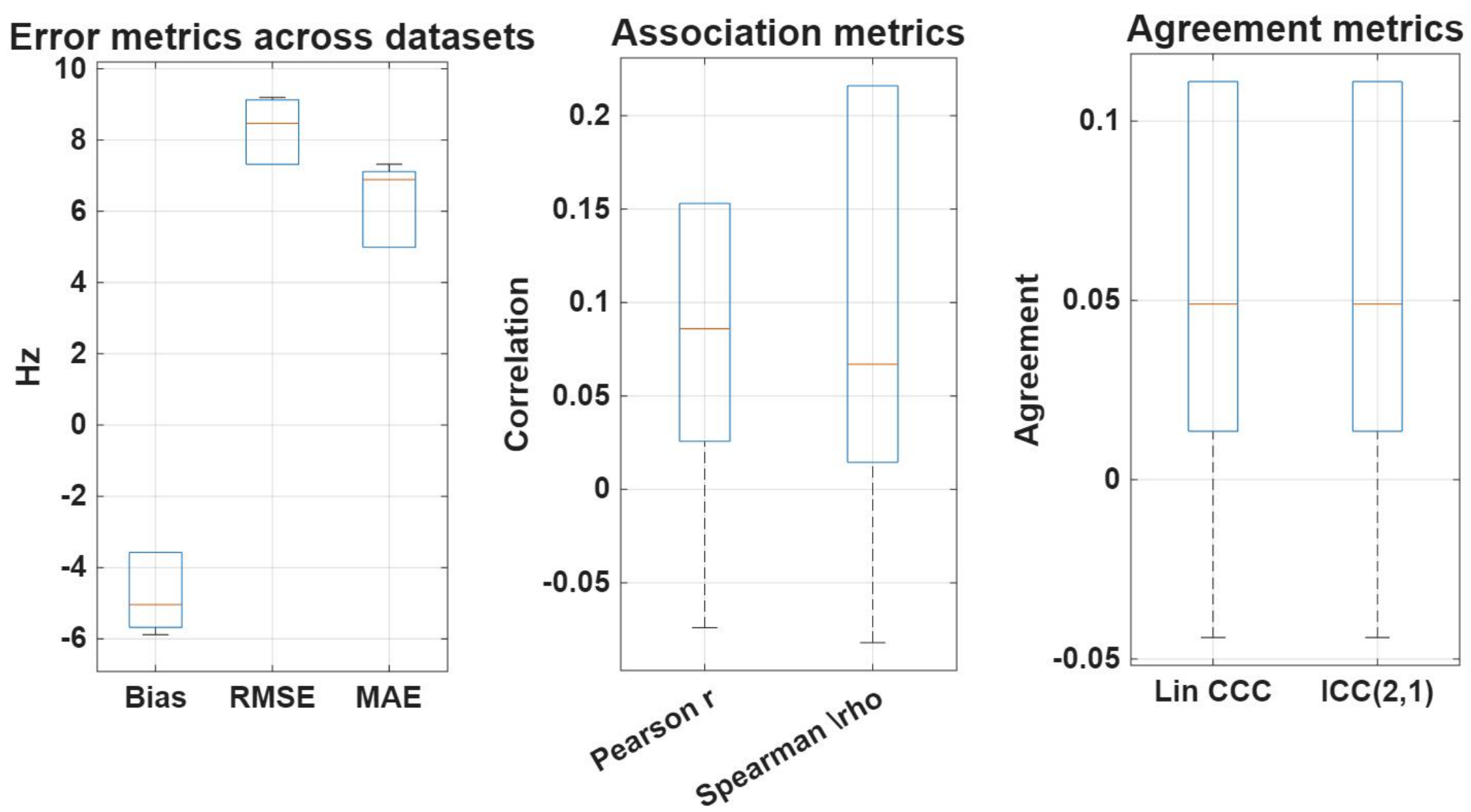

**Figure J. Sliding FFT vs Cycle-by-Cycle agreement across datasets.** Across 9 datasets, the sliding FFT estimator systematically underestimated cycle-by-cycle frequency, with a mean bias of -4.84 Hz (SD 1.00). Agreement between methods was poor, as reflected by low Lin's CCC (mean 0.048) and ICC(2,1) (mean 0.048). Error remained substantial (RMSE 8.28 +- 0.85 Hz; MAE 6.30 +- 1.03 Hz), and correlation between estimators was weak (Pearson r 0.070 +- 0.078). These results indicate that the FFT-based and cycle-by-cycle estimators are not interchangeable.

Table I: Results of Method Comparison across Datasets.

| Dataset | N_pairs | Bias_FFT_minus_Cycle_Hz | RMSE_Hz | MAE_Hz | Pearson_r | Spearman_rho | Lin_CCC | ICC_2_1 | LOA_low_Hz | LOA_high_Hz |
|---|---|---|---|---|---|---|---|---|---|---|
| F0V3N3 | 722 | −3.575 | 7.318 | 4.990 | 0.153 | 0.216 | 0.111 | 0.111 | −16.099 | 8.949 |
| F0V3N2 | 722 | −3.575 | 7.318 | 4.990 | 0.153 | 0.216 | 0.111 | 0.111 | −16.099 | 8.949 |
| F0V3N1 | 744 | −5.039 | 8.872 | 6.928 | −0.074 | −0.082 | −0.044 | −0.044 | −19.360 | 9.281 |
| F0V2S1 | 198 | −4.956 | 7.795 | 6.243 | 0.086 | 0.067 | 0.050 | 0.050 | −16.779 | 6.866 |
| F0V2N4 | 722 | −3.575 | 7.318 | 4.990 | 0.153 | 0.216 | 0.111 | 0.111 | −16.099 | 8.949 |
| F0V1S4 | 547 | −5.798 | 9.110 | 7.058 | 0.089 | 0.095 | 0.049 | 0.049 | −19.583 | 7.988 |
| F0V1N2 | 211 | −5.638 | 9.198 | 7.321 | 0.033 | 0.004 | 0.019 | 0.019 | −19.916 | 8.640 |

| Dataset | N_pairs | Bias_FFT _minus _Cycle_Hz | RMSE _Hz | MAE _Hz | Pearson _r | Spearman_rho | Lin _CCC | ICC_2 _1 | LOA_low _Hz | LOA_ high _Hz |
|---|---|---|---|---|---|---|---|---|---|---|
| FOV1N1 | 323 | -5.525 | 9.170 | 7.275 | 0.010 | 0.018 | 0.006 | 0.006 | -19.892 | 8.843 |

III. FFT amplitude spectra were computed for each whisker trajectory (single whisker labeling) after conversion from Cartesian coordinates to angular displacement. This allowed comparison of spectral consistency across datasets and visualization of whether whisking-related energy was concentrated within the expected physiological band. The spectral properties across multiple recordings were summarized in Figure K.

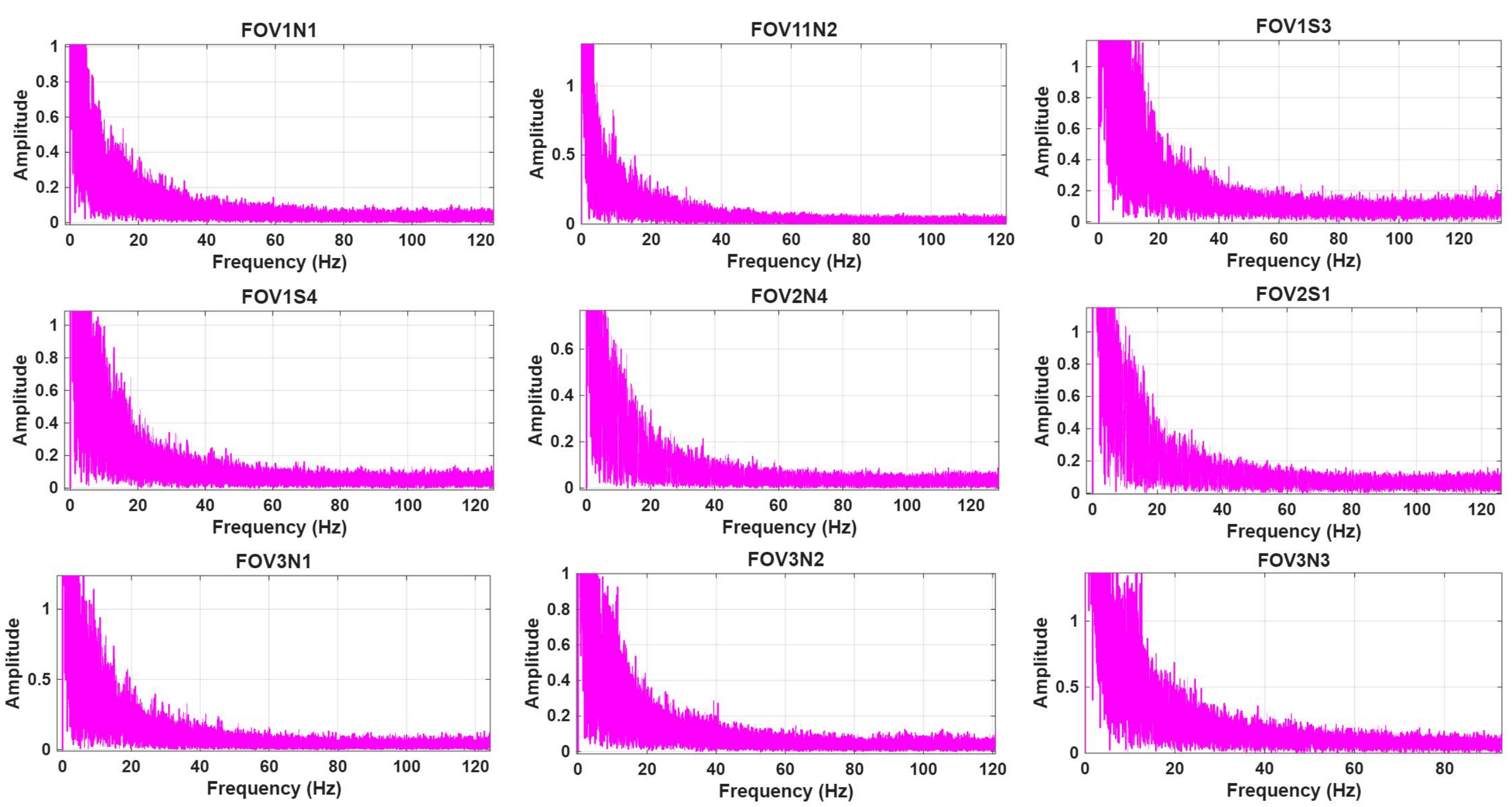

**Figure K. FFT plots to enable comparison of spectral magnitude across 9 recordings.**

IV. A multi-stage ripple-rejection strategy was implemented by progressively relaxing the valley-detection prominence threshold from prom = max(1.0°, 1.0 • IQR) to 0.5°, 0.3°, and finally 0.1°. As shown in Figure A-H, across the datasets, lowering the prominence threshold admits ripple-like micro-deflections that artificially shorten cycle intervals, leading to spurious high-frequency estimates. These findings demonstrate that frequencies above 30 Hz are not physiologically meaningful but instead arise from noise-driven artifacts, and highlight the importance of prominence-based filtering for robust whisking frequency estimation.

**A**

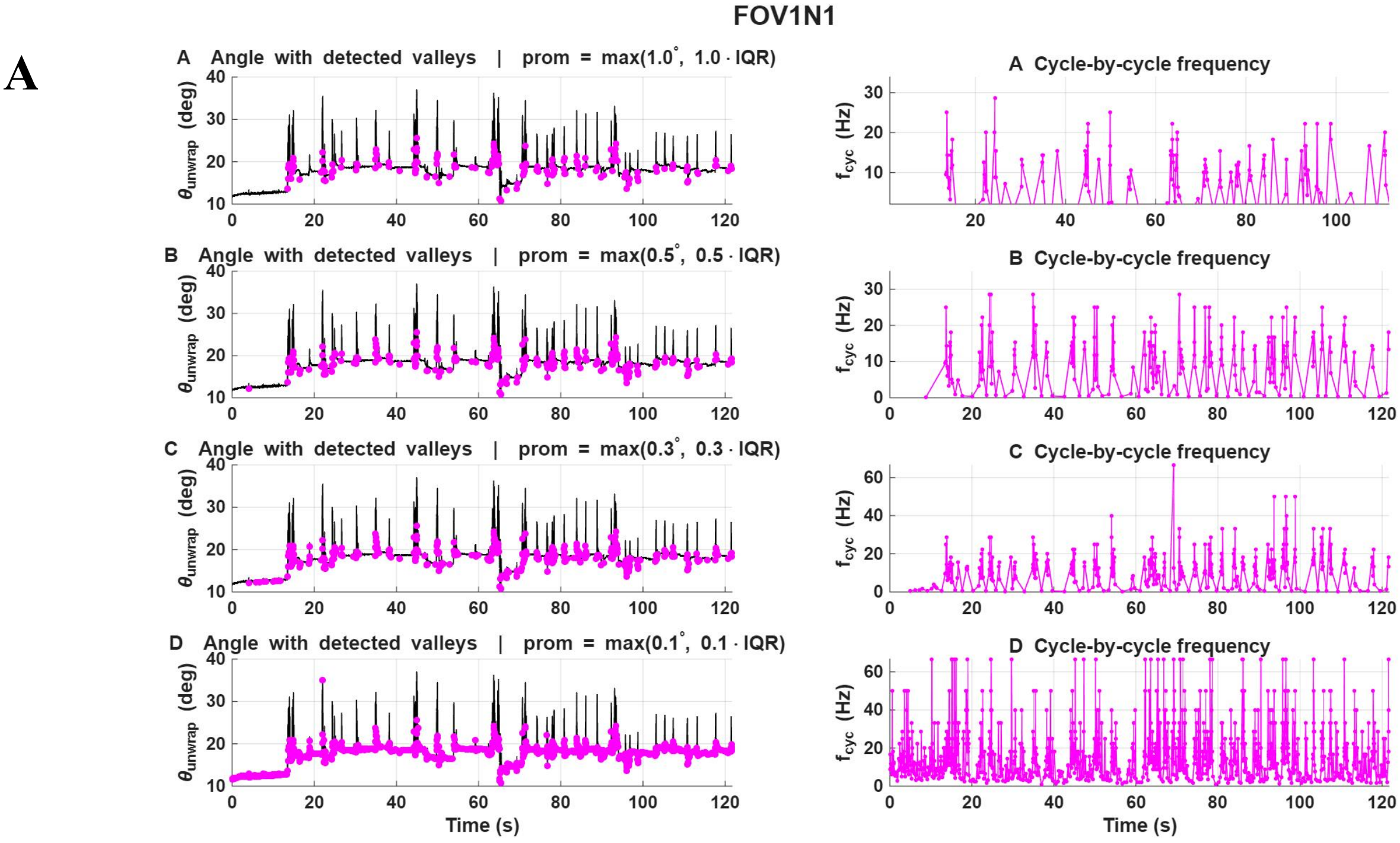

**B**

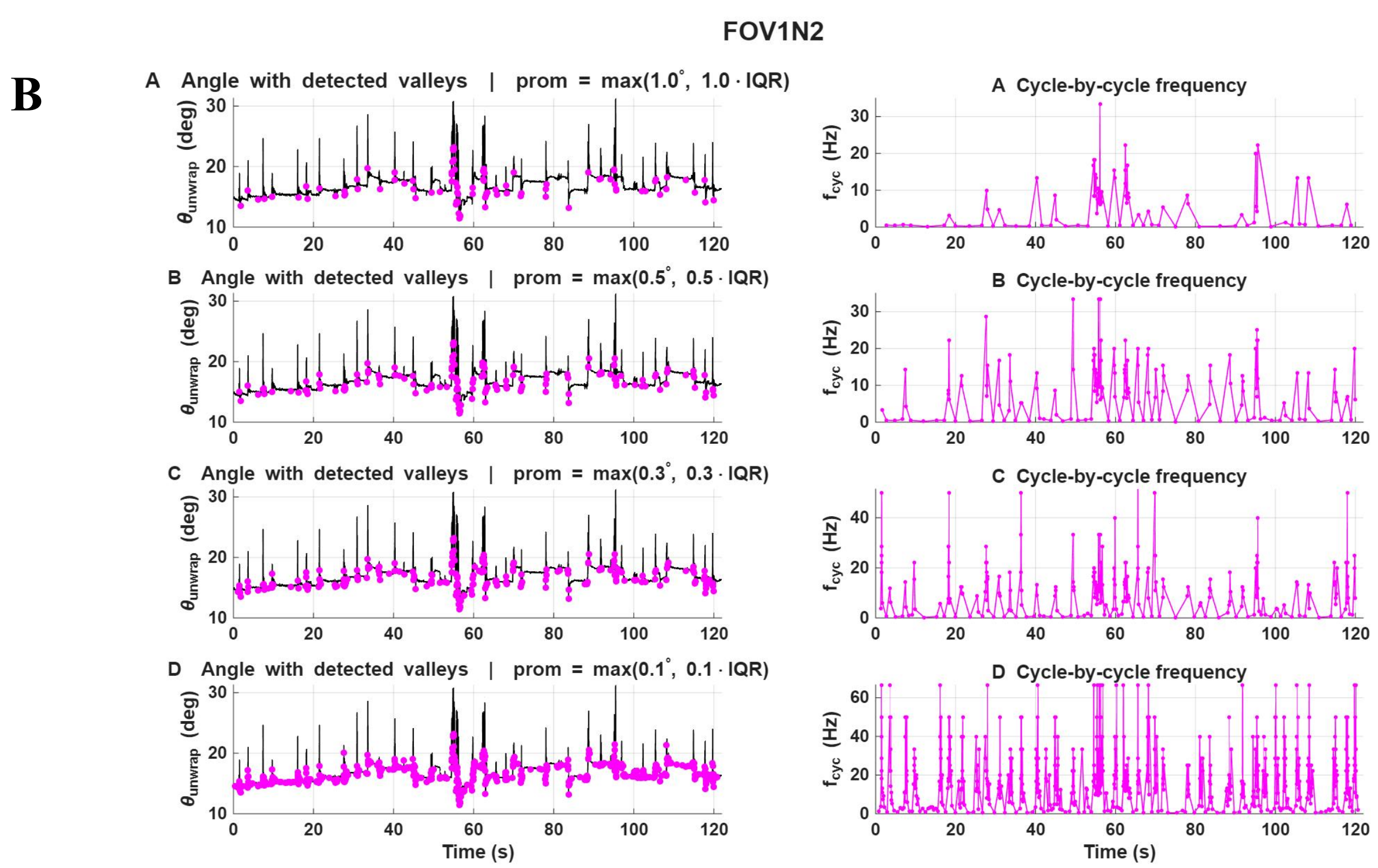

FOV1S3

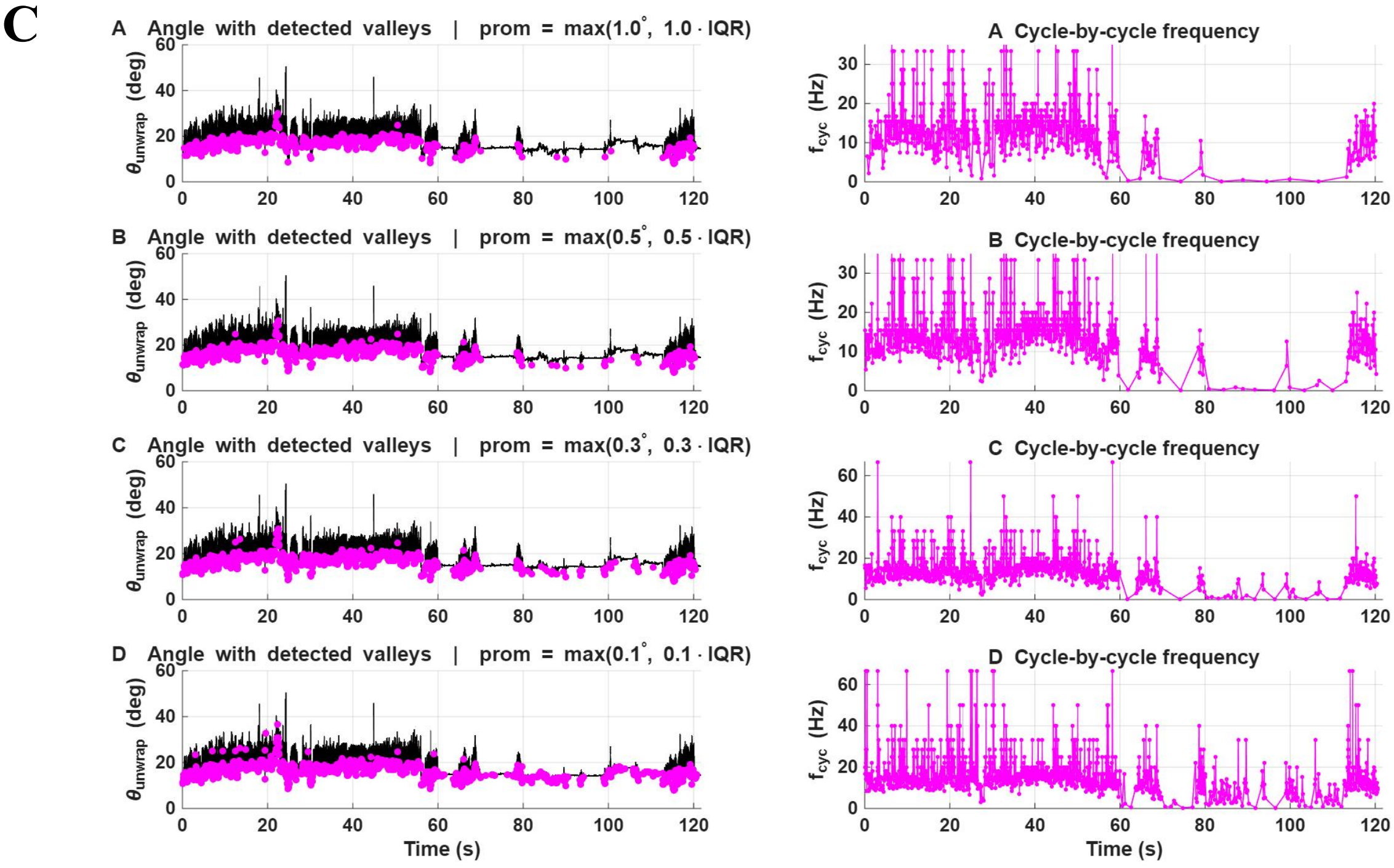
C
A Angle with detected valleys | prom = max(1.0°, 1.0 · IQR)
B Angle with detected valleys | prom = max(0.5°, 0.5 · IQR)
C Angle with detected valleys | prom = max(0.3°, 0.3 · IQR)
D Angle with detected valleys | prom = max(0.1°, 0.1 · IQR)
$\theta_{unwrap}$ (deg)
Time (s)
A Cycle-by-cycle frequency
B Cycle-by-cycle frequency
C Cycle-by-cycle frequency
D Cycle-by-cycle frequency
$f_{cyc}$ (Hz)
Time (s)

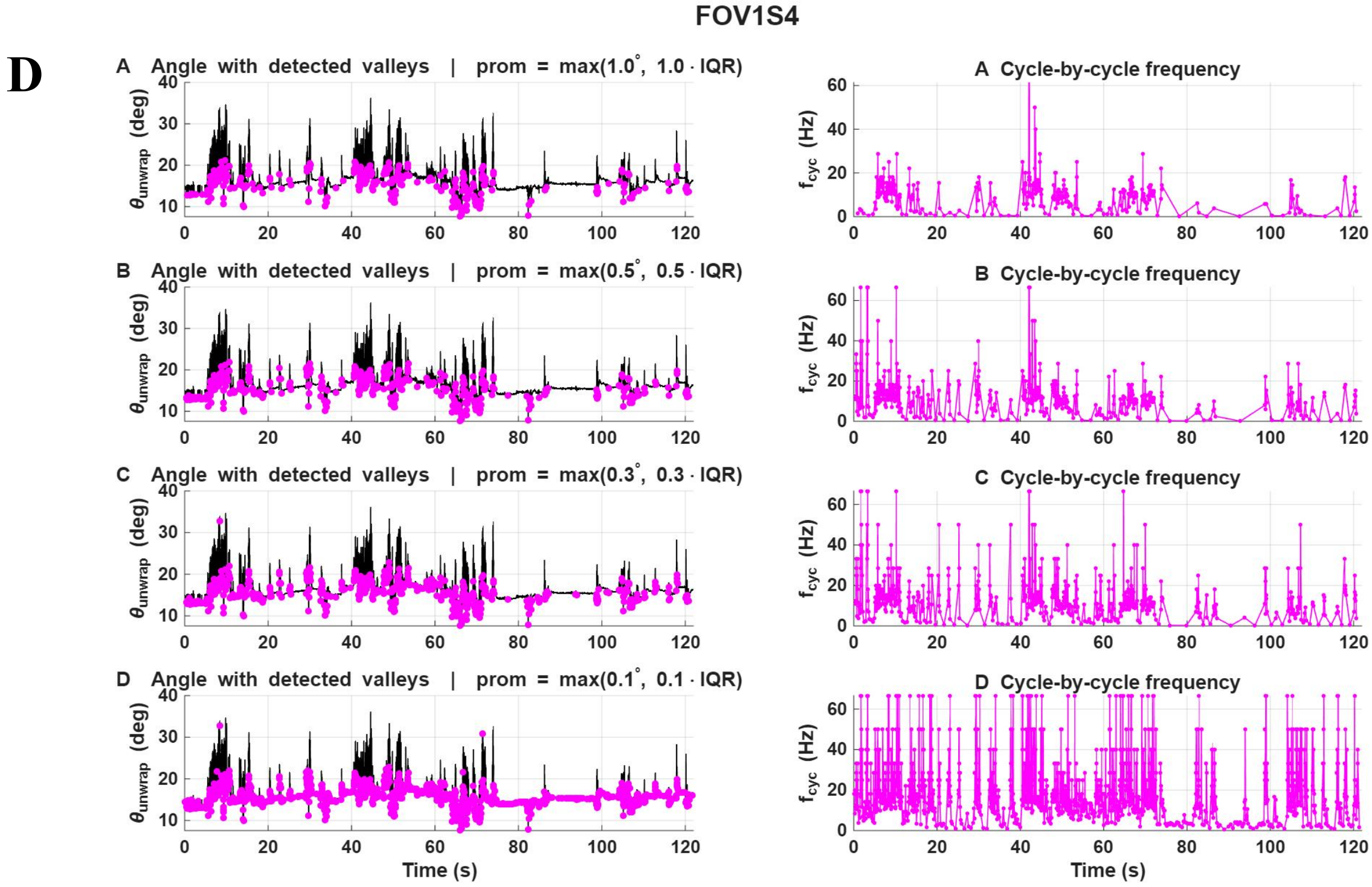
FOV1S4
D
A Angle with detected valleys | prom = max(1.0°, 1.0 · IQR)
B Angle with detected valleys | prom = max(0.5°, 0.5 · IQR)
C Angle with detected valleys | prom = max(0.3°, 0.3 · IQR)
D Angle with detected valleys | prom = max(0.1°, 0.1 · IQR)
$\theta_{unwrap}$ (deg)
Time (s)
A Cycle-by-cycle frequency
B Cycle-by-cycle frequency
C Cycle-by-cycle frequency
D Cycle-by-cycle frequency
$f_{cyc}$ (Hz)
Time (s)

FOV2N4

**E**

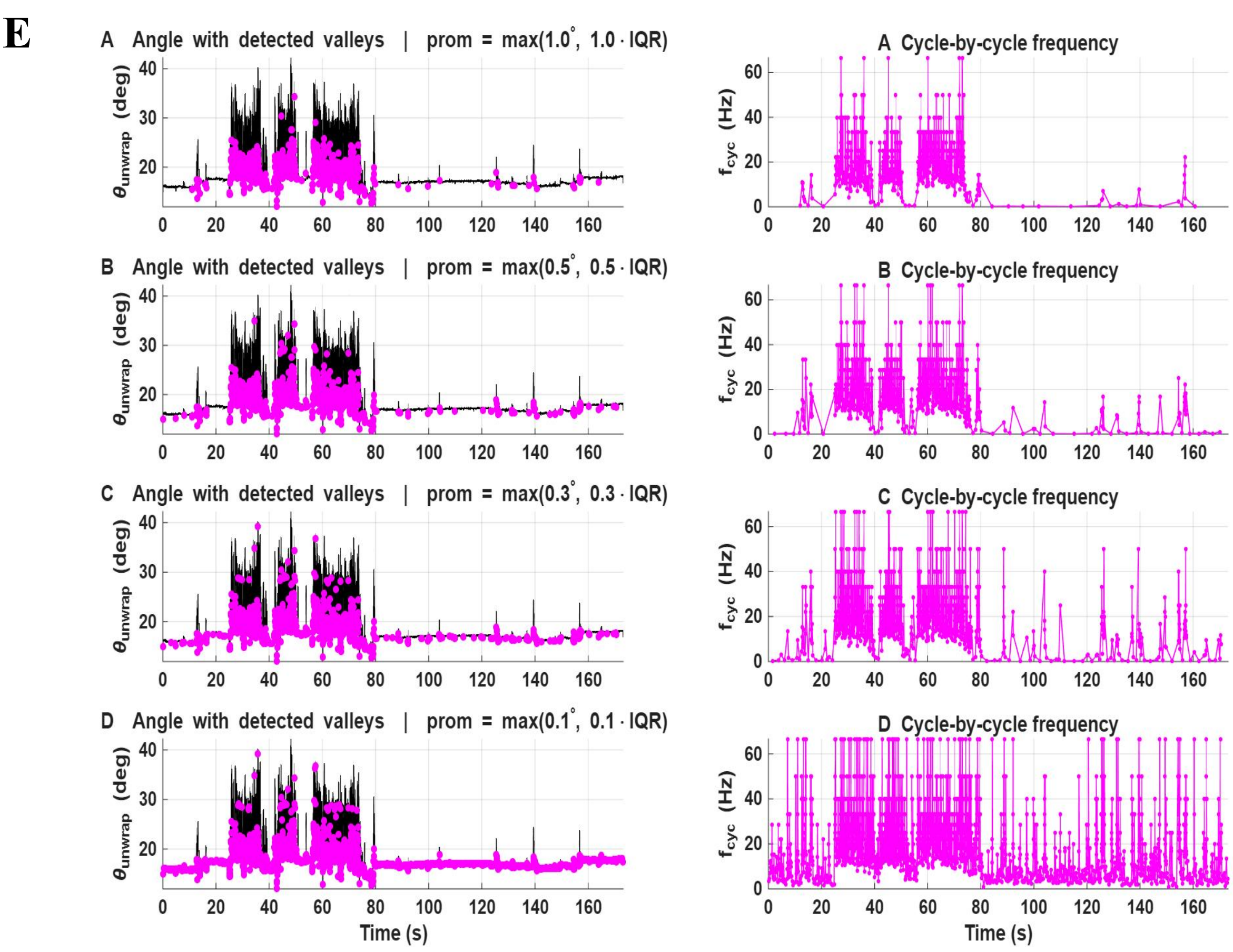

A Angle with detected valleys | prom = max(1.0°, 1.0 · IQR)
B Angle with detected valleys | prom = max(0.5°, 0.5 · IQR)
C Angle with detected valleys | prom = max(0.3°, 0.3 · IQR)
D Angle with detected valleys | prom = max(0.1°, 0.1 · IQR)
θunwrap (deg)
Time (s)
A Cycle-by-cycle frequency
B Cycle-by-cycle frequency
C Cycle-by-cycle frequency
D Cycle-by-cycle frequency
fcyc (Hz)
Time (s)

FOV2S1

**F**

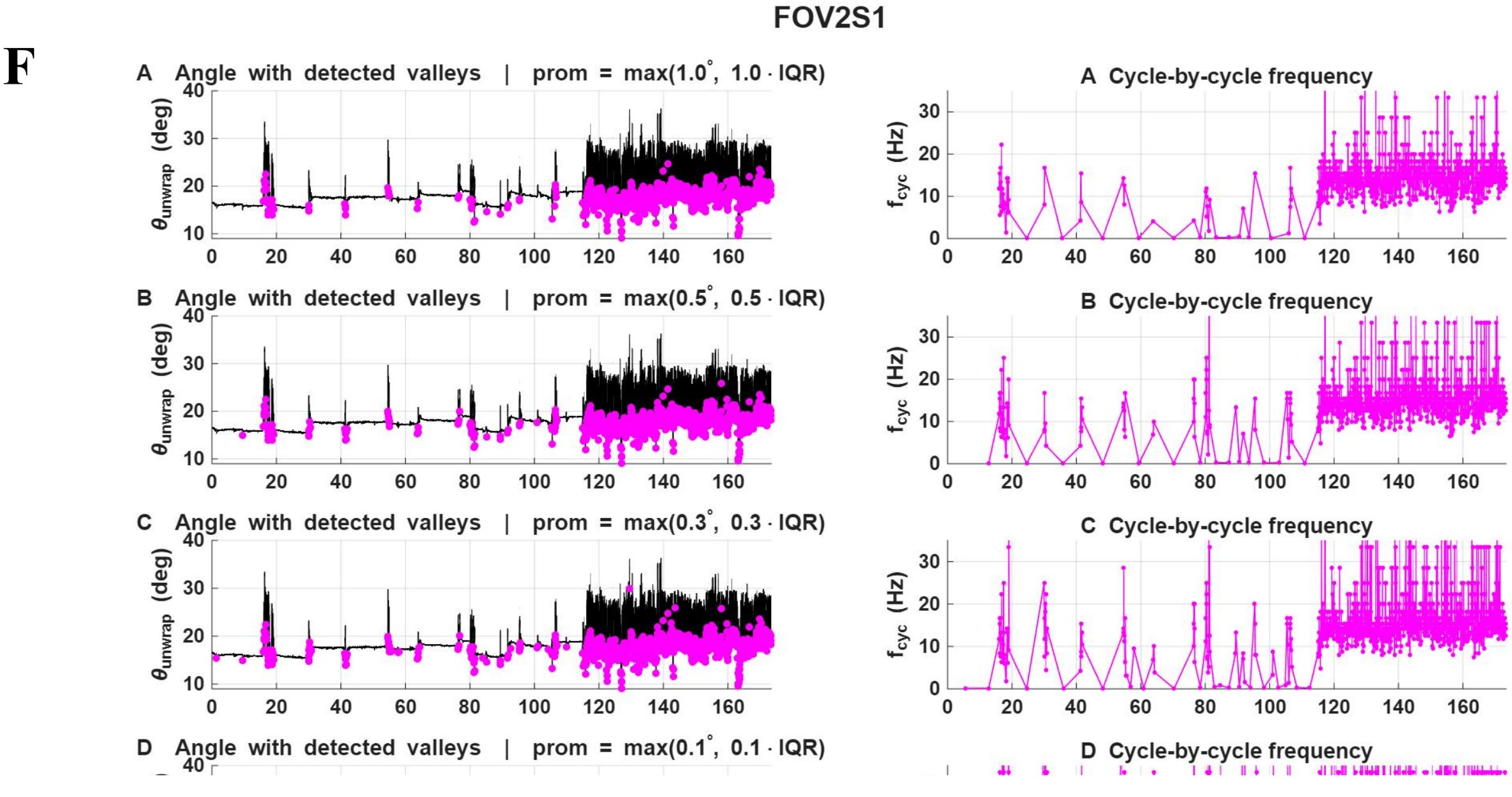

A Angle with detected valleys | prom = max(1.0°, 1.0 · IQR)
B Angle with detected valleys | prom = max(0.5°, 0.5 · IQR)
C Angle with detected valleys | prom = max(0.3°, 0.3 · IQR)
D Angle with detected valleys | prom = max(0.1°, 0.1 · IQR)
θunwrap (deg)
A Cycle-by-cycle frequency
B Cycle-by-cycle frequency
C Cycle-by-cycle frequency
D Cycle-by-cycle frequency
fcyc (Hz)

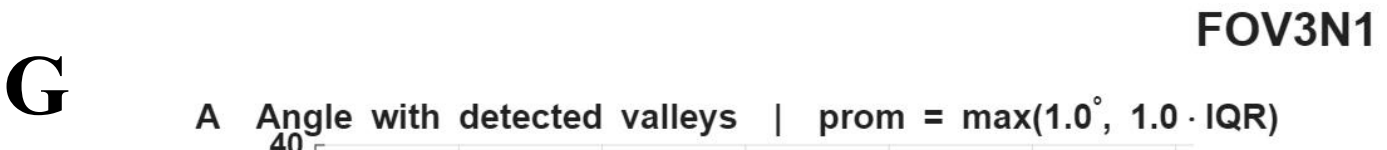

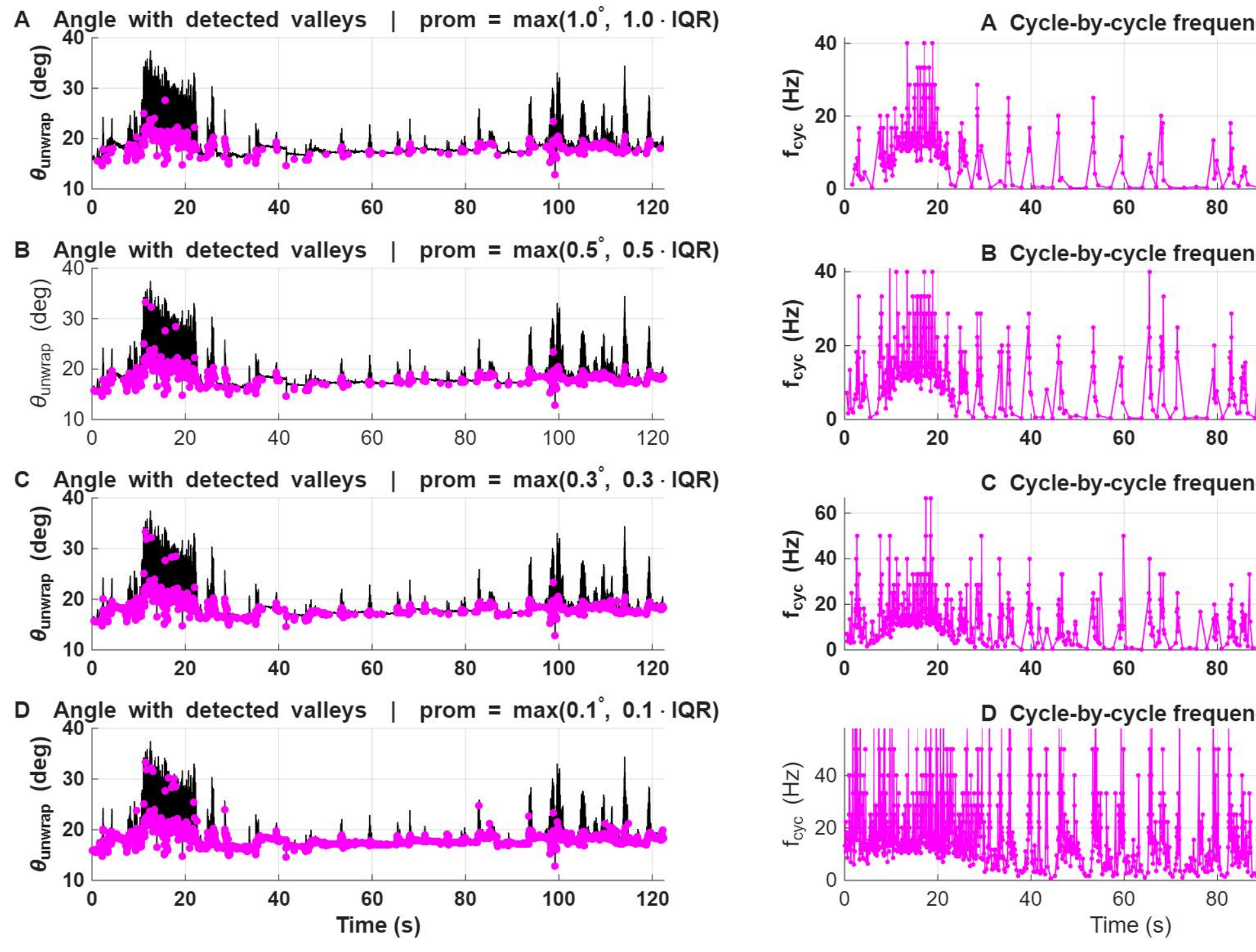
G
FOV3N1
A Angle with detected valleys | prom = max(1.0°, 1.0 · IQR)
B Angle with detected valleys | prom = max(0.5°, 0.5 · IQR)
C Angle with detected valleys | prom = max(0.3°, 0.3 · IQR)
D Angle with detected valleys | prom = max(0.1°, 0.1 · IQR)
A Cycle-by-cycle frequency
B Cycle-by-cycle frequency
C Cycle-by-cycle frequency
D Cycle-by-cycle frequency
θunwrap (deg)
fcyc (Hz)
Time (s)

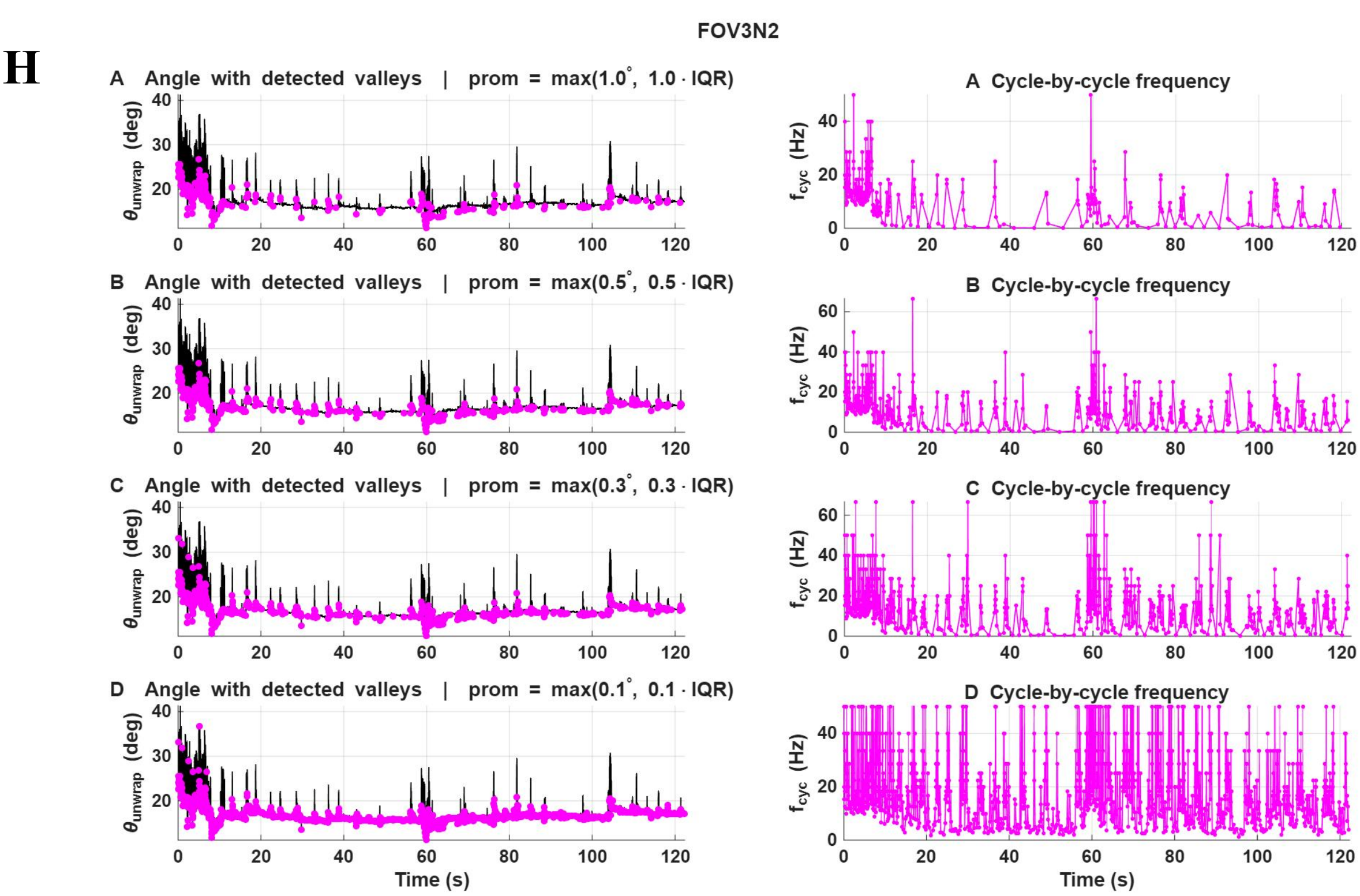
H
FOV3N2
A Angle with detected valleys | prom = max(1.0°, 1.0 · IQR)
B Angle with detected valleys | prom = max(0.5°, 0.5 · IQR)
C Angle with detected valleys | prom = max(0.3°, 0.3 · IQR)
D Angle with detected valleys | prom = max(0.1°, 0.1 · IQR)
A Cycle-by-cycle frequency
B Cycle-by-cycle frequency
C Cycle-by-cycle frequency
D Cycle-by-cycle frequency
θunwrap (deg)
fcyc (Hz)
Time (s)

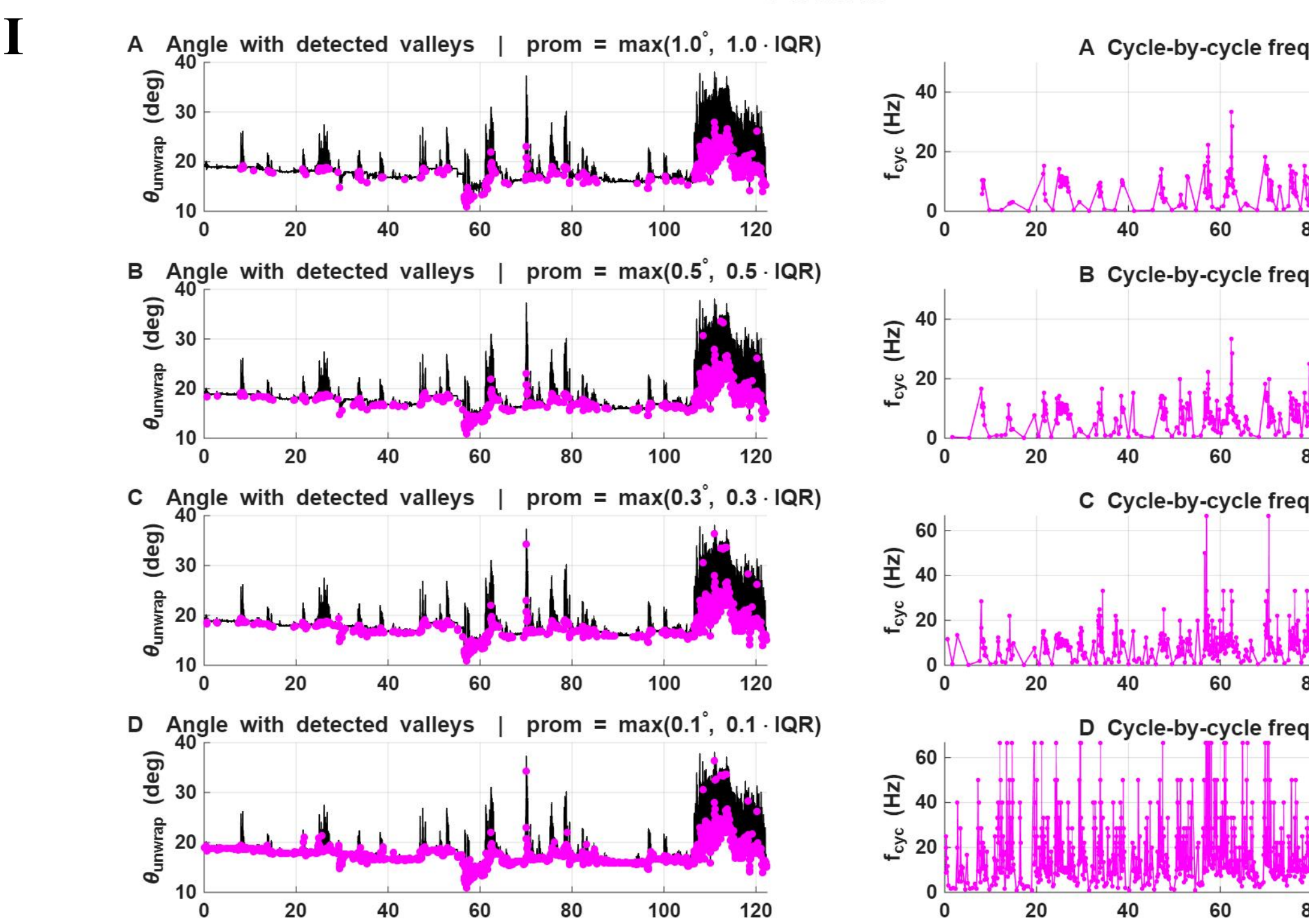

I
FOV3N3
A Angle with detected valleys | prom = max(1.0°, 1.0 · IQR)
B Angle with detected valleys | prom = max(0.5°, 0.5 · IQR)
C Angle with detected valleys | prom = max(0.3°, 0.3 · IQR)
D Angle with detected valleys | prom = max(0.1°, 0.1 · IQR)
$\theta_{unwrap}$ (deg)
Time (s)
A Cycle-by-cycle frequency
B Cycle-by-cycle frequency
C Cycle-by-cycle frequency
D Cycle-by-cycle frequency
$f_{cyc}$ (Hz)
Time (s)

To quantify the contribution of ripple-like fluctuations to apparent high-frequency whisking, we applied the valley-by-valley detector to multiple datasets using four progressively relaxed prominence settings (Figure J). The statistics were summarized in Table I.

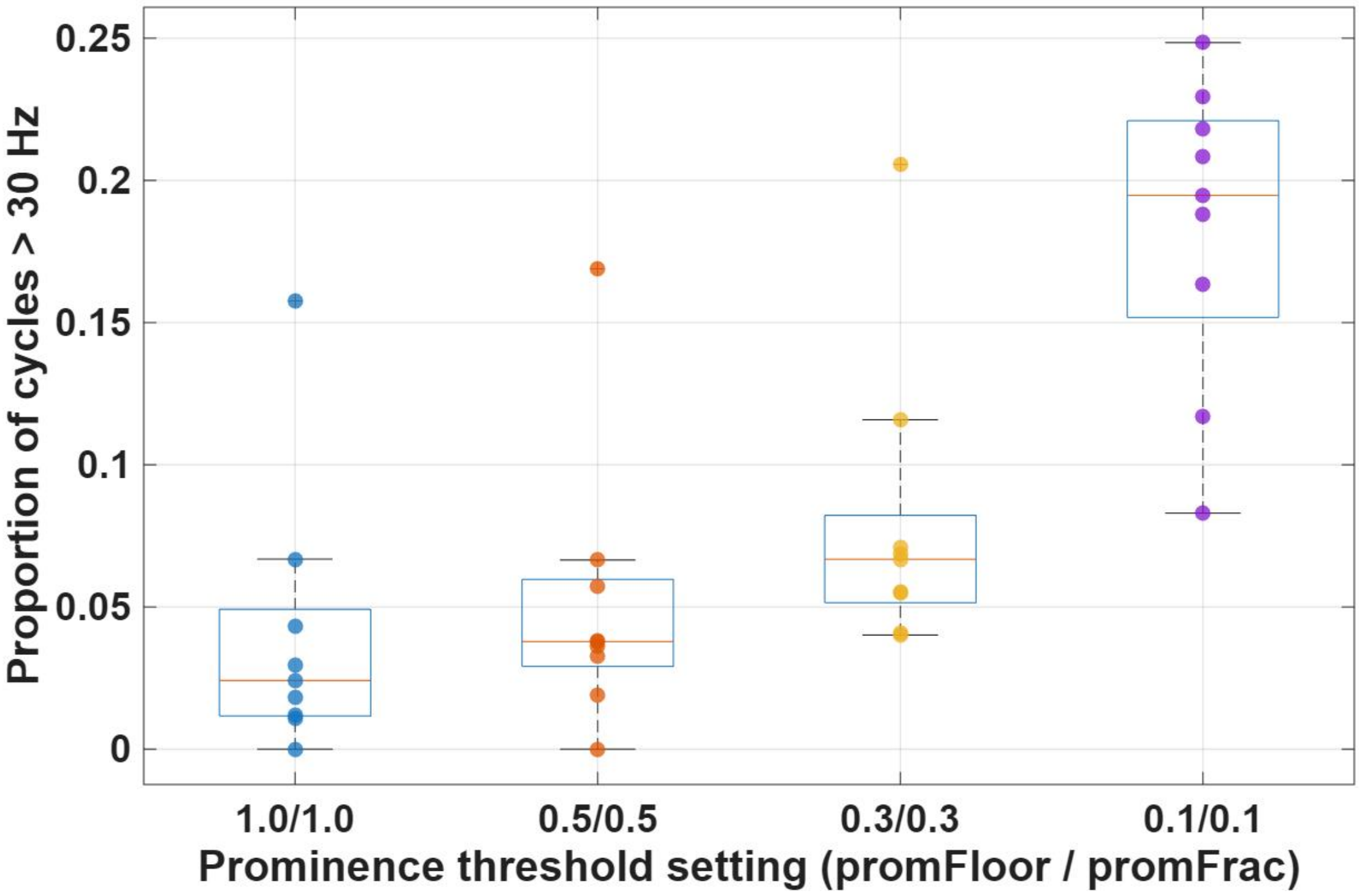

**Figure J. loosening the prominence threshold admits small ripples and increases artifactual high frequency detections.** Across datasets, the proportion of detected frequencies above 30 Hz increased systematically as the prominence threshold was relaxed, rising from approximately 4% under stringent conditions (prom = max(1.0$^0$, 1.0·IQR)) to ~5% (0.5$^0$), ~8% (0.3$^0$), and ~18% under highly permissive settings (0.1$^0$), indicating that high frequency detections are strongly dependent on the inclusion of small-amplitude signal fluctuations.

**Table I**: **Effect of prominence threshold on high-frequency (>30 Hz) detections.**

| Condition | Prominence Definition | Mean_Proportion_gt_30Hz | SD | Median |
|---|---|---|---|---|
| A (Strict) | max(1.0°, 1.0•IQR) | 0.0403 | 0.0483 | 0.0242 |
| B (Moderate) | max(0.5°, 0.5•IQR) | 0.0507 | 0.0483 | 0.0379 |
| C (Relaxed) | max(0.3°, 0.3•IQR) | 0.0799 | 0.0523 | 0.0668 |
| D (Permissive) | max(0.1°, 0.1•IQR) | 0.1834 | 0.0539 | 0.1948 |